\documentclass[12pt]{article}

\usepackage[T1]{fontenc}
\usepackage{tipa}
\usepackage{mathtools} 
\usepackage{semantic}
\usepackage{cancel}
\mathlig{->}{\rightarrow}
\mathlig{|->}{\mapsto}
\usepackage{breqn}
\usepackage{graphicx}
\usepackage{epstopdf}
\usepackage{amsmath}
\usepackage{amssymb}
\usepackage{amsfonts}
\usepackage{amsthm}
\usepackage[usenames]{color}
\usepackage{array}
\usepackage[percent]{overpic}

\usepackage[
colorlinks={},
linkcolor=blue,
urlcolor=blue,
filecolor=blue,
citecolor=blue,
pdfstartview=FitV,
pdftitle={},
pdfauthor={},
pdfsubject={},
pdfkeywords={},
pdfpagemode=None,
bookmarksopen=true
,linktoc=page]{hyperref}

\usepackage{float}
\usepackage[utf8]{inputenc}
\usepackage{epsfig}
\usepackage{textcomp}
\usepackage{setspace}

\usepackage{comment}
\usepackage{amsmath,amssymb,extarrows,mathtools,graphicx,subfigure,setspace}

\usepackage{slashed}
\usepackage{color}
\usepackage{tikz}
\usepackage{fancyhdr}
\usetikzlibrary{decorations.pathmorphing}
\makeatother

\newcommand{\bea}{\begin{eqnarray}}
\newcommand{\eea}{\end{eqnarray}}
\newcommand{\ba}{\begin{array}}
	\newcommand{\ea}{\end{array}}
\newcommand{\ee}{\end{equation}}

\numberwithin{equation}{section}

\begin{document}
\begin{flushright}
	\texttt{}
\end{flushright}

\begin{centering}
\thispagestyle{empty}

	\textbf{\Large{
Firewalls from wormholes in higher genus }}
	
	\vspace{1.4cm}
	
	{\large    Hamed Zolfi }
	
	\vspace{0.9cm}
	
	\begin{minipage}{.9\textwidth}\small
		\begin{center}
			
	{
 School of Particles and Accelerators,
					Institute for Research in Fundamental Sciences (IPM)
					P.O. Box 19395-5531, Tehran, Iran	}

		\vspace{0.9cm}
			{\tt  \ hamedzolphy@ipm.ir }
		\\ 
			
		\end{center}
	\end{minipage}
	\vspace{2cm}

	\begin{abstract}
 An old black hole can tunnel into a white hole/ firewall by emitting large baby universes. This phenomenon was investigated in Jackiw-Teitelboim (JT) gravity for genus one.
In this paper, the focus is on higher genus  corresponding to emitting more than one baby universe ($n > 1$). The probability of encountering a firewall or tunneling into a white hole after emitting $n$ baby universes is proportional to $e^{-2nS(E)}e^{4 \pi \sqrt{E}(n-1)}E^{2n^2-n-9/2}t^{4n^2-2n-5}$, where $t$ is the age of the black hole, and $S$ and $E$ represent the entropy and energy of the black hole, respectively.

	\end{abstract}
\end{centering}
\newpage
\doublespacing
\tableofcontents
\setstretch{1.1}
\setcounter{equation}{0}
\setcounter{page}{2}
\newpage
\section{Introduction}
It has been argued that large AdS black holes in late times must possess some structures at the horizon, often referred to as a fuzzball or firewall \cite{Almheiri:2012rt,Almheiri:2013hfa,Mathur:2009hf,Bousso:2012as,VanRaamsdonk:2013sza,Nomura:2012sw,Verlinde:2012cy,Papadodimas:2012aq,Shenker:2013yza, Czech:2012be}.
Firewalls exhibit similar characteristics to white holes. White holes may not be common in nature, but they are just as abundant as black holes in the Hilbert space that describes systems with entropy $S$.
In the subsequent part of this section, we will review the resemblance between a firewall and a white hole and the phenomenon of tunneling a black hole into a white hole.

\subsection{Firewall and white hole}
\begin{figure}[h]	
	\centering
	\resizebox{9cm}{8cm}{	
		\begin{tikzpicture}
			\draw [black,very thick](0,-4) --(0,4);
			\draw [black,very thick,decoration={zigzag, segment length=3mm,amplitude=1mm},decorate](-8,4) --(0,4);
			\draw [black,very thick,decoration={zigzag, segment length=3mm,amplitude=1mm},decorate](0,-4) --(-8,-4);
			\draw [black,very thick](-8,-4.0) --(-8,4);
			\draw [black,thick](-8,-4) --(0,4);
			\draw [black,thick](-8,4) --(0,-4);
			\node at (.7,1.7) {$\scalebox{1.3}{$\uparrow t_R$}$};
			\node at (-8.6,1.7) {$\scalebox{1.3}{$ t_L \uparrow$}$};
			\node at (-2,1.4) {$\scalebox{1.3}{$r_h$}$};
			\node at (-6,1.4) {$\scalebox{1.3}{$r_h$}$};
			\node at (-3.7,4.6) {$\scalebox{1.3}{$r=0$}$};
			\node at (-3.7,-4.6) {$\scalebox{1.3}{$r=0$}$};
			\node at (-4,1) {$\text{I}$};
			\node at (-4,-1) {$ \text{II}$};
			\node at (1,0) {$\scalebox{1.3}{$r\rightarrow\infty$}$};
			\node at (-9.0,0) {$\scalebox{1.3}{$r\rightarrow\infty$}$};
			\draw [red,very thick](0,4) to [out=190,in=350] (-8,4);
			\draw [red,very thick](0,2) to [out=190,in=350] (-8,2);
			\draw [red,very thick](0,3) to [out=190,in=350] (-8,3);
			\draw [blue,very thick](0,-4) to [in=13,out=167] (-8,-4);
			\draw [blue,very thick](0,-3) to [in=13,out=167] (-8,-3);
			\draw [blue,very thick](0,-2) to [in=13,out=167] (-8,-2);
	\end{tikzpicture}}
	\caption{ Penrose diagram of eternal AdS black/white hole. 
		In region I, located behind the black hole horizon, the spacetime exhibits expansion. As a result, the black hole horizon is transparent. On the other hand, in region II, behind the white hole horizon, the spacetime undergoes contraction, and the horizon is opaque.}
	\label{pen}
\end{figure}
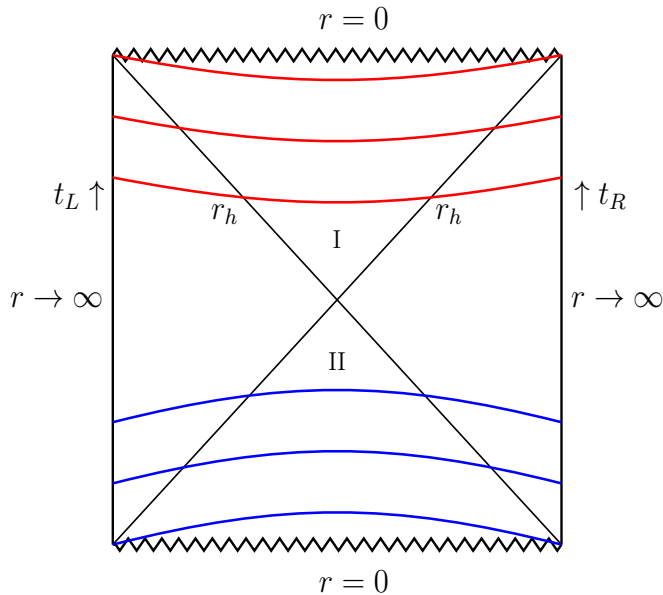

One can distinguish between two types of black hole horizons: transparent and opaque. A transparent horizon can be crossed freely by an in-falling observer, while an opaque horizon does not allow such passage \cite{Susskind:2015toa}.
In classical gravity, there are two criteria to support the notion that certain black holes may have opaque horizons. The expansion criterion states that black hole horizons are transparent when the interior geometry expands, (see figure \ref{pen}). The Page criterion suggests that black holes formed from non-singular Cauchy data, with no past singularities, have transparent horizons.
Based on these criteria, an object assumed to be a black hole with a non-transparent horizon could be a white hole. Because the white hole has a contracting interior geometry, making it impossible to enter its horizon. White holes are essentially geometries formed from singular Cauchy data, and their horizons are not transparent when past singularities are present \cite{page}.
The expansion and Page criteria are based on classical geometric terms and do not directly consider properties of the quantum state. Additionally, there exists a third criterion that solely focuses on the dual quantum state. According to this criterion, the black hole horizon will remain transparent as long as the complexity of the dual  state continues to increase \cite{Susskind:2015toa}. This criterion can be related to the expansion criterion by the complexity=volume
conjecture:
\begin{equation}
	\mathcal{C}=\max\left[ \frac{\mathcal{V}}{Gl}\right].
\end{equation}
Here, $\mathcal{V}$ is the volume of the Einstein-Rosen bridge (ERB), $G$ represents Newton's gravitational constant, and $l$ is a specific length scale associated with the geometry, typically chosen to be the AdS radius of curvature or the Schwarzschild radius.  Another potential explanation for opacity is the presence of a  firewall that effectively seals off the interior geometry \cite{Susskind:2018pmk}. It was  demonstrated that shock waves can generate firewalls \cite{Shenker:2013pqa}.
The volume of the wormhole (ERB) as a function of the times of the left and right boundaries ($t_{\text{L}}, t_{\text{R}}$) in the presence of a firewall, which arises from a shock wave at time  $t_w$ is \cite{Stanford:2014jda}:
\begin{equation}
	\mathcal{V}(t_{\text{L}},t_{\text{R}})=\text{constant} \log\left( \cosh\frac{t_{\text{L}}+t_{\text{R}}}{2}+\exp\left( |t_w|-t_*+\frac{t_{\text{L}}-t_{\text{R}}}{2}\right) \right). 
\end{equation}
Here, $t_*$ is the scrambling 
time. Assuming $t_{\text{L}}$ to be fixed and sufficiently large for simplicity, then $	\mathcal{V}(t_{\text{R}})$, or equivalently the complexity of the perturbed state is a decreasing function of $t_{\text{R}}$ for $t_{\text{R}}<|t_w|-t_*$. Consequently, the horizon is deemed to be opaque. However, for  times when $t_{\text{R}}\gg|t_w|-t_*$, the wormhole grows linearly with time, $	\mathcal{V}(t_{\text{R}},t_{\text{L}})\sim t_{\text{R}}$, and the horizon becomes transparent.
As a result, if the horizon of the black hole becomes opaque, it could indicate the presence of a firewall at the horizon or that the black hole has tunneled into a white hole.
\subsection{Firewalls from wormholes}
In classical general relativity wormhole (ERB) grows forever. 
According to this point, it was introduced a different version of the information paradox that applies to
large, eternal black holes in the context of AdS/CFT \cite{Berti:2009kk,Kokkotas:1999bd,Horowitz:1999jd,Maldacena:2001kr}.\footnote{The information paradox was addressed in the papers \cite{Penington:2019npb, Almheiri:2019psf, Almheiri:2019qdq, Penington:2019kki}.} Let us consider a thermofield double state which is dual to two-sided eternal AdS black hole. In such a system, we would like to understand the long-time behavior of
correlation functions.
In the gravity side, two point function between opposite sides behaves like:
\begin{equation}
	G_{\text{gravity}}(t):=\langle \phi_{\text{R}}(t)\phi_{\text{L}}(0)\rangle \sim e^{- \text{const}\times  \frac{\ell(t)}{\beta}}, \qquad \ell(t)\sim t,
\end{equation} 
where $\ell(t)$ is the length of the wormhole at time $t$.
The cause of the decreasing correlation function is the expanding wormhole  that separates the two sides.
Using the extrapolate dictionary:
\begin{equation}
	\mathcal{O}_{\text{R},\text{L}}(t)  =\lim_{r\rightarrow\infty}r^{-\Delta}\phi_{\text{R},\text{L}}(t,r),
\end{equation}  
the correlation function on the CFT side becomes:
\begin{equation}
	G_{\text{CFT}}(t):=\langle \text{TFD}|\mathcal{O}_{\text{R}}(t)\mathcal{O}_{\text{L}}(0)|\text{TFD}\rangle  =\frac{1}{Z}\sum_{i,j}e^{-\beta E_i+i(E_i-E_j)t}|\mathcal{O}|_{ij}
\end{equation}
In the CFT, any perturbation of the thermal
state (suppose $\mathcal{O}_{\text{L}}(0) $ is a perturbation on thermofield double state)  remains a perturbation of the thermal state forever. Although it undergoes scrambling and appears to thermalize, the initial perturbation is never completely forgotten. In fact, the corrections to the thermal state are finite and suppressed by the entropy for $t \gg \beta$. Therefore, we have:
\begin{equation}
	G_{\text{CFT}}(t)\sim e^{-\text{const}\times S }.
\end{equation}
So, at very late times, gravity forgets the initial perturbation, while 
a unitary CFT does not:
\begin{equation}
	G_{\text{gravity}}(t)\ll G_{\text{CFT}}(t).
\end{equation}
This reveals  where the gravity derivation deviates. Also, if we accept the complexity=volume conjecture, we will run into trouble. The problem arises from the limited number of available mutually orthogonal states, which is of the order of $\exp(S)$ (where $S$ is the entropy of the black hole). After an exponential time, we exhaust all orthogonal states, leading to a saturation of complexity, while the volume of the wormhole continues to grow indefinitely. In other words, 
we expect that 
the length of the wormhole to stop growing when $t \sim \exp (S)$, while
classical geometry does not show this behavior.

 In the context of JT gravity, it has been  demonstrated that the length of wormholes can be diminished through the process of emitting baby universes \cite{Saad:2019pqd}.\footnote{The role of wormholes in quantum gravity has been mysterious and amazing \cite{haw,lav,mal,nim,mar}. In the 1980s, Coleman, Giddings, and Strominger's research established a connection between the physics of spacetime wormholes and the concept of baby universes \cite{gid,col}. } The emission of a baby universe can have different consequences depending on its size. As we will see later; in the case where the size of the emitted baby universe is larger than the age of the black hole, it can lead to the formation of a firewall or cause the black hole to undergo a tunneling process, transforming it into a white hole.
 Stanford and Yang computed the probability of emitting a single baby universe at late times as \cite{Stanford:2022fdt}:
\begin{equation}
	P_{1,\text{firewall}}(t)\approx\frac{e^{-S\left( E\right) }}{2(2\pi)^2}t^2,
\end{equation}
where 
\begin{equation}\label{se}
	e^{S\left( E\right)}=\frac{e^{S_0+2\pi\sqrt{E}}}{2\left(2\pi \right)^{2} }, 
\end{equation}
and the subscript one corresponds to genus, representing the emission of a single baby universe.

This paper aims to extend their work by considering higher genus scenarios. To achieve this, Section \ref{2} provides a  review of the JT gravity wave function. Section \ref{3} focuses on the computations of the probabilities of finding a firewall and smooth geometry through the emission of two baby universes. In Section \ref{sec}, the probability of encountering a firewall for genus two is calculated using an alternative approach.  Section \ref{6} extends the calculations to include an arbitrary number of emitting baby universes. Finally,  concluding remarks are presented in Section \ref{4}.
\section{JT gravity wave function setup}\label{2}
In Euclidean JT gravity,  to compute partition function one should integrate over surfaces of constant negative curvature (usually
chosen to be $R =-2$). These surfaces can be decomposed into ``trumpets'' with asymptotic and geodesic boundaries and surfaces with $g$ handles and a certain number
of geodesic boundaries (which are glued to the trumpets). The partition function of a geometry with one asymptotic boundary, $ n $ geodesic boundaries (referred to as baby universes), and genus $g$,  can be written as \cite{Saad:2019lba,Stanford:2017thb}:
\begin{equation}\label{part1}
Z\left( \beta,\textbf{a}\right) =e^{S_0\chi }\int_{0}^{\infty}Z_{\text{trumpet}}\left( \beta,b\right)V_{g,n+1}(b,\textbf{a})  b\text{d}b, 
\end{equation}
here, $\chi$ and $S_0$  represent the Euler characteristic and  the ground state entropy, respectively. $Z_{\text{trumpet}}$ refers to the trumpet partition function, and $V_{g,n}(b,\textbf{a})$ represents the Weil-Petersson volume of the moduli space of hyperbolic Riemann surfaces with genus $g$ and $n+1$ geodesic boundaries of lengths $(b,a_1,a_2,...,a_n)$. In the following for simplicity, we set $g=0$ (see figure \ref{pants}).   
The aforementioned partition function can be written differently, as follows: 
\begin{align}\label{part2}
Z\left( \beta,\textbf{a}\right) =e^{S_0\chi }\int_{0}^{\infty} \rho\left(E,\textbf{a}\right)e^{-\beta E}\text{d}E,
	\end{align}
where $ \rho\left(E,\textbf{a}\right) $ denotes the density of eigenvalues.
To derive the expression of $\rho\left(E,\textbf{a}\right)$, one can use the substitution of the trumpet partition function $Z_{\text{trumpet}}\left(\beta,b\right)$ defined by:
	\begin{equation}\label{trup}
Z_{\text{trumpet}}\left( \beta,b\right) =\int_{0}^{\infty} \rho_{\text{trumpet}}\left(E,b\right)e^{-\beta E}\text{d}E,\qquad  \rho_{\text{trumpet}}\left(E,b\right)=\frac{\cos b\sqrt{E}}{2\pi \sqrt{E}},
	\end{equation}
into the relation \eqref{part1} as follows:
\begin{equation}\label{rho}
Z\left( \beta,\textbf{a}\right) =e^{S_0\chi }\int_{0}^{\infty} \int_{0}^{\infty}\rho_{\text{trumpet}}\left(E,b\right)V_{0,n+1}(b,\textbf{a})e^{-\beta E}  b\text{d}b\text{d}E.
\end{equation} 
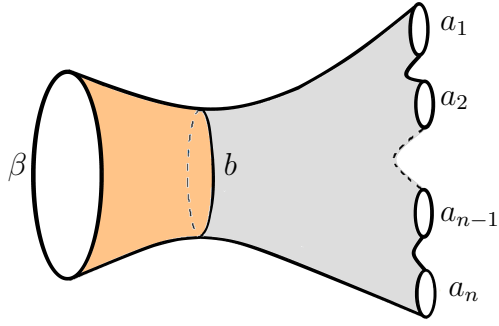
\begin{figure}[h]
	\centering
	\begin{overpic}
		[width=0.32\textwidth]{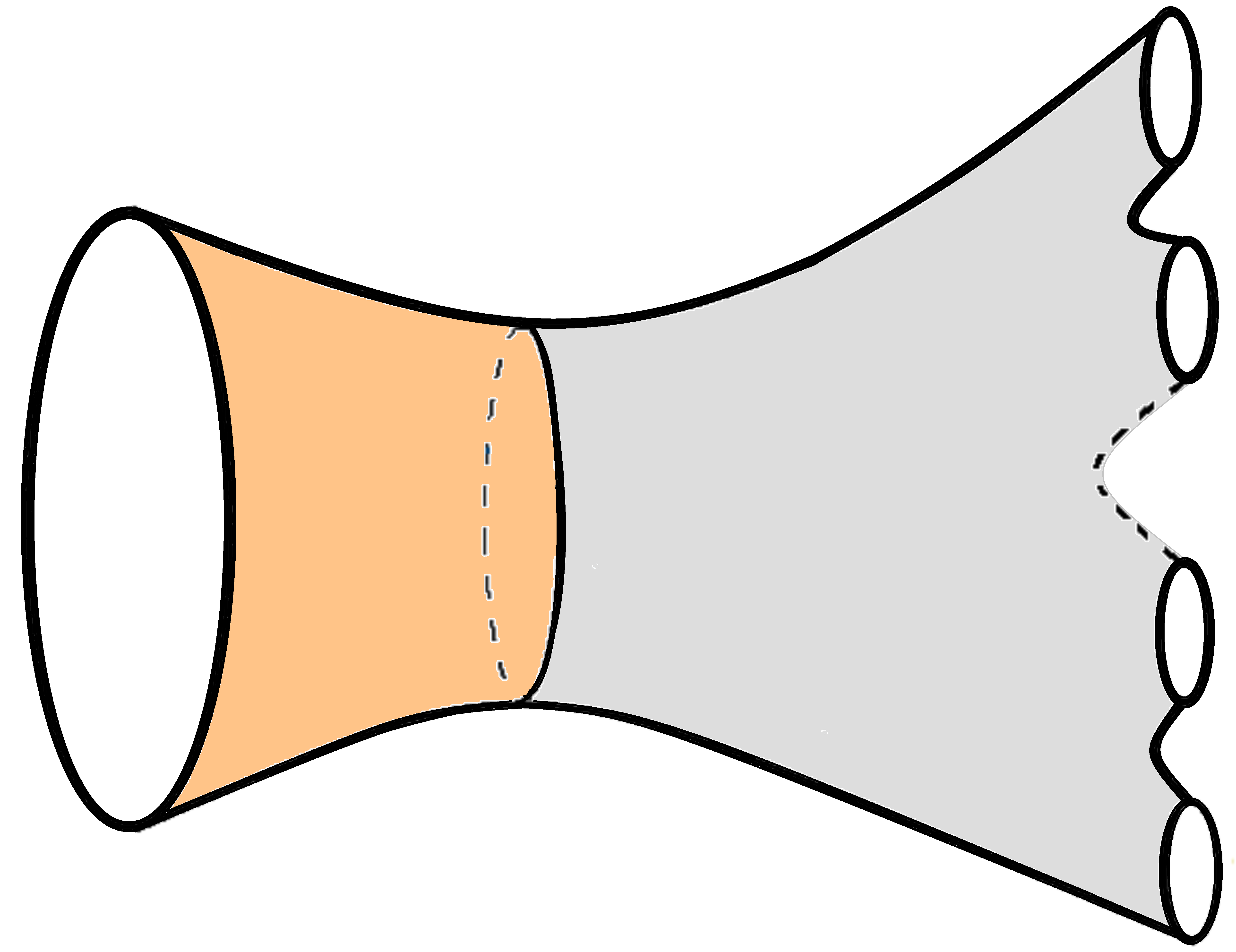}
		\put (100,52) {$\displaystyle a_{2}$}
		\put (100,24) {$\displaystyle a_{n-1}$}
		\put (102,6) {$\displaystyle a_{n}$}
		\put (100,70) {$\displaystyle a_{1}$}
		\put (48,35) {$\displaystyle b$}	
		\put (-4,35) {$\displaystyle\beta$}	
	\end{overpic}
	\caption{A Riemann surface with genus $g=0$ and an asymptotically AdS boundary $\beta$, along with $n$ geodesic boundaries or baby universes $\left(a_1, a_2, \ldots, a_n\right)$. The loop $b$, minimized within its homology class, is homologous to $\beta$. This geometric structure can be reconstructed by joining a trumpet, marked in orange, and a multi-holed sphere, shown in gray, along the geodesic $b$.}	
	\label{pants}
\end{figure}

\begin{figure}[h]
	\centering
	\begin{overpic}
		[width=0.375\textwidth]{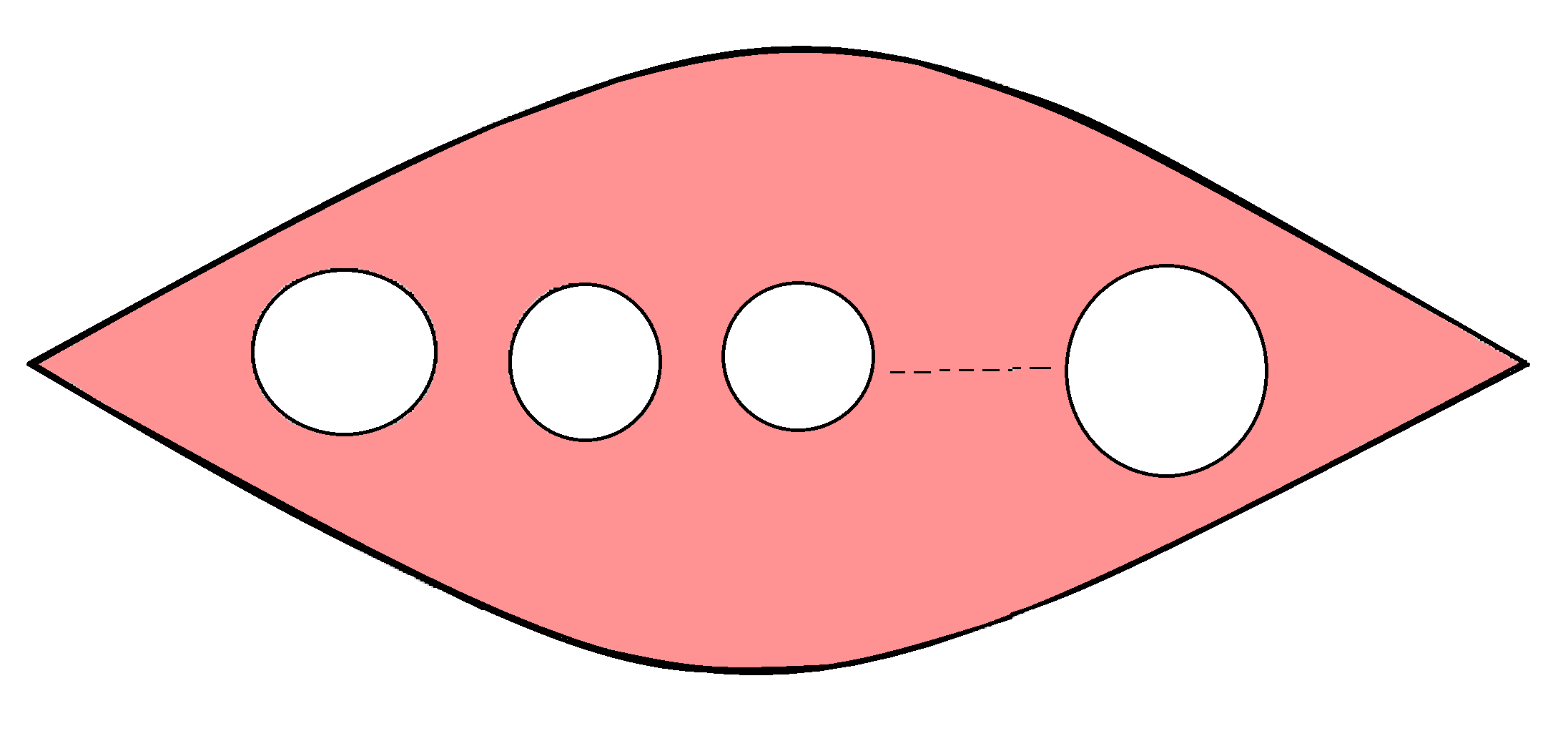}
		\put (35,23) {$\displaystyle a_{2}$}
		\put (49,23) {$\displaystyle a_{3}$}
		\put (72,23) {$\displaystyle a_{n}$}
		\put (20,23) {$\displaystyle a_{1}$}
		\put (48,45) {$\displaystyle\ell$}	
		\put (48,-2) {$\displaystyle\ell_{1}$}	
	\end{overpic}
	\caption{The amplitude $\langle\ell,\textbf{a}|\ell_{1}\rangle$ is illustrated, representing a wormhole with length $\ell_1$ emits $n$ baby universes and transforms into another wormhole with length $\ell$. The colored region   is referred to as the tunneling amplitude region.}	
	\label{vol}
\end{figure}
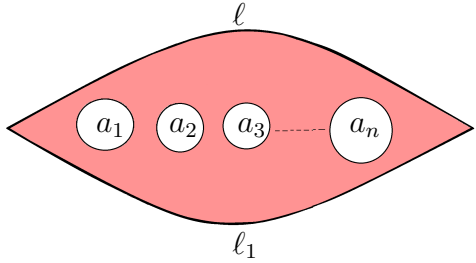

The expression for $\rho\left(E,\textbf{a}\right)$ is obtained by comparing equations \eqref{rho} and \eqref{part2}. The resulting expression is: 
\begin{equation}\label{density}
\rho\left(E,\textbf{a}\right)=\int_{0}^{\infty}\rho_{\text{trumpet}}\left(E,b\right)V_{0,n+1}(b,\textbf{a})b\text{d}b.
\end{equation}
The partition function \eqref{part1} can be alternatively expressed as the gluing of two wave functions of Hartle-Hawking states with a specific amplitude, as follows:
\begin{equation}\label{partw}
Z\left( \beta,\textbf{a}\right) =e^{S_0\chi }\int_{-\infty}^{\infty}\langle\frac{\beta}{2}|\ell\rangle\langle\ell,\textbf{a}|\ell_{1}\rangle\langle\ell_{1}|\frac{\beta}{2}\rangle \text{d}\ell_{1}\text{d}\ell,
\end{equation} 
where the amplitude $\langle\ell,\textbf{a}|\ell_{1}\rangle$ is given by:
\begin{equation}\label{a}
\langle\ell,\textbf{a}|\ell_{1}\rangle=\int_{0}^{\infty}\langle\ell|E\rangle\langle E|\ell_{1}\rangle \rho\left(E,\textbf{a}\right)\text{d}E.
\end{equation}
The expression $\langle\ell,\textbf{a}|\ell_{1}\rangle$ represents the amplitude for a wormhole with length $\ell_1$ to emit $n$ baby universes and transition into another wormhole with length $\ell$ ( see figure  \ref{vol}). 
For latter calculations, it would be useful to replace $ \rho\left(E,\textbf{a}\right) $ from expression \eqref{density} into \eqref{a}, so one gets: 
\begin{equation}\label{amp}
	\langle\ell,\textbf{a}|\ell_{1}\rangle=\int_{0}^{\infty} \int_{0}^{\infty}\rho_{\text{trumpet}}\left(E,b\right)V_{0,n+1}(b,\textbf{a})\langle\ell|E\rangle\langle E|\ell_{1}\rangle b\text{d}b\text{d}E.
\end{equation}

In the framework of JT gravity, the wave functions $\langle \ell|E\rangle$ and the density of states $\rho\left(E \right)$ are defined as \cite{yang,kit,Bagrets:2017pwq,har}:
\begin{equation}
	\langle \ell|E\rangle = 2^{3/2}K_{2i\sqrt{E}}\left(2e^{-\ell/2} \right), \quad \rho\left(E \right) = \frac{\sinh\left(2\pi\sqrt{E} \right)}{\left(2\pi \right)^{2}}.
\end{equation}
They satisfy  the orthogonality and completeness relations:
\begin{align}\label{orl}
	\int_{-\infty}^{\infty} \text{d}\ell \langle E|\ell\rangle \langle\ell|E'\rangle = \frac{\delta\left(E-E' \right)}{\rho\left(E \right)},
	\\
	\int_{0}^{\infty} \text{d}E \rho\left(E \right) \langle \ell|E\rangle \langle E|\ell'\rangle = \delta(\ell-\ell').
\end{align}
By  substituting \eqref{amp} into \eqref{partw} and integrating over $\ell$ and $\ell_1$ one can obtain \eqref{rho}.
\section{Probability distributions in genus 2}\label{3} 
In this section, we investigate a bulk topology with two handles. The goal is to calculate the probability distribution for the length of a spatial slice connecting two boundary points as shown in figure \ref{volppp}.
 The  focus is on the scenario where $ S$  and $t$ are large.
The probability distribution can be written as:
\begin{equation}\label{bet}
\hat{P}_{2,\beta}\left( \ell\right)=e^{-3S_{0}} \int_{-\infty}^{\infty}\text{d}\ell_{1}\text{d}\ell_{2}\int \text{d}a_{2}\text{d}s_{2}\text{d}a_{1}\text{d}s_{1}\langle\frac{\beta}{2}+it|\ell_{2}\rangle\langle\ell_{2}|\ell,a_{1},a_{2}\rangle\langle\ell,a_{1},a_{2}|\ell_{1}\rangle\langle\ell_{1}|\frac{\beta}{2}+it\rangle.
\end{equation}
The notation $\hat{P}$ represents an un-normalized probability distribution. To achieve proper normalization, we must divide $\hat{P}$ by the partition function. The subscript two refers to the genus.
The inverse Laplace transformation  yields the fixed-energy version of the probability distribution as follows: 
\begin{equation}\label{prob2}
\hat{P}_{2}\left( \ell\right)=e^{-3S_{0}}\int\frac{\text{d}\beta}{2\pi i}e^{\beta E} \int_{-\infty}^{\infty}\text{d}\ell_{1}\text{d}\ell_{2}\int \text{d}a_{2}\text{d}s_{2}\text{d}a_{1}\text{d}s_{1}\langle\frac{\beta}{2}+it|\ell_{2}\rangle\langle\ell_{2}|\ell,a_{1},a_{2}\rangle\langle\ell,a_{1},a_{2}|\ell_{1}\rangle\langle\ell_{1}|\frac{\beta}{2}+it\rangle.
\end{equation}
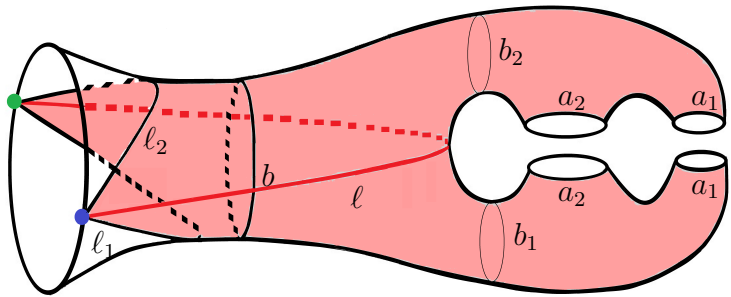
\begin{figure}[h]
	\centering
	\begin{overpic}
		[width=0.57\textwidth]{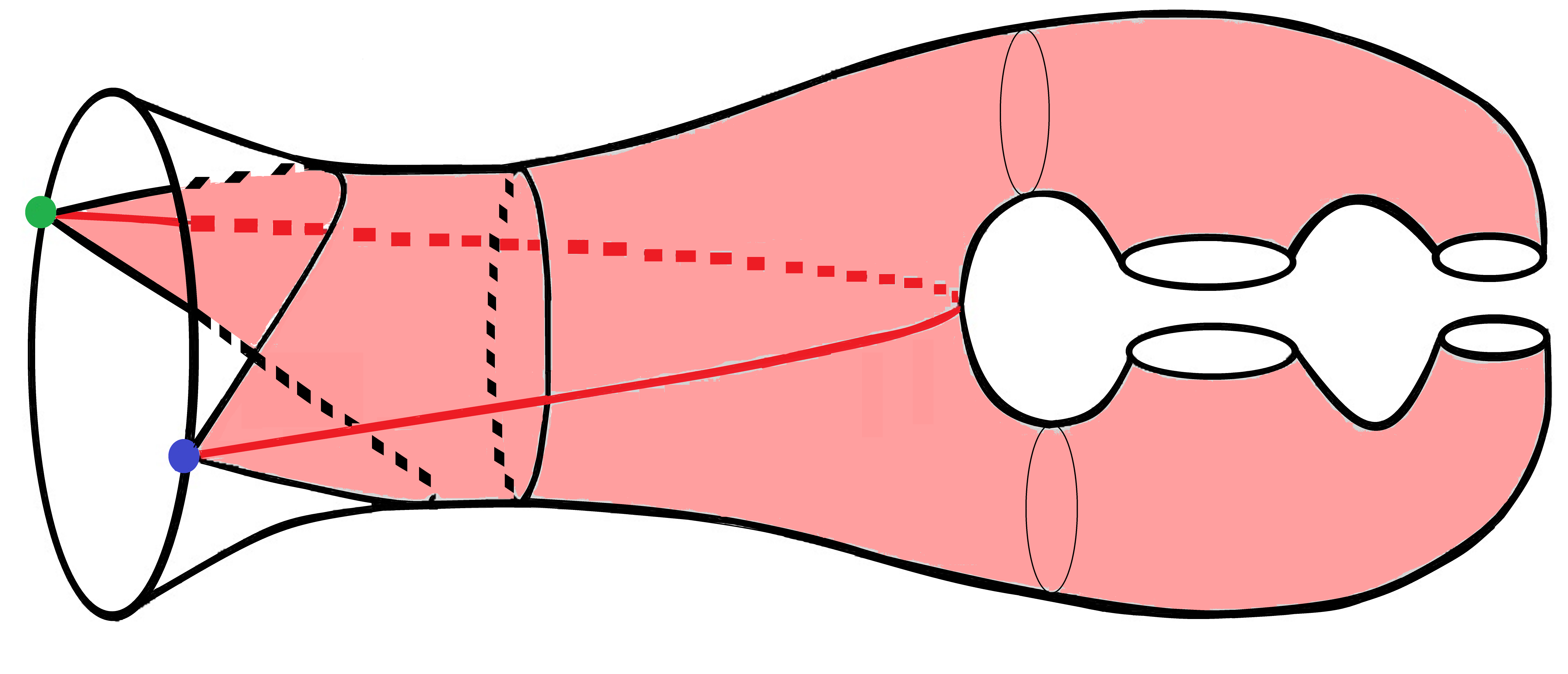}
		\put (48,15) {$\displaystyle\ell$}
		\put (94,17) {$\displaystyle a_{1}$}
		\put (94,29) {$\displaystyle a_{1}$}
		\put (76,29) {$\displaystyle a_{2}$}
		\put (76,16) {$\displaystyle a_{2}$}
		\put (20,23) {$\displaystyle\ell_{2}$}	
		\put (13,9) {$\displaystyle\ell_{1}$}
		\put (68,35) {$\displaystyle b_{2}$}
		\put (70,10) {$\displaystyle b_{1}$}
		\put (36,18) {$\displaystyle b$}	
	\end{overpic}
	\caption{The configuration with $\mathbb{Z}_2$ symmetry represents the length of the spatial slice connecting two boundary points, accompanied by two baby universes $a_1$ and $a_2$. The closed geodesic $b$ is homologous to the asymptotic boundary.	
		The colored region indicates the tunneling amplitude region. }	
	\label{volppp}
\end{figure}
\\
From \eqref{amp} one can get:
\begin{equation}\label{amp2}
\langle\ell,a_{1},a_{2}|\ell_{1}\rangle=\int_{0}^{\infty} \int_{0}^{\infty}\rho_{\text{trumpet}}\left(E,b\right) V_{0,3}\left(b,a_{1},a_{2} \right)\langle\ell|E\rangle\langle E|\ell_{1}\rangle b\text{d}b\text{d}E,
\end{equation}
where $ V_{0,3}\left(b,a_{1},a_{2} \right)=1 $ \cite{mir}.
By substituting  \eqref{amp2} into \eqref{prob2}  and integrating over $ \ell_1 $ and $ \ell_2 $, the following equation is obtained:
\begin{align}
\hat{P}_{2}\left( \ell\right)=e^{-3S_{0}}\int&\frac{\text{d}\beta}{2\pi i}e^{\beta E} \int \text{d}a_{2}\text{d}s_{2}\text{d}a_{1}\text{d}s_{1}\int_{0}^{\infty}b_1\text{d}b_1b_2\text{d}b_2\text{d}E_1\text{d}E_2\nonumber \\
&\rho_{\text{trumpet}}\left(E_1,b_1\right)\rho_{\text{trumpet}}\left(E_2,b_2\right)\langle\frac{\beta}{2}+it|E_{2}\rangle\langle E_{2}|\ell\rangle\langle\ell|E_1\rangle\langle E_1|\frac{\beta}{2}+it\rangle.
\end{align}
 By substituting inner product $ \langle E_{1,2}|\frac{\beta}{2}+it\rangle\nonumber=e^{(-\beta/2+it)E_{1,2}} $ we have: 
\begin{align}\label{prob3}
	\hat{P}_{2}\left( \ell\right)=e^{-3S_{0}}\int&\frac{\text{d}\beta}{2\pi i}e^{\beta \left( E-\frac{E_1+E_2}{2}\right) } \int \text{d}a_{2}\text{d}s_{2}\text{d}a_{1}\text{d}s_{1}\int_{0}^{\infty}b_1\text{d}b_1b_2\text{d}b_2\text{d}E_1\text{d}E_2\nonumber \\
	&\rho_{\text{trumpet}}\left(E_1,b_1\right)\rho_{\text{trumpet}}\left(E_2,b_2\right)e^{-it\left( E_2-E_1\right) }\langle E_{2}|\ell\rangle\langle\ell|E_1\rangle.
\end{align}
The integral over $ \beta $ sets $ E_1 + E_2 = 2E $.  By performing a change of variables:
\begin{equation}\label{change}
E_1=E-\sqrt{E}\omega \qquad E_2=E+\sqrt{E}\omega,  
\end{equation}
expression \eqref{prob3} reduces to:
\begin{align}\label{b}
	\hat{P}_{2}\left( \ell\right)=e^{-3S_{0}} \hspace{-.2cm}\int \text{d}a_{2}\text{d}s_{2}\text{d}a_{1}\text{d}s_{1}\int_{0}^{\infty}b_1\text{d}b_1b_2\text{d}b_2\int\hspace{-.1cm} \text{d}\omega\left(2\sqrt{E}\right) \frac{\cos b_{1}\sqrt{E_{1}}}{2\pi \sqrt{E_{1}}}\frac{\cos b_{2}\sqrt{E_{2}}}{2\pi \sqrt{E_{2}}}e^{-i2t\sqrt{E}\omega }\langle E_{2}|\ell\rangle\langle\ell|E_1\rangle.
\end{align}
Notice that the explicit value of $\rho_{\text{trumpet}}$ is replaced from \eqref{trup}. Using the following integral:
\begin{align}\label{int}
	\int_{0}^{\infty}\frac{b^k \cos \left(b \sqrt{E}\right)}{2 \sqrt{E}}\text{d}b&=b^{k}_{\infty}\left(  \frac{\sin \left(b_{\infty}\sqrt{E} \right)}{2 E}+\frac{k
		\cos \left(b_{\infty} \sqrt{E}\right)}{2b_{\infty} E^{3/2}}\right)-\frac{1}{2} E^{-\frac{k}{2}-1} \sin \left(\frac{\pi 
		k}{2}\right) \Gamma (k+1)\nonumber\\&\approx b^{k}_{\infty} \frac{\sin \left(b_{\infty}\sqrt{E} \right)}{2 E},
\end{align}
and considering the conditions $b_{\infty}, E \gg 1$, after setting $k=1$, the relation \eqref{b} can be approximated as:\footnote{In the following, the subscript $``\infty"$ is thrown   to save writing.}
\begin{align}\label{step}
	\hat{P}_{2}\left( \ell\right)\approx e^{-3S_{0}} \int \text{d}a_{2}\text{d}s_{2}\text{d}a_{1}\text{d}s_{1}\int \text{d}\omega\left(2\sqrt{E}\right) \frac{b_1\sin b_{1}\sqrt{E_{1}}}{2\pi E_{1}}\frac{b_2\sin b_{2}\sqrt{E_{2}}}{2\pi E_{2}}e^{-i2t\sqrt{E}\omega }\langle E_{2}|\ell\rangle\langle\ell|E_1\rangle.
\end{align}
 Our calculation is restricted to the semi-classical regime. In this regime, $E$ is large and the wave functions $\langle E|\ell\rangle$ are approximated as:
\begin{equation} \label{volt}
\langle E|\ell\rangle\approx\frac{\left( 8\pi\right) ^{1/2}}{2E^{1/4}}e^{-\pi \sqrt{E}}\left( e^{i\left( \sqrt{E}L-\frac{\pi}{4}\right) }+ e^{-i\left( \sqrt{E}L-\frac{\pi}{4}\right) }\right),\qquad L=\ell +2\log\left(2 E \right)-2.
\end{equation}
Also in this limit one should set
$
\sqrt{E_1}\simeq\sqrt{E}-\frac{\omega}{2}$ and $ \sqrt{E_2}\simeq\sqrt{E}+\frac{\omega}{2}$.
Therefore, the probability distribution \eqref{step} in semi-classical regime  becomes: 
\begin{align}
	\hat{P}_{2}\left( \ell\right)\approx-e^{-3S_{0}-2 \pi  \sqrt{E}}\frac{1}{4 \pi  E^2}& \int \text{d}a_{2}\text{d}s_{2}\text{d}a_{1}\text{d}s_{1}\int \text{d}\omega b_1 b_2  e^{-i \ell_t \omega } \left(e^{-i b_1 \left(\sqrt{E}-\frac{\omega }{2}\right)}-e^{i b_1
		\left(\sqrt{E}-\frac{\omega }{2}\right)}\right)
\nonumber	\\	
		& \left(e^{-i b_2 \left(\sqrt{E}+\frac{\omega }{2}\right)}-e^{i
		b_2 \left(\sqrt{E}+\frac{\omega }{2}\right)}\right) \left(e^{-i \left(\ell \left(\sqrt{E}-\frac{\omega
		}{2}\right)-\frac{\pi }{4}\right)}+e^{i \left(\ell \left(\sqrt{E}-\frac{\omega }{2}\right)-\frac{\pi }{4}\right)}\right)
	\nonumber\\
	&\left(e^{-i \left(\ell \left(\sqrt{E}+\frac{\omega }{2}\right)-\frac{\pi }{4}\right)}+e^{i \left(\ell \left(\sqrt{E}+\frac{\omega
		}{2}\right)-\frac{\pi }{4}\right)}\right),
\end{align}
where $ \ell_{t}=2t\sqrt{E} $ and   some unimportant $ 2\log(2E) $ terms were omitted. 
The integration over $\omega$ yields:
\begin{dmath}\label{lol}	
	\hat{P}_{2}\left( \ell\right)\approx e^{-3S_{0}-2 \pi  \sqrt{E}}\frac{1}{2  E^2} \int \text{d}a_{2}\text{d}s_{2}\text{d}a_{1}\text{d}s_{1}b_1 b_2\bigg\lbrace e^{i \sqrt{E} (b_1-b_2)}\delta\left( \frac{b_1+b_2}{2}  - \ell+\ell_t\right)  	+e ^{\ -i \sqrt{E} (b_1-b_2)}\left( \delta\left( \frac{b_1+b_2}{2}- (\ell+\ell_t)\right) +\delta\left( \frac{b_1+b_2}{2}+\ell-\ell_t\right) \right)\\+	\Delta\left( b_1,b_2,\ell_t\right) +	\tilde{\Delta}\left( b_1,b_2,\ell,\ell_t\right)
	\bigg\rbrace, 
\end{dmath}
where
\begin{dmath}\label{del}
	\Delta\left( b_1,b_2,\ell_t\right) =i\left( e^{i \sqrt{E} \left(-b_1-b_2+2 \ell\right)} -e^{-i \sqrt{E} \left(b_1+b_2+2 \ell\right)}\right) \delta\left(\frac{b_1-b_2}{2}-\ell_t \right)
	+i\left(  e^{i \sqrt{E} \left(b_1+b_2+2 \ell\right)}-e^{-i \sqrt{E} \left(-b_1-b_2+2 \ell\right)}\right) \delta\left(\frac{b_1-b_2}{2}+\ell_t \right)
	+i\left(e^{-i \sqrt{E} \left(b_1-b_2+2 \ell\right)}- e^{i \sqrt{E} \left(-b_1+b_2+2\ell\right)}\right) \delta\left(\frac{b_1+b_2}{2}-\ell_t \right),
\end{dmath}
and
\begin{dmath}\label{tdel}
	\tilde{\Delta}\left( b_1,b_2,\ell,\ell_t\right) =e^{i\sqrt{E} \left(b_1-b_2\right) }\delta\left(\frac{b_1+b_2}{2}+\ell_t+\ell \right)	- e^{-i \sqrt{E} (b_1+b_2)}\left( \delta \left( \frac{b_1-b_2}{2} - (\ell+\ell_t)\right) + \delta \left( \frac{b_1-b_2}{2}+ \ell-\ell_t\right) \right)-e ^{i \sqrt{E} (b_1+b_2)}\left( \delta\left( \frac{b_1-b_2}{2}  +\ell+\ell_t\right) +\delta\left( \frac{b_1-b_2}{2}-\ell+ \ell_t\right) \right).
\end{dmath}
 In the expression $\Delta\left( b_1,b_2,\ell_t\right)$, the absence of $\ell$ in the delta functions indicates that these terms do not impose any constraint between the length of the wormhole and the baby universes.
Although the delta functions in the expression $\tilde{\Delta}\left( b_1, b_2, \ell, \ell_t\right)$ establish relations between $b_1$, $b_2$, $\ell$, and $\ell_t$, its terms do not affect the amplitude.
The first term in the expression $\tilde{\Delta}\left( b_1, b_2, \ell, \ell_t\right)$ is zero due to the conditions that $b_1$, $b_2$, and $\ell_t$ are greater than zero, and $\ell$ is greater than or equal to zero.
Similarly, the other terms in the expression $\tilde{\Delta}\left( b_1, b_2, \ell, \ell_t\right)$ do not contribute because they would lead to non-physical consequences $b_1=b_2\pm2\ell\pm2\ell_t$, while it is expected that $b_1$ and $b_2$ are approximately equal $(b_1\approx b_2:=\tilde{b})$, as indicated in figure \ref{volppp} the probability amplitude geometry has $\mathbb{Z}_2$ symmetry. In other words, after integrating over $E$ with smooth window functions, it becomes evident that $b_1=b_2$. 
As a result, within the probability expression \eqref{lol}, only three terms are relevant for our specific purposes, leading to a simplified expression as presented below:
\begin{dmath}\label{proo}	
	\hat{P}_{2}\left( \ell\right)\approx e^{-3S_{0}-2 \pi  \sqrt{E}}\frac{1}{2  E^2} \int \text{d}a_{2}\text{d}s_{2}\text{d}a_{1}\text{d}s_{1}\tilde{b}^{2}\left\lbrace \delta\left( \tilde{b}  - \ell+ \ell_t\right) + \delta\left(\tilde{b}- \ell-\ell_t\right) +\delta\left( \tilde{b}+ \ell-\ell_t\right) 
	\right\rbrace .
\end{dmath}
These three terms correspond to three different types of geometry that  appeared in  \cite{Stanford:2022fdt}.
In the late-time limit, the lengths $\ell_1$ and $\ell_2$ will be significantly large. Furthermore, according to equation \eqref{int}, both $b_1$ and $b_2$ (or $\tilde{b}$) are large, and for most of the support of the probability distribution, $a_1$, $a_2$, and $\ell$ are expected to be large.  In accordance with the Gauss-Bonnet theorem, the area of the region corresponding to tunneling amplitudes remains small, allowing us to approximate the geometry as thin strips (see subsection \ref{tori}). By employing the thin strip approximation, the length $\tilde{b}$ can be expressed in terms of the lengths of the baby universes $a_1$ and $a_2$, as well as their common region $\lambda$, as follows:
\begin{dmath}
\tilde{b}\approx a_1+a_2-2\lambda .
\end{dmath}
Using the above relation, we can establish a connection between the lengths of the baby universes and the length of the wormhole.

Integration over baby universes is crucial in our computation, as it  relates to the fate of the infalling observer.  It seems possible to choose the  correct interior using a very simple criterion: it should be ensured that the spatial slice does not have shortcuts, meaning that distant points along the spatial slice or wormhole should not be close together in the spacetime geometry or bulk. This criterion is designed to avoid overcounting geodesics that cover portions of the same fundamental spatial slice more than once. This criterion will be discussed further in the following.

\subsection{Firewall-free geometry}\label{i}
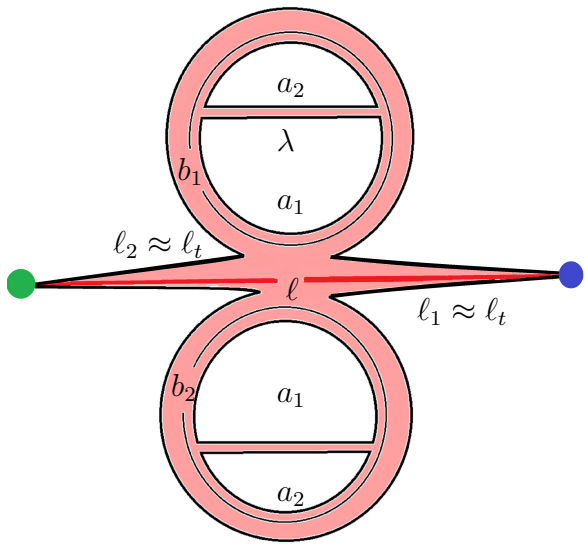
\begin{figure}[h]
 	\centering
 	\begin{overpic}
 		[width=0.45\textwidth]{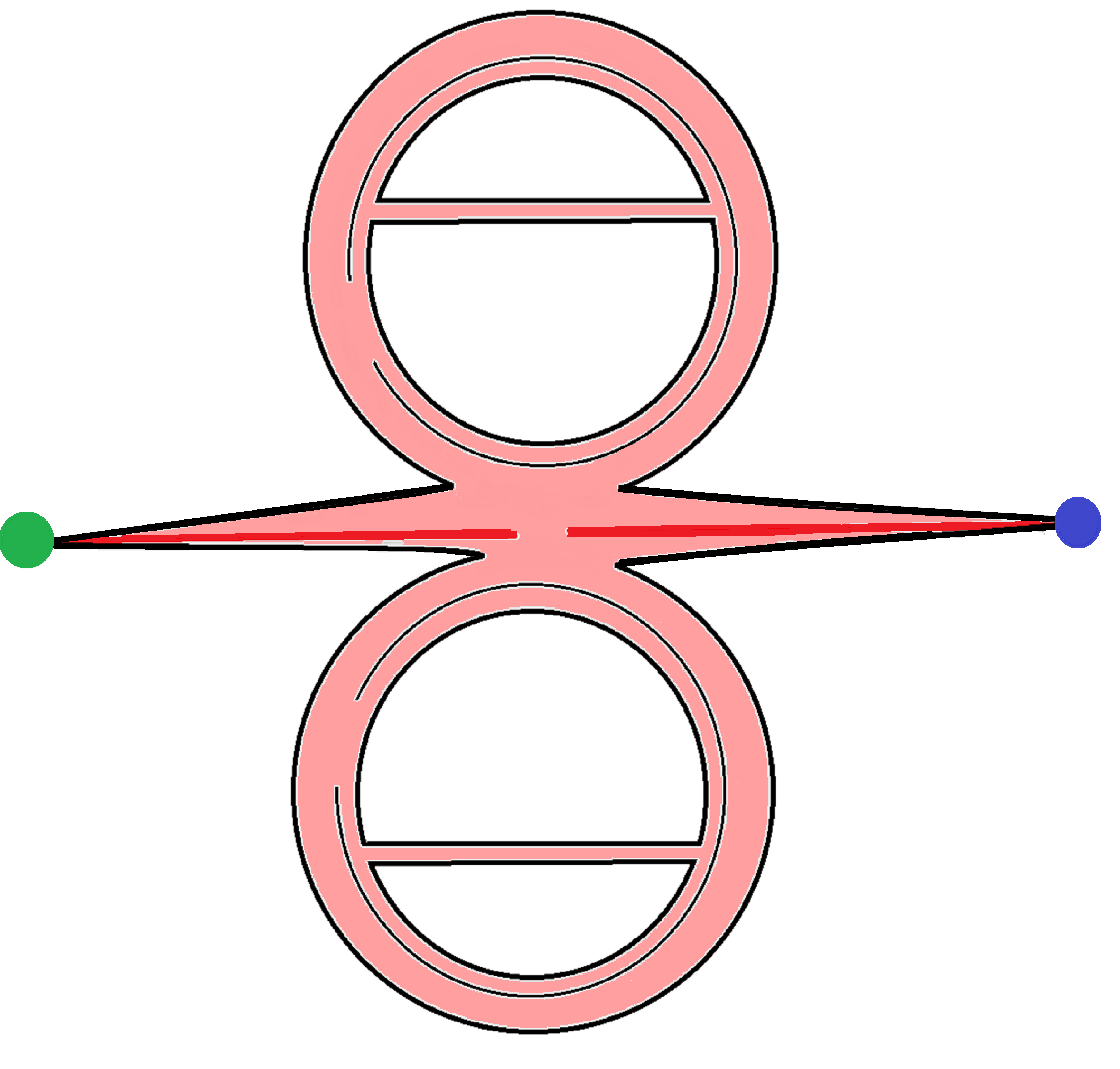}
 		\put (29.,65) {$\displaystyle b_{1}$}
 		\put (46,80) {$\displaystyle a_{2}$}
 		\put (27.93,28) {$\displaystyle \scalebox{1}{$b_{2}$}$}
 		\put (46,60) {$\displaystyle a_{1}$}
 		\put (46,70) {$\displaystyle \lambda$}
 		\put (46,10) {$\displaystyle a_{2}$}
 		\put (46,27) {$\displaystyle a_{1}$}
 		\put (47.54,44.5) {$\displaystyle\ell$}	
 		\put (70,41) {$\ell_1\approx\displaystyle\ell_{t}$}	
 		\put (18,52.5) {$\displaystyle\ell_2\approx\ell_{t}$}	
 	\end{overpic}
 	\caption{
 		The geometry that corresponds to $ \ell=\ell_{t}-\tilde{b} $ depicts a growing wormhole within the black hole regions, resulting in a transparent horizon. The parameter $\lambda$ represents the length of the region between two baby universes.}	
 	\label{g}
 \end{figure}

The wormhole of the geometry corresponding to  $ \ell=\ell_{t}-\tilde{b}=2\sqrt{E}t-\tilde{b} $ is expanding with time, so our black hole has a smooth horizon. The thin strip diagram in figure \ref{g} clearly shows that points on the $\ell$ slice are not close to each other, regardless of the assigned values of $a_1$, $a_2$, twists $s_1$, and $s_2$. Thus, the ``no shortcut'' criterion does not impose any restrictions on the moduli space.
Therefore, twists
$s_1$ and $s_2$ should  be integrated over the entire fundamental region $0 < s_1 < a_1$ and  $0 < s_2 < a_2$. 
The integration ranges for $a_{1}$ and $a_{2}$ extend from $0$ to some upper limits, where the upper limits are constrained by the condition that $\ell = \ell_t + 2\lambda - a_{1} - a_{2}$ must be positive. This constraint can be enforced by incorporating a theta function and setting the upper limit of integration to $\infty$.
So, the probability of a firewall-free geometry with a wormhole length of $\ell$ and a shared region $\lambda$ between baby universes is
\begin{align}	
\nonumber	\hat{P}_{2,\text{smooth}}\left( \ell,\lambda\right)
	 &= e^{-3S_{0}-2 \pi  \sqrt{E}}\frac{\left(\ell_t-\ell\right)^{2}}{2  E^2} \int_{0}^{\infty} \text{d}a_{2} \text{d}a_{1}\theta\left(\ell_t-a_1-a_2+2\lambda\right)\\&\hspace{4cm}\int_{0}^{a_2} \text{d}s_{2}\int_{0}^{a_1} \text{d}s_{1}  \delta\left(a_1+a_2-2\lambda +\ell-\ell_t\right)
\nonumber\\&=e^{-3S_{0}-2 \pi  \sqrt{E}}\frac{\left(\ell_t-\ell\right)^{2}}{12  E^2}
(\ell_t+2 \lambda -\ell)^3.
\end{align}
\subsection{Firewall geometry first method }\label{three}

\begin{figure}[h]
	\centering
	\begin{overpic}
		[width=0.35\textwidth]{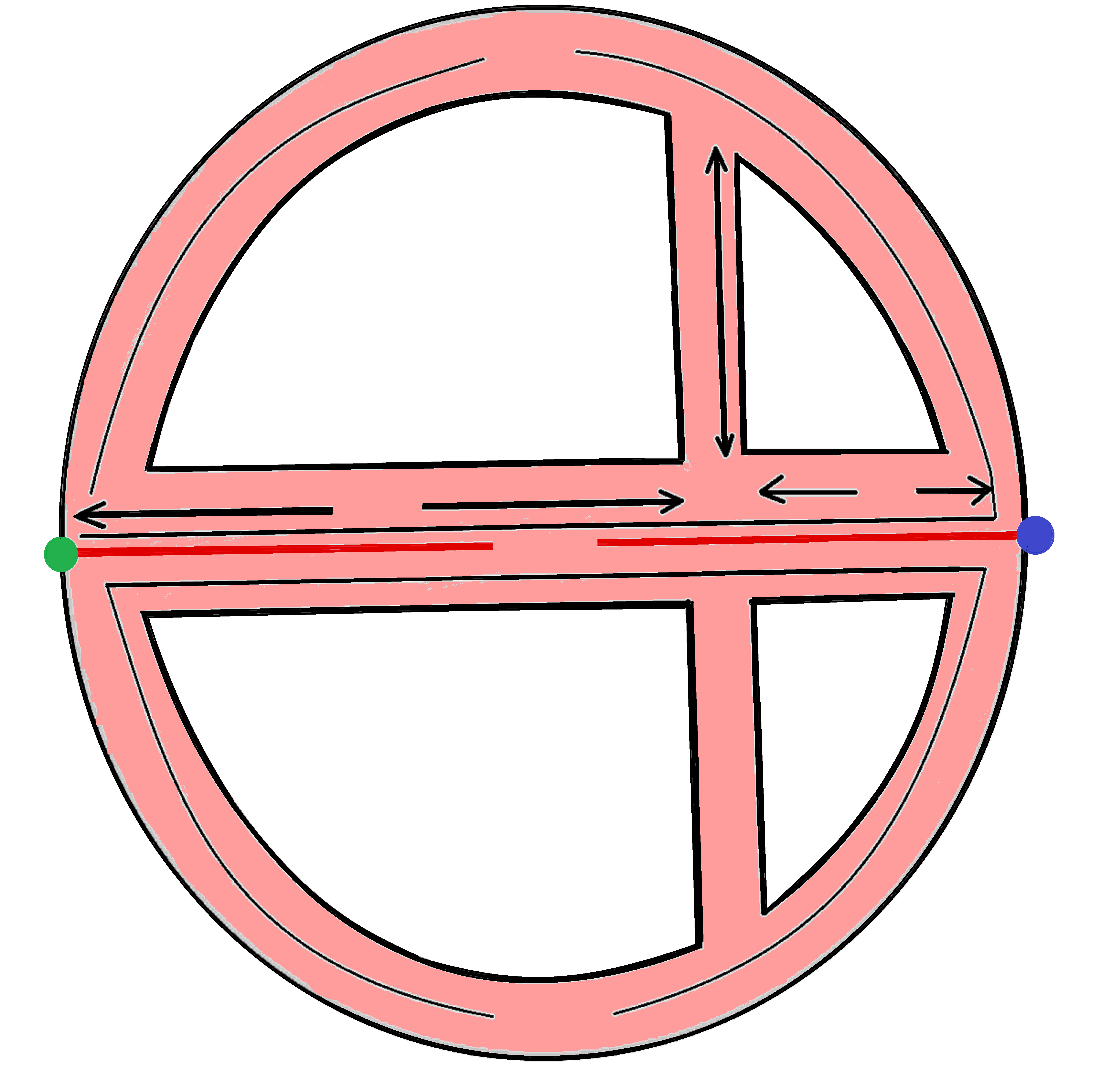}
		\put (47,92) {$\displaystyle b_{1}$}
		\put (61.5,70) {$\displaystyle \lambda $}
		\put (64.5,27) {$\displaystyle \lambda $}
		\put (40,70) {$\displaystyle a_{1}$}
		\put (72,70) {$\displaystyle a_{2}$}
		\put (40,27) {$\displaystyle a_{1}$}
		\put (73,27) {$\displaystyle a_{2}$}
		\put (33,52.5) {$\displaystyle\gamma$}	
		\put (79,52.8) {$\displaystyle\tilde{\gamma}$}	
		\put (47.22,46.8) {$\displaystyle\ell$}	
		\put (46,-5) {$\displaystyle \ell_1\approx\ell_{t}$}	
		\put (46,102) {$\displaystyle \ell_2\approx\ell_{t}$}	
		\put (49,5) {$\displaystyle b_{2}$}
	\end{overpic}
	\caption{The geometry characterized by $ \ell=\tilde{b}-\ell_{t} $, corresponds to a contracting 
		branch of the wave functions (firewall). In this configuration, both baby universes share a section with the wormhole $\ell$, denoted as $\gamma$ for baby universe $a_1$ (and $\tilde{\gamma}$ for $a_2$). The  region between the two baby universes is represented by $\lambda$.}	
	\label{k}
\end{figure}
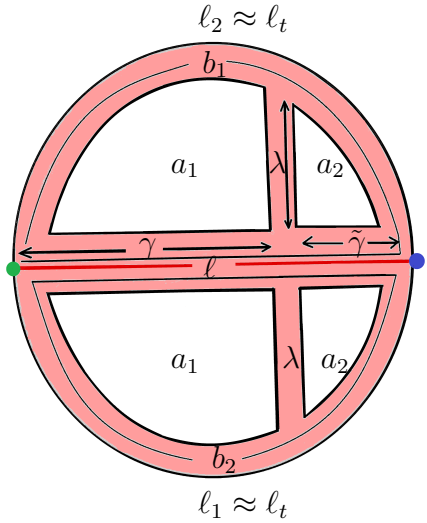

To calculate the probability of finding a firewall, one should consider the geometry in which $\ell$ shrinks over time. In this case, the delta functions that establish the relationship $\ell=\tilde{b}-2\sqrt{E}t$ can potentially contribute to the firewall probability.
In this particular geometry, the wormhole is overlapped by segments of both baby universes, denoted as $a_1$ and $a_2$ in figure \ref{k}. The portion of the wormhole overlapping with baby universe $a_1$ is labeled as $\gamma$, while the overlapping region with baby universe $a_2$ is referred to as $\ell-\gamma:=\tilde{\gamma}$. The shared region between the two baby universes is represented by $\lambda$. In this case, the ``no shortcut'' condition affects the moduli space.
To see this point, let us assume initially that  twists satisfy $0 < s_1 < \gamma$ and $0 < s_2 < \tilde{\gamma}$. Then for example the four points marked with ``$\times$''
in the left hand side of the figure \ref{r} will be identified with each other, respectively. In this case, there exists a more fundamental geodesic $\ell^{\prime} = \gamma^{\prime} + \tilde{\gamma}^{\prime}$, which serves as a better candidate for the correct spatial slice.
This geodesic indicated with thick black line in the right hand side of the figure \ref{r}. In other words, the mapping class group element that transforms the left sketch into the right in  figure \ref{r} can be understood, in the thin strip approximation, as mapping:
\begin{dmath}\label{dom}
{\left( a_1, s_1, \gamma+\lambda\right) \longrightarrow \left( a_1^{\prime}, s_1^{\prime}, \gamma^{\prime}+ \lambda^{\prime}\right)=\left( a_1-s_1,s_1, \gamma+\lambda-s_1\right), }\\
{\left( a_2, s_2, \tilde{\gamma}+\lambda\right) \longrightarrow \left( a_2^{\prime}, s_2^{\prime}, \tilde{\gamma}^{\prime}+\lambda^{\prime}\right) =\left( a_2-s_2, s_2 , \tilde{\gamma}+\lambda-s_2\right). }
\end{dmath}
As this process is continued,  we will  reach a point where $\gamma+\lambda < s_1$ and $\tilde{\gamma}+\lambda < s_2$, and there will be no shortcuts. Additionally, after applying the mapping class group, we expect that $a_{1}^{\prime}>\gamma$ and $a_{2}^{\prime}>\tilde{\gamma}$. Consequently, the constraints on the twists are
$\gamma + \lambda < s_1 < a_1 - \gamma$ and $\ell - \gamma + \lambda < s_2 < a_2 - \ell + \gamma$. These constraints can be written as:
\begin{dmath}\label{ineq}
	{a_1 + a_2 - \ell_t < s_1 + s_2 < a_1 + a_2 - \ell},\end{dmath}
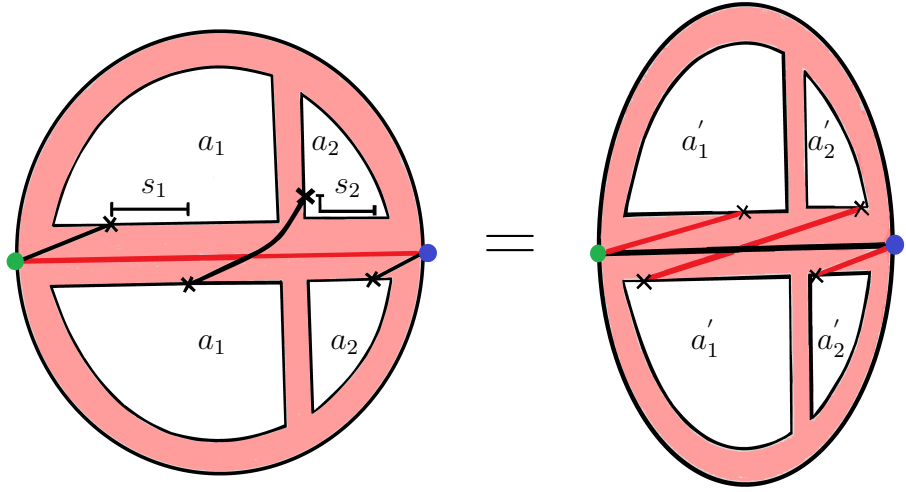
\begin{figure}[h]
	\centering
	\begin{overpic}
		[width=0.73\textwidth]{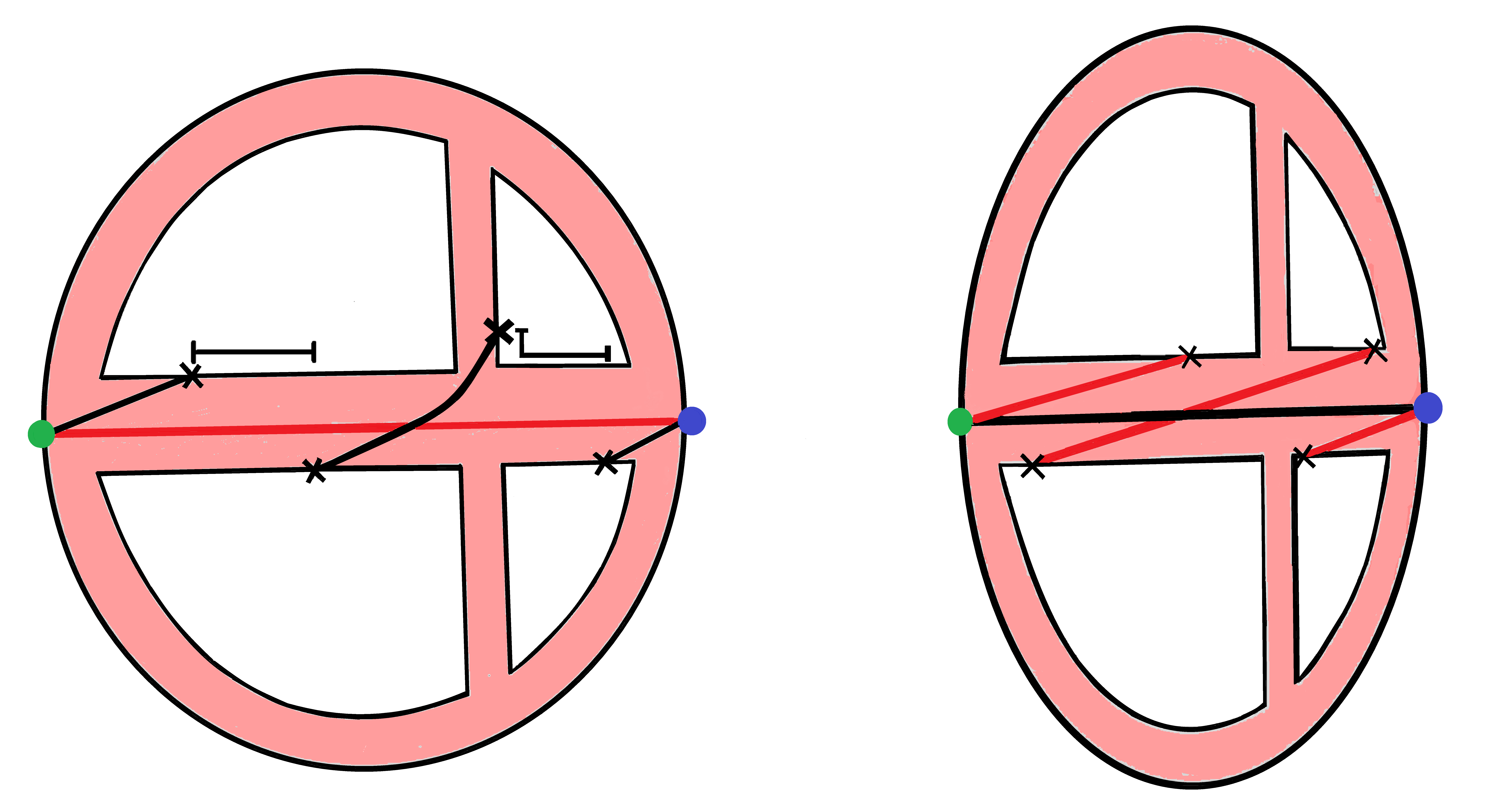}
		\put (86.3,37) {$\displaystyle a_{2}^{'}$}
		\put (34,37) {$\displaystyle a_{2}$}
		\put (22,37) {$\displaystyle a_{1}$}
		\put (73,37) {$\displaystyle a_{1}^{'}$}
		\put (87.3,16) {$\displaystyle a_{2}^{'}$}
		\put (36,16) {$\displaystyle a_{2}$}
		\put (22,16) {$\displaystyle a_{1}$}
		\put (74,16) {$\displaystyle a_{1}^{'}$}
		\put (16,32.5) {$\displaystyle s_1$}		
		\put (36.4,32.5) {$\displaystyle s_{2}$}
		\put (52,25.5) {$\displaystyle \scalebox{2}{\textbf{=}}$}
	\end{overpic}
	\caption{ 
		The left and right geometries can be related by an action of the mapping class group. Points marked with $\times$ are identified. The geodesic $\ell^{\prime}$, depicted as a thick black line, might be a more suitable candidate for the correct spatial slice, as the red line geodesic $\ell$ is essentially a repetition of $\ell^{\prime}$. In other words, left picture shows that points along the geodesic $\ell$ that appear distant are, in fact, close to each other within the entire geometry or bulk. }	
	\label{r}
\end{figure}
and	
\begin{align}\label{giv}	
	2\gamma+\lambda<a_1, \qquad	2\tilde{\gamma}+\lambda<a_2, \quad \text{or} \quad	2\ell+2\lambda<a_1+a_2.
\end{align}
Using the delta function that sets $\ell=a_1+a_2-2\lambda-\ell_t$, one can
rewrite the constraint \eqref{giv} as:
\begin{align}\label{given }	
	a_1+a_2<2\ell_t+2\lambda.
\end{align}
To ensure the positivity of $\ell$, it is necessary that 
\begin{align}\label{givn}	
	a_1+a_2>\ell_t+2\lambda.
\end{align}
 So, the probability of  emitting two baby universes, leading to a firewall, with   the wormhole length $\ell$, and both baby universes having a common region $\lambda$, is:
\begin{align}\label{pro2}	
\nonumber\hat{P}_{2,\text{firewall}}\left( \ell,\lambda\right)&\approx e^{-3S_{0}-2 \pi  \sqrt{E}}\frac{\left(\ell+\ell_t \right)^{2}}{2  E^2}\int_{0}^{\infty} \text{d}a_1\text{d}a_2~\bigg\lbrace\theta\left(a_1+a_2-2\lambda-\ell_t \right) \theta(2\ell_t+2\lambda-a_1-a_2) \\\nonumber&\hspace{-2cm}\delta\left(\ell+\ell_t-a_1-a_2+2\lambda \right)\int_{0}^{a_2} \int_{0}^{a_1}\hspace{-.2cm}\text{d}s_1\text{d}s_2 \theta\left(a_1+a_2-\ell-s_1-s_2 \right)  \theta\left(s_1+s_2+\ell_t-a_1-a_2 \right)\bigg\rbrace\\&=e^{-3S_{0}-2 \pi  \sqrt{E}}\frac{\left(\ell+\ell_t \right)^{2}}{12  E^2}(\ell_t-\ell) (\ell^2+6 \lambda  (\ell+\ell_t)+4 \ell \ell_t+\ell_t^2) .
\end{align}
Note that,
the constraints from inequalities \eqref{ineq}, \eqref{giv} and \eqref{givn} are enforced through the use of theta functions.
\subsection{Negative probability contribution  }
\begin{figure}[H]
	\centering
	\begin{overpic}
		[width=0.46\textwidth]{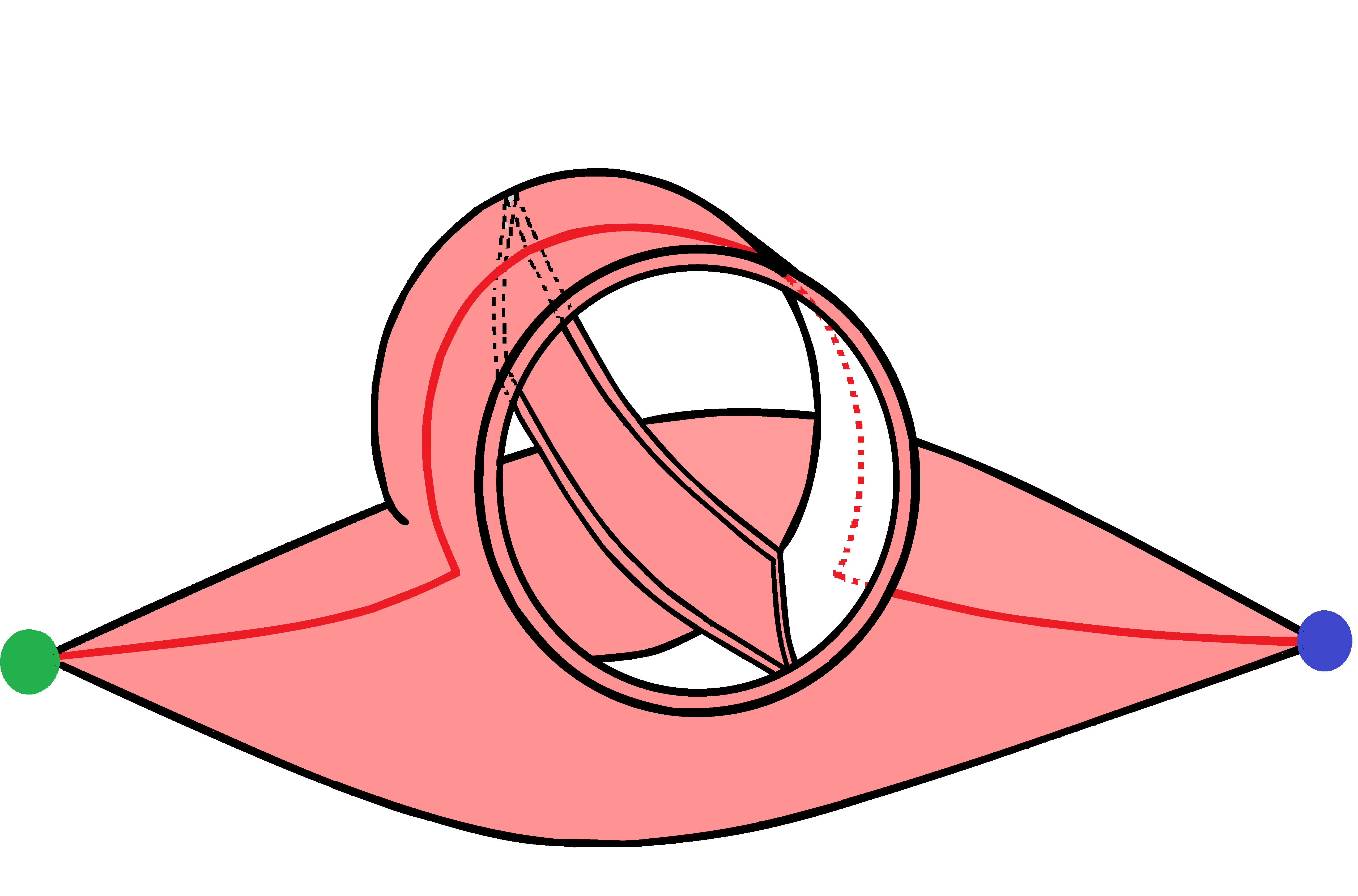}
		\put (30.1,53) {$\displaystyle b_{1}$}
		\put (41,20) {$\displaystyle a_{2}$}
		\put (53,31) {$\displaystyle a_{1}$}
		\put (48,8) {$\displaystyle b_{2}$}
		\put (46,27) {$\displaystyle \lambda$}
		\put (22,14) {$\displaystyle\ell$}	
		\put (70,5) {$\displaystyle\ell_1\approx\ell_{t}$}	
		\put (6.8,27.5) {$\displaystyle\ell_2\approx\ell_{t}$}	
	\end{overpic}
	\caption{
		In the geometry associated with $\ell = \ell_{t}+\tilde{b}$,  any value of the twists $s_1$ and $s_2$ induces a nonlocal identification along the $\ell$ geodesic. The $\ell$ geodesic closely tracks either the $\ell_1$ or $\ell_2$ geodesics, deviating slightly around the $b_1$ or $b_2$ cycles.}	
	\label{pu}
\end{figure}
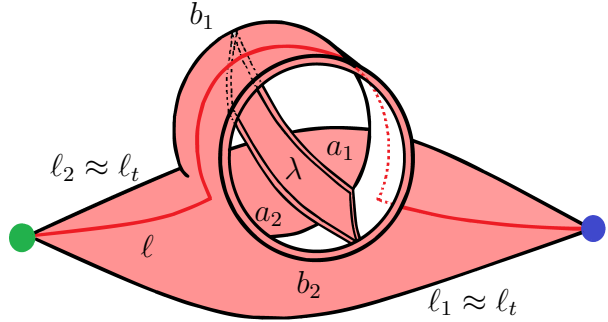

The wormhole in the geometry corresponding to $\ell = \ell_{t} + \tilde{b}$ is also expanding. Within this specific class of thin-strip geometry, illustrated in figure \ref{pu}, any values of the twists $s_1$ and $s_2$ induce nonlocal identifications along the $\ell$ geodesic. Another way to say this is
that the $\ell$ geodesic closely traces the paths of either the $\ell_1$ or $\ell_2$ geodesics, exhibiting a slight deviation around the $b_1$ or $b_2$ cycles. This behavior suggests that the correct interior is defined by the $\ell_1$ or $\ell_2$ geodesics \cite{Stanford:2022fdt}.
Calculating the total probability within this region poses difficulties due to the small size of the closed geodesic homologous to the asymptotic boundary, referred to as $b$ in figure \ref{five}. Additionally, the straightforward gluing procedure mentioned earlier does not include the mapping class group restriction on the corresponding twist parameter $\tau$. In the next section, we demonstrate that the small $b$ region contributes negatively to the probability.

\section{Firewall geometry second method}\label{sec}
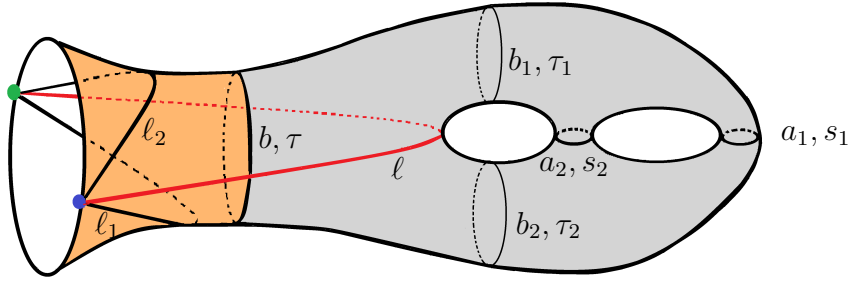
\begin{figure}[H]
	\centering
	\begin{overpic}
		[width=.6\textwidth,tics=6]{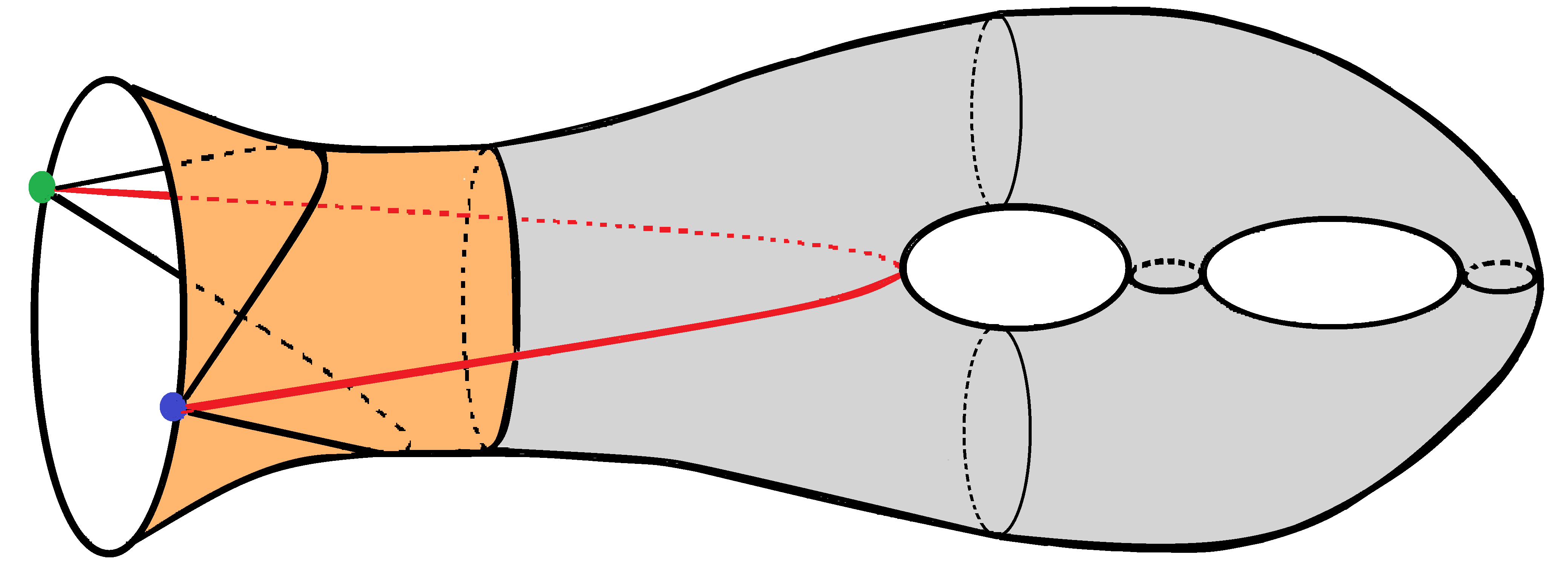}	
		\put (67,6) {$\displaystyle b_2,\tau_2$}
		\put (66,27) {$\displaystyle b_1,\tau_1$}
		\put (51,13) {$\displaystyle\ell$}
		\put (19,18) {$\displaystyle\ell_{2}$}
		\put (13,6) {$\displaystyle\ell_{1}$}
		
		\put (34,17) {$\displaystyle b,\tau$}
		\put (70,14) {$\displaystyle a_{2},s_{2}$}
		\put (101,18) {$\displaystyle a_{1},s_{1}$}
	\end{overpic}
	\caption{A trumpet attached to a double handle disk, which has a single geodesic boundary with a length of $b$. The $ \ell $ is the spatial slice connecting the two boundary points.}
	\label{gin}
\end{figure}

The surface of interest is a trumpet attached to a double torus, which has a single geodesic boundary with a length of $b$, as shown in figure \ref{gin}. 
The flat Weil-Petersson measure $\text{d}b_2$$\text{d}\tau_2$$\text{d}b_1$$\text{d}\tau_1 $$\text{d}a_{2}$$\text{d}s_{2}$$\text{d}a_{1}$$\text{d}s_{1}$ is the only factor related to the handle parts of the spacetime. As we observe later, at long times, the most
important contribution comes from double torus geometries with large $b$. Therefore, it is expected that the volume in the large $b$ limit is proportional to (see relation \eqref{volairy}):
\begin{dmath}\label{airy}
	{	V_{2,1}\left( b\right)\propto\int \text{d}b_2\text{d}\tau_2 \text{d}b_1  \text{d}\tau_1 \text{d}a_2 \text{d}s_2 \text{d}a_1 \text{d}s_1 	\propto b^8 }.
\end{dmath}  
However, our goal is not to calculate the partition function of the double handle disk. Instead, we are interested in the fate of the infalling observer. To achieve this, the inclusion of a delta function into the partition function is required. This delta function is employed to choose the $\ell$ slice that defines the correct interior. So one can write:
\begin{align}\label{prt}
	\hat{P}_{2}(\ell)=e^{-3S_0}\int \frac{d\beta}{2\pi i}&\text{d}\ell_1\text{d}\ell_2\text{d}b \text{d}\tau  \text{d}b_2\text{d}\tau_2\text{d}b_1\text{d}\tau_1 \text{d}a_{2}\text{d}s_{2}\text{d}a_{1}\text{d}s_{1} \nonumber\\& e^{\beta E}\langle\frac{\beta}{2}+it|\ell_{2}\rangle\langle \ell_{2},b|\ell_{1}\rangle\langle \ell_1|\frac{\beta}{2}+it\rangle  \delta\left(\ell-\ell\left(\text{moduli} \right)  \right), 
\end{align}
where the $\ell(\text{moduli})$ is the length of the geodesic labeled $\ell$ in the figure \ref{gin}.
\subsection{Five-holed sphere with $a_1+a_2>\ell_t$}
\begin{figure}[h]
	\centering
	\begin{overpic}
		[width=.45\textwidth,tics=6]{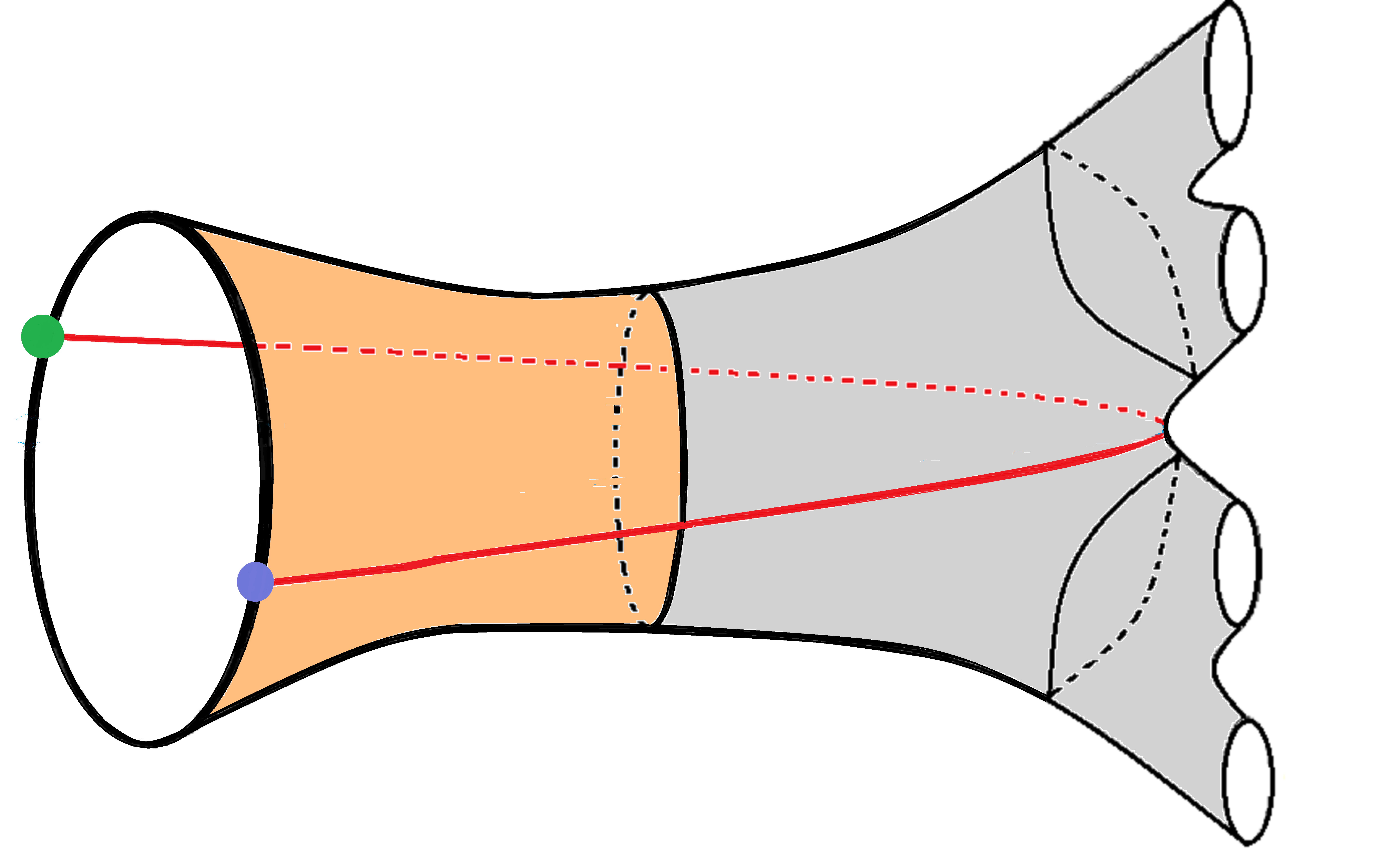}	
		\put (92,56) {$\displaystyle a_{1}$}	
		\put (63,20) {$\displaystyle\ell$}
		\put (71,20) {$\displaystyle b_2$}
		\put (71,40) {$\displaystyle b_1$}
		\put (51,27) {$\displaystyle b$}
		\put (92,42) {$\displaystyle a_2$}
		\put (92,21) {$\displaystyle a_2$}
		\put (93,5) {$\displaystyle a_1 $}
	\end{overpic}
	\caption{A five-holed sphere attached to a trumpet. The endpoints of the geodesic $\ell$ are at the boundary of the trumpet region.}
	\label{five}
\end{figure}
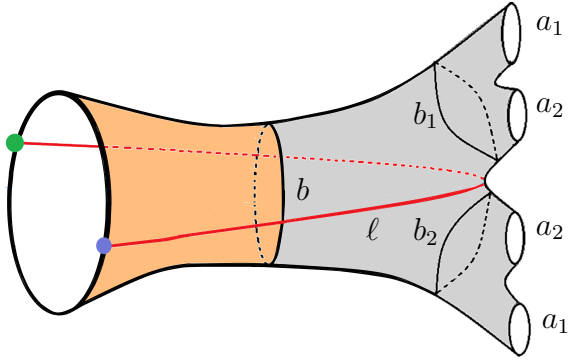

Let us begin by considering the geometry, which consists of a trumpet with a closed geodesic of length $b$ and a five holed sphere, as shown in figure \ref{five}.
The  length distribution $ 	\hat{p}(\ell,\textbf{a}) $  is described by the following expression:
\begin{align}\label{exp}
	\hat{p}(\ell,\textbf{a})=e^{-3S_0}\int &\frac{\text{d}\beta}{2\pi i E}\text{d}\ell_1\text{d}\ell_2\text{d}b \text{d}\tau  \text{d}b_2\text{d}\tau_2\text{d}b_1\text{d}\tau_1 e^{\beta E}\langle\frac{\beta}{2}+it|\ell_{2}\rangle\langle \ell_{2},b|\ell_{1}\rangle\langle \ell_1|\frac{\beta}{2}+it\rangle \nonumber\\&\underbrace{\delta\left(\ell-\ell\left(\text{moduli} \right)  \right)}_{\text{The constraint that} ~\ell~ \text{is held fixed.}} \underbrace{\delta\left(b_1
	+2\lambda-a_1-a_2 \right)\delta\left(b_2+2\lambda-a_1-a_2 \right)}_{\text{To fix  other parts of the geometry.}}.
\end{align}
In comparison to  the case of a three-holed sphere examined in \cite{Stanford:2022fdt}, expression \eqref{exp} includes two additional delta functions. For the three-holed sphere, the $\ell\left(\text{moduli} \right)$  can be expressed as a function of two holes or baby universes \cite{Stanford:2022fdt}. In other words, in the case of a three-holed sphere, the insertion of a delta function in the partition function restricts  us to choose a geometry where the wormhole's length is determined by the lengths of baby universes and  $\ell_t$. However, in the case of a five-holed sphere, the length of the wormhole can be expressed as a function of geodesics $b_1$ and $b_2$, as computed in Appendix \ref{apen}, and this length is independent of geodesic boundaries or baby universes. Hence, to choose a geometry in the partition function where baby universes determine the wormhole, it is necessary to insert two more delta functions. These delta functions fix the lengths of $b_1$ and $b_2$ in terms of baby universes $a_1$ and $a_2$. Note that the presence of $``E"$ in the expression \eqref{exp} is related to dimensional analysis.

In the trumpet region, one can take the  metric:
\begin{equation}\label{metric}
	\text{d}s^2=\text{d}\sigma^{2}+\cosh^{2}\left(\sigma \right) \text{d}\tau^{2},\qquad \tau\sim\tau+b.
\end{equation}
Let us first integrate over $ \ell $, this removes the delta function and one can use the orthogonality of wave function to get:
\begin{align}
	\int_{-\infty}^{\infty}	\hat{p}(\ell,\textbf{a})\text{d}\ell
	&=e^{-3S_0}\int \frac{\text{d}\beta}{2\pi iE}\text{d}b\text{d}\tau  \text{d}E^{'}e^{\beta E}\langle\frac{\beta}{2}+it|E^{'}\rangle\langle E^{'}|\frac{\beta}{2}+it\rangle \frac{\cos b\sqrt{E^{'}}}{2\pi \sqrt{E^{'}}}\left(a_1+a_2-2\lambda \right)^{2}\nonumber\\
	&=e^{-3S_0}\int \frac{\text{d}\beta}{2\pi iE}\text{d}b~\text{d}\tau  \text{d}E^{'}e^{\beta\left(E-E^' \right) }  \frac{\cos b\sqrt{E^{'}}}{2\pi \sqrt{E^{'}}}\left(a_1+a_2-2\lambda \right)^{2}
	\nonumber\\	
	&=e^{-3S_0} \int_{0}^{\infty} \text{d}b~\int_{0}^{b/2}\text{d}\tau \frac{\cos b\sqrt{E}}{2\pi E \sqrt{E}}\left(a_1+a_2-2\lambda \right)^{2}.
\end{align}	
The twist $ \tau $ was integrated from zero to $ b/2 $ instead of $ b $ because of the $ \pi $ rotation symmetry of
the five-holed sphere. By slightly smearing over $E$, the contribution from $b = \infty$ can be eliminated, resulting in the following:
\begin{align}
	\int_{-\infty}^{\infty}	\hat{p}(\ell,\textbf{a})\text{d}\ell	=  \frac{-e^{-3S_0}}{4\pi E^{5/2} }\left(a_1+a_2-2\lambda \right)^{2},
\end{align}
 and the integration over $\lambda$ will give:
 \begin{align}\label{neg}
 \frac{-e^{-3S_0}}{4\pi E^{5/2}}	\int_{0}^{\ell_t}\left(a_1+a_2-2\lambda \right)^{2}\text{d}\lambda=  \frac{-e^{-3S_0}}{4\pi E^{5/2} }\left(\ell_t\left(a_1+a_2\right) \left(a_1+a_2-2 \ell_t\right) +\frac{4 \ell_{t}^{3}}{3}\right).
 \end{align}
The above result will be used for the contribution of small $b$, and, similar to the three-holed sphere answer, it is also negative. For the negative terms, it is crucial to consider the appropriate spatial slice, which should be either the $\ell_1$ or $\ell_2$ slice, not $\ell$ itself. Therefore, the no shortcut condition for $\ell$ should not be imposed \cite{Stanford:2022fdt}.
For simplicity, the calculations in the following will focus on large $ b $ region with $ b_{*}<b<\infty $, where $ b_{*} $ is lower cutoff satisfying $ 1\ll b_{*}\ll t $.
To calculate $ 	\hat{p}(\ell,\textbf{a}) $ for fixed $ \ell $, it is convenient to change variables $ \left( \ell_{1},\ell_2,b,\tau\right) \rightarrow 
\left( x_{12},\hat{\sigma}_{+},b,\frac{x_{1}+x_{2}}{2}\right) $ as following:
\begin{equation}
	e^{\ell_{1}}=e^{\hat{\sigma}_{+}}\sinh^{2}\left(\frac{b}{4}-\frac{x_{12}}{2} \right),\qquad e^{\ell_{2}}=e^{\hat{\sigma}_{+}}\sinh^{2}\left(\frac{b}{4}+\frac{x_{12}}{2} \right),
\end{equation}
where $ x_{12} = x_{1}-x_{2} $, and $ \hat{\sigma}_{+} = \hat{\sigma}_{1}+\hat{\sigma}_{2} $ is the sum of the regularized $ \sigma $ variables $ e^{\hat{\sigma}} = \epsilon e^{\sigma} $. The measure in the new
coordinates is  $\text{d}\ell_1\text{d}\ell_2~\text{d}b=2\text{d}x_{1,2}\text{d}b~ \text{d}\hat{\sigma}_{+} $, so:
\begin{align}\label{prob}
\nonumber	\hat{p}(\ell,\textbf{a})\supset&2e^{-3S_0}\int \text{d}b_1\text{d}\tau_1\text{d}b_2\text{d}\tau_2 \delta\left(b_1
	+2\lambda-a_1-a_2 \right)\delta\left(b_2+2\lambda-a_1-a_2 \right)\\&\hspace{.5cm}\int \frac{\text{d}\beta}{2\pi iE}e^{\beta E}\text{d}x_1\text{d}x_2~\text{d}b~ \text{d}\hat{\sigma}_{+}\langle\frac{\beta}{2}+it|\ell_{2}\rangle\langle \ell_{2},b|\ell_{1}\rangle\langle \ell_1|\frac{\beta}{2}+it\rangle\delta\left(\ell-\ell\left(\text{moduli} \right)  \right).
\end{align}
The pieces that depend on $ \beta $ can be extracted, and the integral can be performed as following:
\begin{align}
\nonumber	\int\frac{\text{d}\beta}{2\pi i}e^{\beta E} & \langle\frac{\beta}{2}+it|\ell_{2}\rangle\langle \ell_1|\frac{\beta}{2}+it\rangle\\\nonumber&=\int \frac{\text{d}\beta}{2\pi i}e^{\beta E}\text{d}E_1\text{d}E_2\rho\left(E_1 \right)\rho\left(E_2 \right)\langle\frac{\beta}{2}+it|E_2\rangle\langle E_2|\ell_{2}\rangle\langle \ell_1|E_1\rangle\langle E_1|\frac{\beta}{2}+it\rangle\\\nonumber&=\int \frac{\text{d}\beta}{2\pi i}\text{d}E_1\text{d}E_2\rho\left(E_1 \right)\rho\left(E_2 \right)e^{\beta \left( E-\frac{E_1+E_2}{2}\right) }e^{-it(E_{2}-E_1)}\langle E_2|\ell_{2}\rangle\langle \ell_1|E_1\rangle\\&=\int \text{d}E_1\text{d}E_2\rho\left(E_1 \right)\rho\left(E_2 \right)\delta\left( E-\frac{E_1+E_2}{2}\right) e^{-it(E_{2}-E_1)}\langle E_2|\ell_{2}\rangle\langle \ell_1|E_1\rangle, 
\end{align}
by using \eqref{change} and \eqref{volt} one can get:
\begin{align}
&	\int  \text{d}\omega 2\sqrt{E}\rho\left(E_1 \right)\rho\left(E_2 \right)e^{-i2t\sqrt{E}\omega}\langle E_2|\ell_{2}\rangle\langle \ell_1|E_1\rangle\approx\int  \frac{ \text{d}\omega}{\left( 2\pi\right)^{3} }\bigg\lbrace 2e^{-2\pi \sqrt{E}} \sinh\left(2\pi \left( \sqrt{E}-\frac{\omega}{2}\right)  \right)\nonumber\\ &\sinh\left(2\pi \left( \sqrt{E}+\frac{\omega}{2}\right)  \right)e^{-i2t\sqrt{E}\omega}	
\left( e^{i \left( \sqrt{E}-\frac{\omega}{2}\right) \ell_{1} }+ e^{-i\left( \sqrt{E}-\frac{\omega}{2}\right) \ell_{1} }\right)
	\left( e^{i\left(\sqrt{E}+\frac{\omega}{2}\right)\ell_{2}}+ e^{-i \left( \sqrt{E}+\frac{\omega}{2}\right) \ell_{2} }\right)\bigg\rbrace,  
\end{align}
after simplifying and omitting unimportant terms, the above relation becomes:
\begin{align}\label{beta}	
\nonumber\int  \frac{ \text{d}\omega}{2\left( 2\pi\right)^{3} }e^{2 \pi  \sqrt{E}} e^{-i2t\sqrt{E}\omega}\left( e^{i \left( \sqrt{E}-\frac{\omega}{2}\right) \ell_{1} }+ e^{-i\left( \sqrt{E}-\frac{\omega}{2}\right) \ell_{1} }\right)
	\left( e^{i\left(\sqrt{E}+\frac{\omega}{2}\right)\ell_{2} }+ e^{-i \left( \sqrt{E}+\frac{\omega}{2}\right) \ell_{2} }\right)\\\approx \frac{e^{\left( 2 \pi+i\left(\ell_{1}-\ell_{2} \right)  \right)  \sqrt{E}}}{2\left( 2\pi\right)^{2}}\int  \frac{ \text{d}\omega}{2\pi} e^{i\omega\left(\frac{\ell_{1}+\ell_{2}}{2}- \ell_{t}\right) }\nonumber\\\hspace{-5mm}=\frac{e^{\left( 2 \pi+i\left(\ell_{1}-\ell_{2} \right)  \right)  \sqrt{E}}}{2\left( 2\pi\right)^{2}} \delta\left(\frac{\ell_{1}+\ell_{2}}{2}- \ell_{t}\right)\nonumber\\\hspace{3mm}\approx
	\frac{e^{2\pi\sqrt{E}}}{2\left( 2\pi\right)^{2}}e^{-2i\sqrt{E}x_{12}} \delta\left(\hat{\sigma}_{+}+\frac{b}{2}-\ell_{t}\right).  
\end{align}
The factor $ 	\langle \ell_{2},b|\ell_{1}\rangle $  in the integrand of \eqref{prob}  is given by:
\begin{align}\label{d}
	\langle \ell_{2},b|\ell_{1}\rangle=\int_{0}^{\infty}\text{d}E\frac{\cos\left( b\sqrt{E}\right) }{2\pi \sqrt{E}}	\langle \ell_{2}|E\rangle\langle E|\ell_{1}\rangle=2K_0\left(2\sqrt{e^{-\ell_{2}}+e^{-\ell_{1}}+2e^{-\left(\ell_{1}+ \ell_{2}\right)/2 }\cosh\left(\frac{b}{2} \right) } \right),
\end{align}	
and in the first approximation at large $ b $, Bessel function becomes: 
\begin{align}\label{approx}
2K_0\left(4e^{-\frac{\hat{\sigma}_{+}}{2}}\frac{\sinh\frac{b}{2}}{\cosh\frac{b}{2}-\cosh x_{12}}\right)\approx 2K_0\left(4e^{-\frac{\hat{\sigma}_{+}}{2}}\right).
\end{align}
 Replacing \eqref{beta} and \eqref{approx} into \eqref{prob} results in:
\begin{dmath}\label{probb}	
	\hat{p}(\ell,\textbf{a})\supset e^{-3S_0}	\frac{e^{2\pi\sqrt{E}}}{\left( 2\pi\right)^{2}E}\int \text{d}b_1\text{d}\tau_1\text{d}b_2\text{d}\tau_2 \delta\left(b_1
	+2\lambda-a_1-a_2 \right)\delta\left(b_2+2\lambda-a_1-a_2 \right)\\\int_{b_*}^{\infty}\text{d}b 2K_0\left(4e^{-\frac{\hat{\sigma}_{+}}{2}}\right)\int \text{d}x_1\text{d}x_2~ e^{-2i\sqrt{E}x_{12}}\delta\left(\ell-\ell\left(\text{moduli} \right)  \right)|_{\hat{\sigma}_{+}=\ell_{t}-\frac{b}{2}},
\end{dmath}
and by substituting $ \ell\left(\text{moduli} \right) $ from \eqref{mwod}, \eqref{probb} becomes:
\begin{align}\label{prb}	
	\hat{p}(\ell,\textbf{a})\supset e^{-3S_0}	\frac{2e^{2\pi\sqrt{E}}}{\left( 2\pi\right)^{2}E}&\int \text{d}b_1\text{d}\tau_1\text{d}b_2\text{d}\tau_2~ \delta\left(b_1
	+2\lambda-a_1-a_2 \right)\delta\left(b_2+2\lambda-a_1-a_2 \right)\nonumber\\&\hspace*{-2.8cm}\int_{b_*}^{\infty}\text{d}b K_0\left(4e^{\frac{b-2\ell_t}{4}}\right)\hspace*{-.1cm}\int \text{d}x_1\text{d}x_2~ e^{-2i\sqrt{E}x_{12}}
	\delta\left(\ell-\ell_t+\frac{b}{2}-\gamma_{2}-2\log\cosh\frac{x_1}{2}-2\log\cosh\frac{x_2}{2} \right).
\end{align}
To carry out the integration over  $x_1$ and $ x_2$, one can expand the $\delta$ function in powers of $\log\cosh\frac{x_i}{2}$:
\begin{align}\label{delta}
	\delta\left(\ell-\ell\left(\text{moduli} \right)  \right) =&\delta\left( \ell-\ell_t+\frac{b}{2}-\gamma_{2} \right)-2\left( \log\cosh\frac{x_1}{2}+\log\cosh\frac{x_2}{2}\right) \delta^{'}\left(\ell-\ell_t+\frac{b}{2}-\gamma_{2} \right)\nonumber\\&+4\log\cosh\frac{x_1}{2}\log\cosh\frac{x_2}{2}\delta^{''}\left(\ell-\ell_t+\frac{b}{2}-\gamma_{2} \right)\nonumber\\&+2\left( ( \log\cosh\frac{x_1}{2})^{2}+(  \log\cosh\frac{x_2}{2})^{2}\right)  \delta^{''}\left(\ell-\ell_t+\frac{b}{2}-\gamma_{2} \right)+\ldots,
\end{align}
Only the third term of expansion \eqref{delta} contributes to the integral \eqref{prb} and other terms become zero after integrating over
$ x_1$ and $x_2 $. Using the integral:
\begin{equation}\label{x}
	\int_{-\infty}^{\infty}\text{d}x~e^{-2i\sqrt{E}x}\log\cosh\frac{x}{2}=-\frac{\pi}{2\sqrt{E}\sinh\left(2\pi\sqrt{E} \right) }\approx -\frac{\pi}{\sqrt{E} }e^{-2\pi\sqrt{E}}, 
\end{equation}
and the expansion \eqref{delta}, \eqref{prb} becomes: 
\begin{align}\label{ppp}
\nonumber	\hat{p}(\ell,\textbf{a})\supset e^{-3S_0}	\frac{e^{-2\pi\sqrt{E}}}{ E^{2}}\int \text{d}b_1\text{d}\tau_1\text{d}b_2\text{d}\tau_2 \delta\left(b_1
	+2\lambda-a_1-a_2 \right)\delta\left(b_2+2\lambda-a_1-a_2 \right)\nonumber\\\int_{b_*}^{\infty}\text{d}b 2K_0\left(4e^{\frac{b-2\ell_t}{4}}\right) \delta^{''}\left(\ell-\ell_t+\frac{b}{2}-\gamma_{2} \right)\nonumber\\
	= e^{-3S_0}	\frac{e^{-2\pi\sqrt{E}}}{ E^{2}}\int \text{d}b_1\text{d}\tau_1\text{d}b_2\text{d}\tau_2 \delta\left(b_1
	+2\lambda-a_1-a_2 \right)\delta\left(b_2+2\lambda-a_1-a_2 \right)\nonumber\\\int_{b_*}^{\infty}\text{d}b  \partial_{\ell}^{2}2K_0\left(4e^{\frac{b-2\ell_t}{4}}\right)\delta\left(\ell-\ell_t+\frac{b}{2}-\gamma_{2} \right).
\end{align}
In the above expression, the terms from the $b_{\star}$ limit were dropped. It would be more convenient to approximate the Bessel function as a delta function:
\begin{equation}\label{delt}
	\partial_{\ell}^{2}2K_0\left(4e^{\frac{b-2\ell_t}{4}}\right) \rightarrow \frac{1}{2}\delta\left(b-2\ell_t \right).
\end{equation}
Therefore, \eqref{ppp} simplifies to:
\begin{align}\label{pp}
	\hat{p}(\ell,\textbf{a})&\supset
 e^{-3S_0}	\frac{e^{-2\pi\sqrt{E}}}{2 E^{2}}\int \text{d}b_1\text{d}\tau_1\text{d}b_2\text{d}\tau_2 \delta\left(b_1
		+2\lambda-a_1-a_2 \right)\delta\left(b_2+2\lambda-a_1-a_2 \right)\nonumber\\&\hspace*{6.2cm}\int_{b_*}^{\infty}\text{d}b~\delta\left(b-2\ell_t \right)\delta\left(\ell-\ell_t+\frac{b}{2}-\gamma_{2} \right)
\nonumber	\\&= e^{-3S_0}	\frac{e^{-2\pi\sqrt{E}}}{ 2E^{2}}\int \text{d}b_1\text{d}\tau_1\text{d}b_2\text{d}\tau_2 \delta\left(b_1
	+2\lambda-a_1-a_2 \right)\delta\left(b_2+2\lambda-a_1-a_2 \right)\nonumber\\&\hspace*{0.2cm} \times\delta\left(\ell- 2\sinh ^{-1}\left(\frac{2}{e^{\ell_t}}\sqrt{\cosh ^2\left( \frac{b_1}{2}\right) +\cosh
			^2\left(\frac{b_2}{2}\right)+e^{\ell_t}
			\cosh \frac{b_2}{2} \cosh
			\frac{b_1}{2}}
		\right)  \right),
\end{align}
where in the last step,  $\gamma_2$ was replaced from \eqref{gd}.	
Due to the condition $\ell_t \gg 1$, and both $b_1$ and $b_2$ being significantly larger than $\ell_t$, we can approximate the above expression as follows:
\begin{align}		
\nonumber&	\hat{p}(\ell,\textbf{a})\supset
e^{-3S_0}	\frac{e^{-2\pi\sqrt{E}}}{ 2E^{2}}\int \text{d}b_1\text{d}\tau_1\text{d}b_2\text{d}\tau_2 \delta\left(b_1
	+2\lambda-a_1-a_2 \right)\delta\left(b_2+2\lambda-a_1-a_2 \right)\\&\hspace{6.5cm}\times \delta\left(\ell- 2\sinh ^{-1}\left(e^{-\ell_t/2+b_1/4+b_2/4}\right) 
	\right)\nonumber\\&\approx
	e^{-3S_0}	\frac{e^{-2\pi\sqrt{E}}}{ 2E^{2}}\int \text{d}b_1\text{d}\tau_1\text{d}b_2\text{d}\tau_2 \delta\left(b_1
	+2\lambda-a_1-a_2 \right)\delta\left(b_2+2\lambda-a_1-a_2 \right)\delta\left(\ell+\ell_t-\frac{b_1+b_2}{2} \right),
	\end{align}
after performing the integration, the expression above yields:
\begin{dmath}\label{f5}
	\hat{p}(\ell,\textbf{a})\approx
	e^{-3S_0} \frac{e^{-2\pi\sqrt{E}}}{ 2E^{2}}\left(\ell+\ell_t\right)^{2} \delta\left(\ell+\ell_t-a_1-a_2+2\lambda \right),
\end{dmath}
this corresponds to the delta function in \eqref{proo}, which characterizes the firewall geometry. However, the current method clarifies that additional terms  \eqref{neg} originated from the small $b$ region.
At this stage, we need to integrate over $a_1$, $a_2$, $s_1$, and $s_2$, ensuring adherence to the no shortcut rule for the $\ell$ geodesic. Therefore, we delve into the details of the geometry of a disk with two handles (a double torus with one geodesic boundary) in the following.
\subsection{Double handle disk}\label{tori}	
Let us  consider a more general surface with geodesic boundaries, in the limit that
the geodesic boundaries are long. The Gauss-Bonnet theorem,
\begin{dmath}
	\int_{\mathcal{M}}RdA+\int_{\partial\mathcal{M}}kds=2\pi\chi(\mathcal{M}),
\end{dmath}
  makes a connection between the integrals of the curvature of the bulk and the integral of the curvature along the boundary of the surface.
In the case we are interested in, the bulk curvature is  constant  $(R=-2)$, and the extrinsic curvature of the boundaries is zero. So, the total area of the surface is connected to the Euler characteristic as: 
\begin{equation}
A=-\pi\chi.
\end{equation}
In the limit where the boundaries become long (Airy limit), considering that the area is fixed, the surface must become thin. This implies that any segment of the boundary approaches closely to another segment of the boundary. It can be helpful to maintain the intuitive notion that the hyperbolic surface with long boundaries resembles the ribbon graph or strip diagram \cite{kon,do, ander}.\footnote{Special thanks to Phil Saad for providing insights into the concept of ribbon graphs.}
\begin{figure}[H]
	\centering
	\begin{overpic}
		[width=.4\textwidth,tics=6]{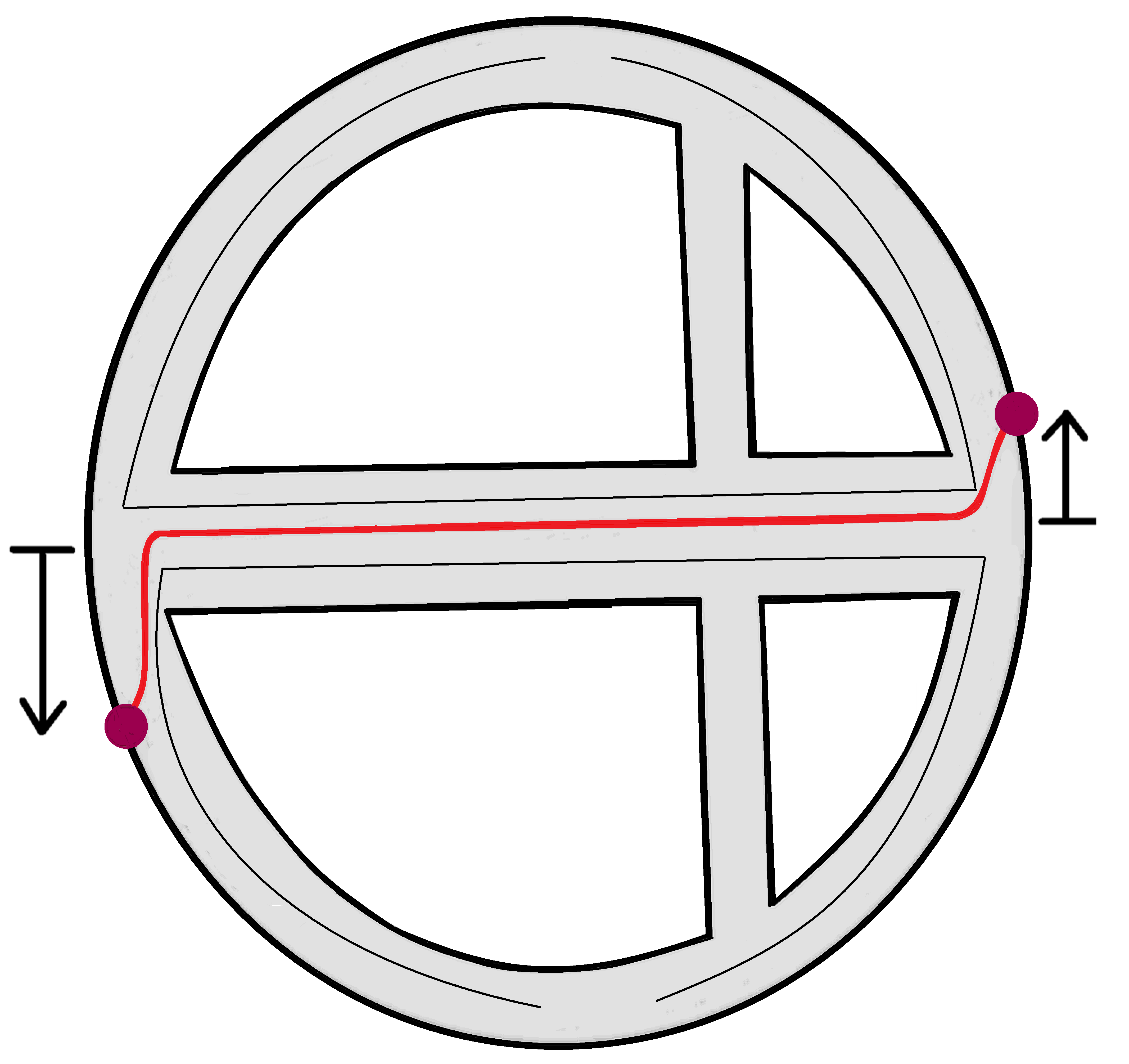}	
		\put (-6,37) {$\displaystyle x_{1}$}	
		\put (33,67) {$\displaystyle a_{1}$}	
		\put (69,66) {$\displaystyle a_{2}$}	
		\put (35,20) {$\displaystyle a_{1}$}	
		\put (70,25) {$\displaystyle a_{2}$}
		\put (99,53) {$\displaystyle x_{2}$}
		\put (49,86.5) {$\displaystyle b_{1}$}	
		\put (50,3) {$\displaystyle b_{2}$}
		\put (5,65) {$\displaystyle b$}
	\end{overpic}
	\caption{The Ribbon graph of a five-holed. The red line depicts the portion of the wormhole located within the five holed sphere.}
	\label{ribon5}
\end{figure}
\begin{figure}[H]
	\centering
	\begin{overpic}
		[width=.7\textwidth,tics=6]{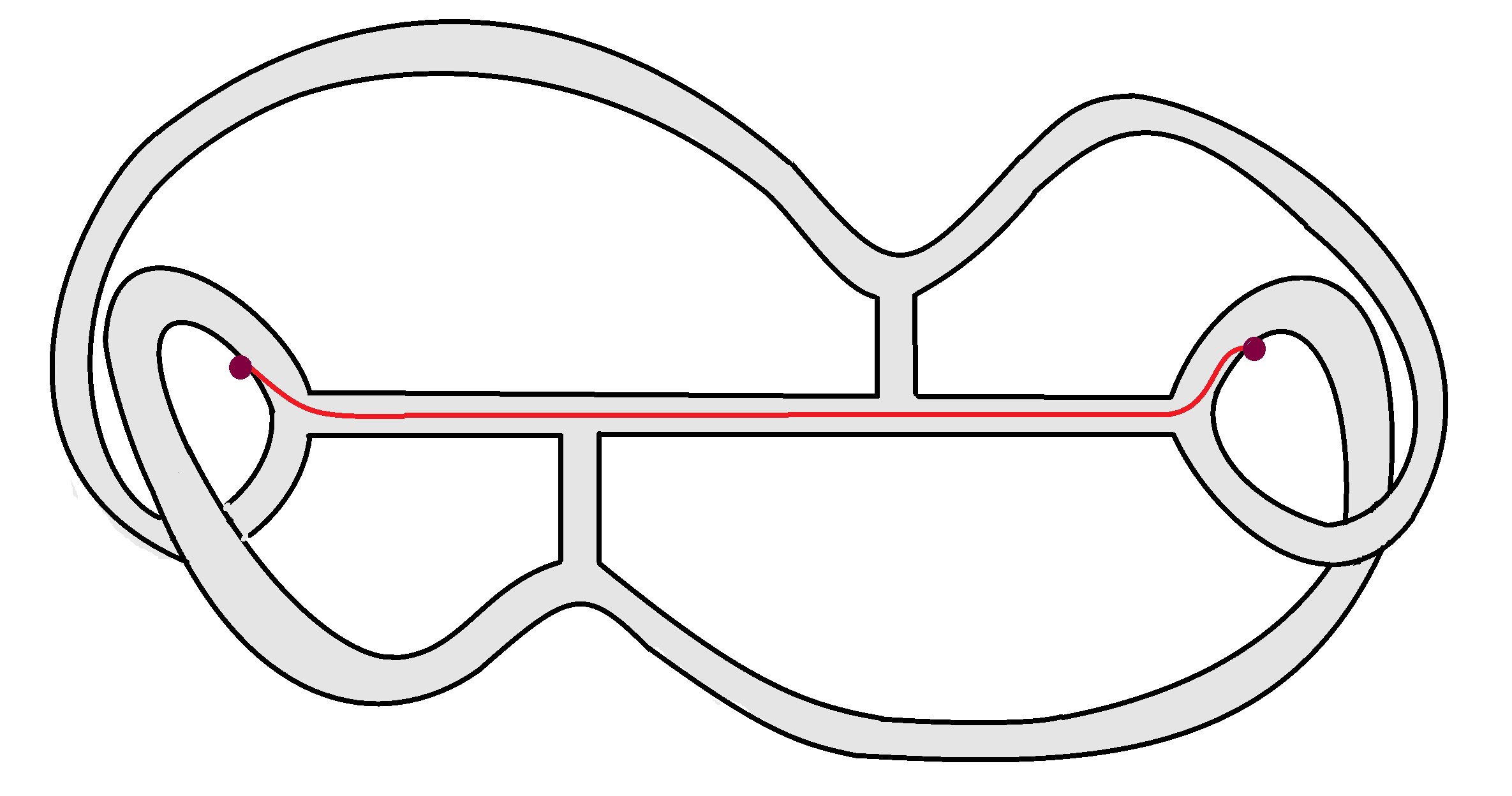}
		\put (80,47) {$\displaystyle y_{1}$}	
		\put (33,19) {$\displaystyle y_6$}
		\put (26,21) {$\displaystyle y_7$}
		\put (54,30) {$\displaystyle y_5$}
		\put (46,21) {$\displaystyle y_4$}
		\put (10,49) {$\displaystyle y_9 $}
		\put (72.6,0) {$\displaystyle y_2 $}
		\put (27,4) {$\displaystyle y_8 $}
		\put (67.3,21) {$\displaystyle y_3$}		
		\put (12,26.5) {$\displaystyle x_2 $}
		\put (83.7,27) {$\displaystyle x_1$}	
	\end{overpic}
	\caption{ Ribbon graph of a double torus with one boundary constructed by gluing geodesics of the same length together in figure \ref{ribon5}. The boundary of the double handle disk is depicted by the thick black closed line, while the red line represents the portion of the wormhole located within the double handle disk.}
	\label{doubler}
\end{figure}
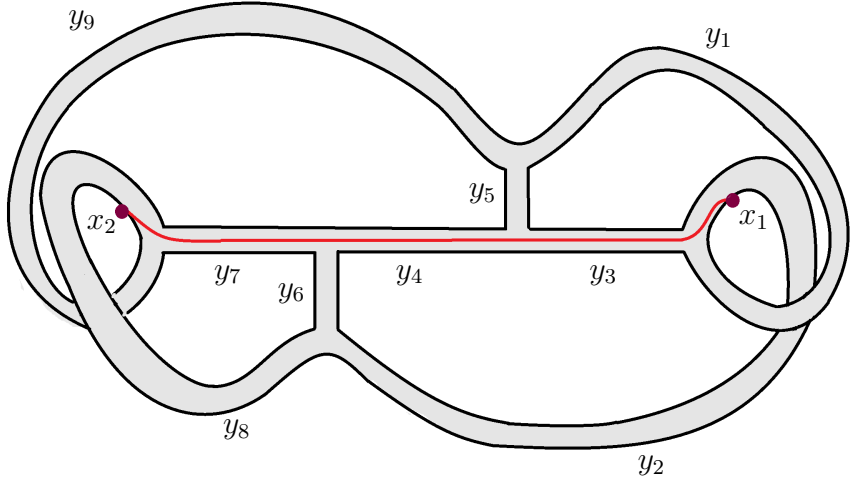
Figure \ref{ribon5} represents the ribbon graph of five holed sphere.
Also, figure \ref{doubler}  represents the ribbon graph of a double handle disk. This ribbon graph is constructed by gluing geodesics of the same length together in figure \ref{ribon5}. The resulting strip geometry can be depicted as a trivalent band structure, assembled from nine strips of lengths $y_1, y_2, \ldots, y_9$, and connected with two twists, resulting in a geometry with a single boundary.
Sometimes, the actual surface is represented solely by a trivalent graph, where the edges are assigned lengths, and the bulk geometry is not depicted. This graph exhibits $E=6g-6+3n$ edges and $V=4g-4+2n$ vertices. In other words, a graph with $E$ edges and $V$ vertices is associated to a  ribbon graph with genus $g$ and $n$ boundaries through the following relation:
\begin{equation}
	V-E=2-2g-n.
\end{equation}
By thickening the graph one can obtain the ribbon graph.
For a double handle disk (with $g=2$ and $n=1$), the graph has nine edges and six vertices, as depicted in figure \ref{graph}.
\begin{figure}[h]
	\centering
	\begin{overpic}
		[width=.71\textwidth,tics=6]{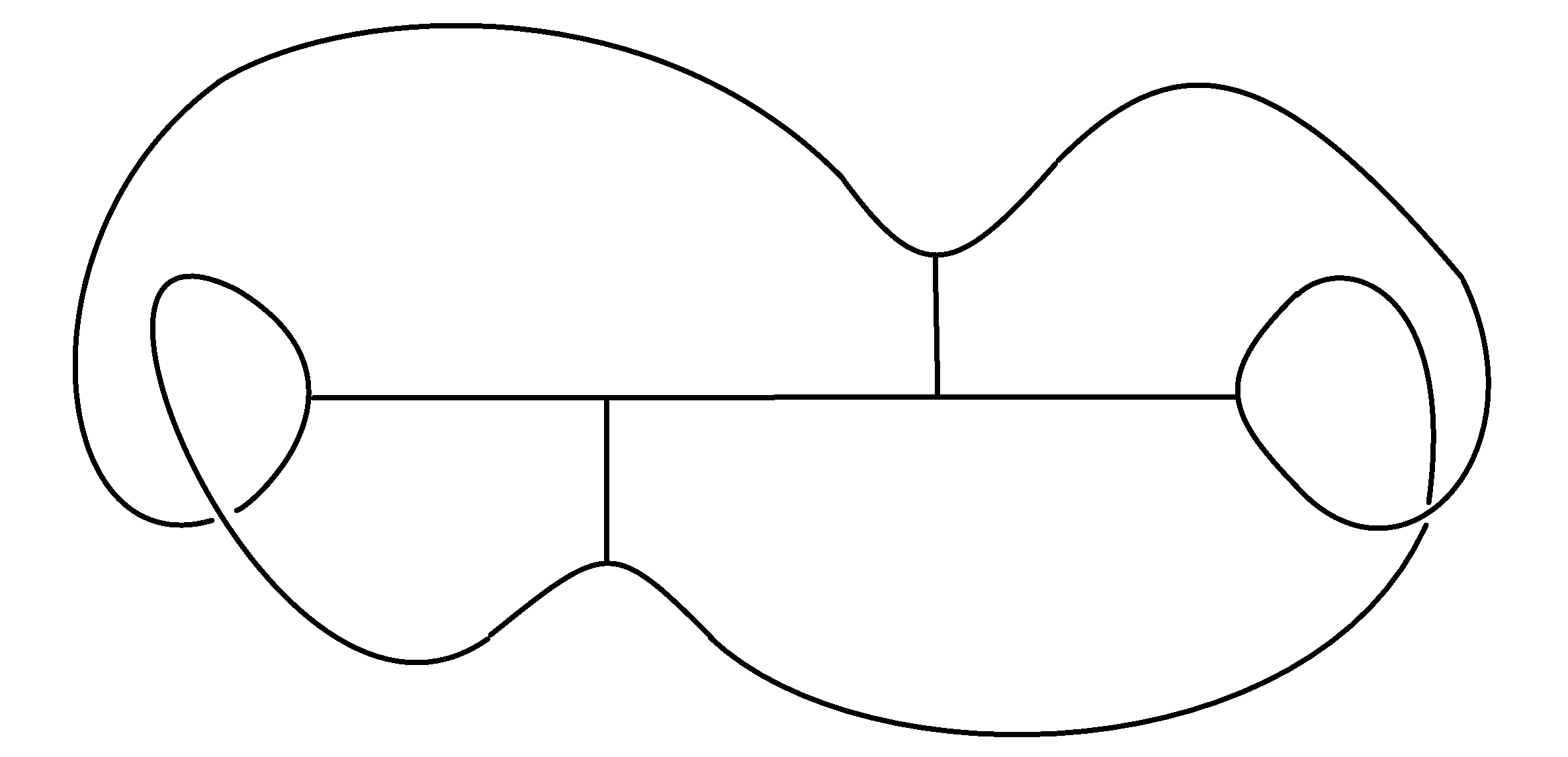}
		\put (80,45) {$\displaystyle y_{1}$}	
		\put (35,17) {$\displaystyle y_6$}
		\put (26,25) {$\displaystyle y_7$}
		\put (56,29) {$\displaystyle y_5$}
		\put (46,20) {$\displaystyle y_4$}
		\put (10,45) {$\displaystyle y_9 $}
		\put (72.6,-1) {$\displaystyle y_2 $}
		\put (27,3) {$\displaystyle y_8 $}
		\put (67.3,20) {$\displaystyle y_3$}		
		
	\end{overpic}
	\caption{The graph corresponding to the ribbon graph in figure \ref{ribon5}. This graph can be embedded within the ribbon graph, or, by thickening the graph, one can obtain the associated ribbon graph.}
	\label{graph}
\end{figure}
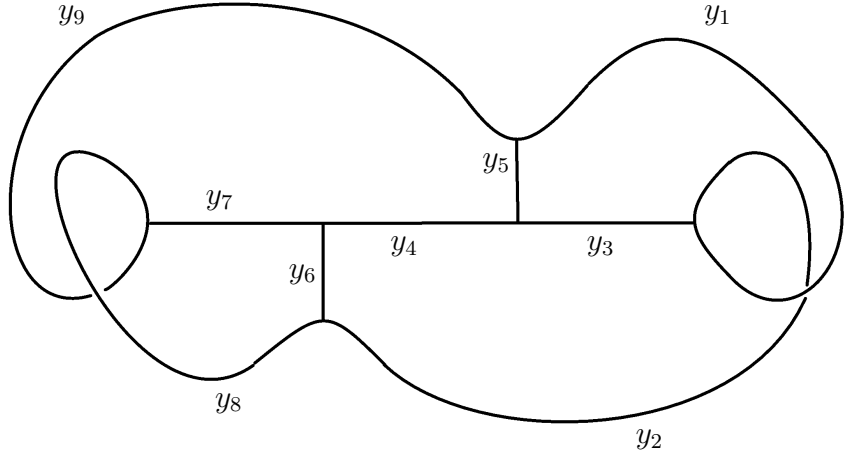
The application of the thin strip limit  simplifies the decomposition of the moduli space related to hyperbolic surfaces. Within this thin strip context, the moduli space can be expressed as a summation over trivalent ribbon graphs. This summation is coupled with an integral over the lengths of the edges constituting these graphs, with the condition that the boundaries have lengths $a_i$. The resulting expression is given by:
\begin{equation}\label{moduli}
	V^{\text{Airy}}_{g,n}(\textbf{a})=\sum_{\Gamma_{g,n}}\frac{2^{2g-2+n}}{\left|\text{Aut}\left(\Gamma_{g,n} \right)  \right| }\prod_{k=1}^{E}\int_{0}^{\infty}dy_{k}\prod_{i=1}^{n}\delta\left( a_i-\sum_{k=1}^{n}n^{i}_{k}y_{k}\right). 
\end{equation}
Here, $\Gamma_{g,n}\in\Gamma$, where $\Gamma$ is the set of trivalent ribbon graphs with genus $g$ and $n$ boundaries. These graphs are constructed from
$E = 6g - 6 + 3n$ edges and $V = 4g - 4 + 2n$ trivalent vertices. In the above expression $y_k$ is the length of edge $k$, and
$n_i\in \left\lbrace 0,1,2 \right\rbrace $
denotes the number of sides of edge $k$ that belong to boundary $i$.

\begin{figure}[H]
	\centering
	\begin{overpic}
		[width=.7\textwidth,tics=6]{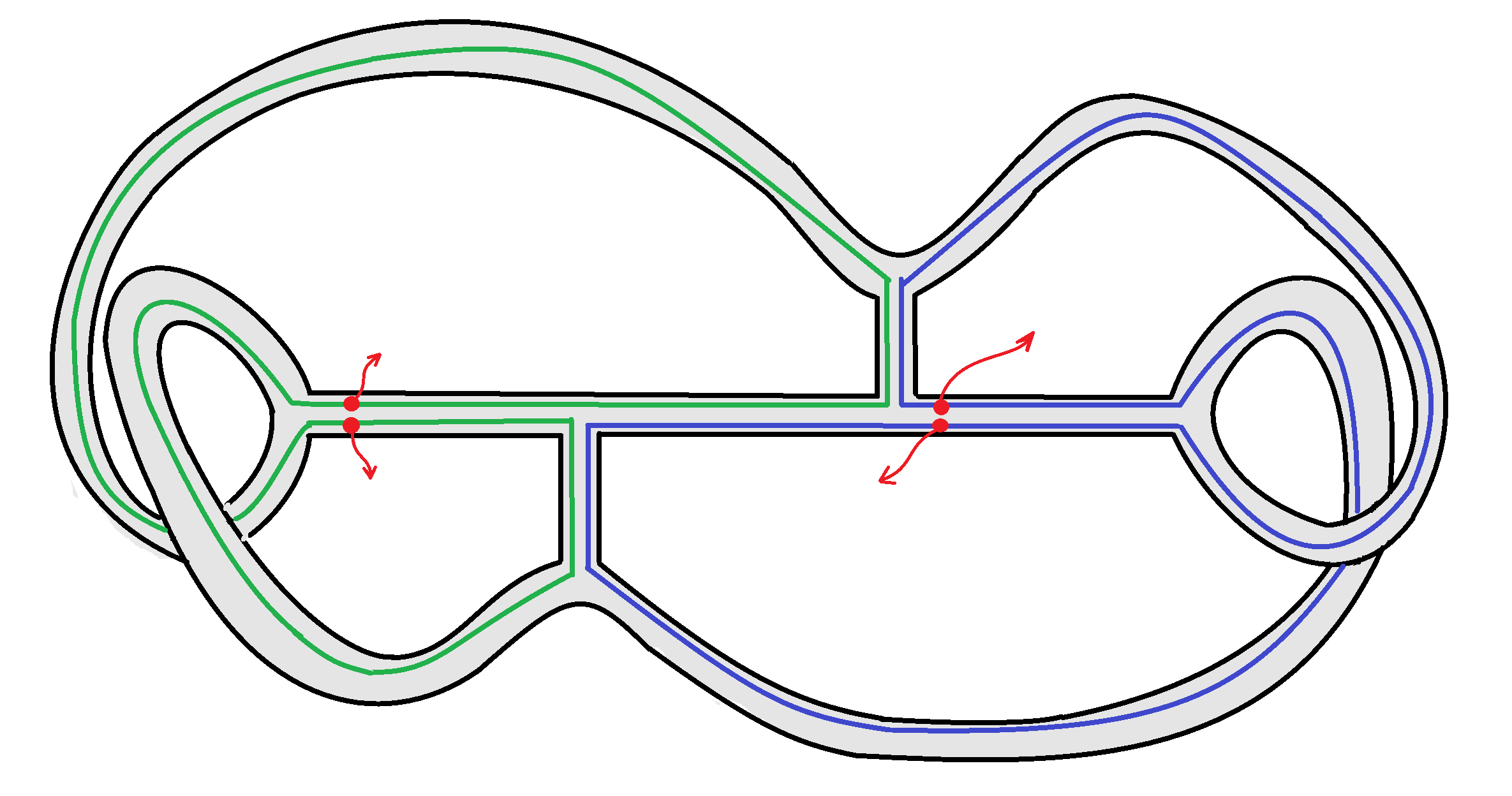}
		\put (80,47) {$\displaystyle y_{1}$}	
		\put (33,19) {$\displaystyle y_6$}
		\put (32,29) {$\displaystyle y_7$}
		\put (54,30) {$\displaystyle y_5$}
		\put (46,21) {$\displaystyle y_4$}
		\put (10,49) {$\displaystyle y_9 $}
		\put (72.6,0) {$\displaystyle y_2 $}
		\put (27,4) {$\displaystyle y_8 $}
		\put (67.3,21) {$\displaystyle y_3$}
		\put (55.3,17) {$\displaystyle A_{1}^{\prime}$}	
		\put (22.3,17) {$\displaystyle A_{2}^{\prime}$}
		\put (69.3,30) {$\displaystyle B_{1}$}	
		\put (22.3,31) {$\displaystyle B_{2}$}			
	\end{overpic}
	\caption{The blue and green curves correspond to  $a_1$ and $a_2$ geodesics, respectively.
		$A_{1}^{\prime}$ ($A_{2}^{\prime}$) and $B_1$ ($B_2$) represent typical points that are far apart along the $a_1$ ($a_2$) geodesic but are close within the bulk geometry. The distance they are situated along the $a_1$ ($a_2$) characterizes the twist $s_1$ ($s_2$).}
	\label{aa}
\end{figure}

\begin{figure}[h]
	\centering
	\begin{overpic}
		[width=.4\textwidth,tics=6]{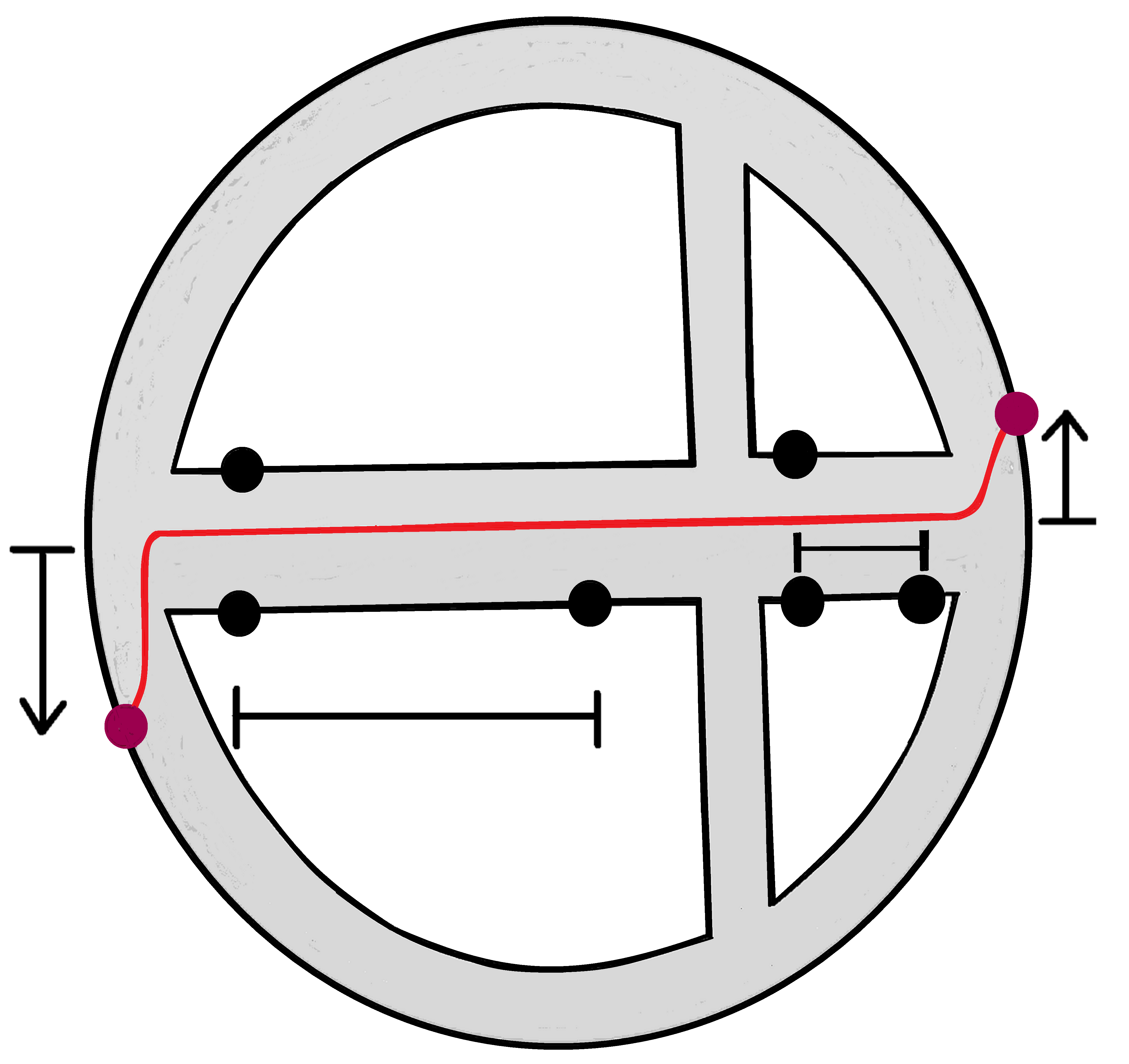}	
		\put (-6,37) {$\displaystyle x_{1}$}	
		\put (33,67) {$\displaystyle a_{1}$}	
		\put (69,66) {$\displaystyle a_{2}$}	
		\put (45,18) {$\displaystyle a_{1}$}	
		\put (70,21) {$\displaystyle a_{2}$}
		\put (35,27) {$\displaystyle s_{1}$}	
		\put (73,42.5) {$\displaystyle s_{2}$}			
		\put (5,65) {$\displaystyle b$}	
		\put (99,53) {$\displaystyle x_{2}$}	
		\put (20,56) {$\displaystyle A_{1}$}	
		\put (21,33.5) {$\displaystyle A^{\prime}_{1}$}	
		\put (50,33.5) {$\displaystyle B_{1}$}
		\put (70,56.6) {$\displaystyle A_{2}$}	
		\put (68.5,33.4) {$\displaystyle A^{\prime}_{2}$}	
		\put (76,33) {$\displaystyle B_{2}$}
	\end{overpic}
	\caption{ $A_1$ ($A_2$) is a typical point along $a_1$ ($a_2$) geodesic, which should be identified to $B_1$ ($B_2$) point.  $A_{1}^{\prime}$ ($A_{2}^{\prime}$) is the mirror image of  $A_1$ ($A_2$) point and it is very close to it. The twist $s_1$ ($s_2$) is distance between $A_{1}^{\prime}$ ($A_{2}^{\prime}$) and  $B_1$ ($B_2$).}
	\label{ribont}
\end{figure}
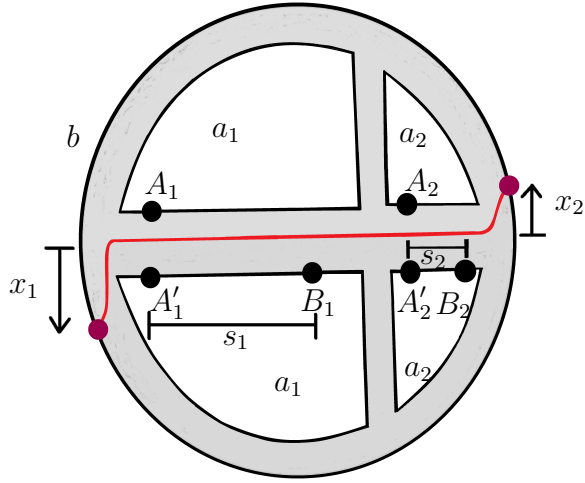

For a double handle disk, the volume of the moduli space is \cite{mir1}:
 \begin{equation}\label{volairy}
 	V_{2,1}\left( b\right) =\frac{	1}{2211840}\left(4\pi^2+b^2 \right) \left(12\pi^2+b^2 \right)\left(6960\pi^2+384\pi^2b^2+5b^4 \right), 
 \end{equation}
in the limit of large 
$b$, it approximates to:
 \begin{equation}
 	V_{2,1}^{\text{Airy}}\left( b\right) \approx\frac{b^8}{442368}.
 \end{equation}
 A portion of the moduli space volume, corresponding to the ribbon graph depicted in figure \ref{doubler}, can be calculated using equation \eqref{moduli}. The length of the boundary of double handle disk according to figure \ref{doubler} is:
 \begin{equation}\label{.1}
 	b=2\left( y_1+y_2+y_3+y_4+y_5+y_6+y_7+y_8+y_9\right),
 \end{equation}
for more explanation, one can begin from an arbitrary point on the thick black line and follow it. After tracing both sides of each edge, one will eventually return to the starting point. So, the contribution of this ribbon graph for the volume
of the moduli space of double torus with one boundary is:
 \begin{dmath}\label{ribairy}
 	{	V_{2,1}^{\text{Airy}}\left( b\right) \supset2^3 \int_{0}^{\infty}\delta\left( 2\sum_{i=1}^{9}y_i-b \right) \prod_{i=1}^{9} \text{d}y_i=\frac{6}{35}\frac{b^{8}}{442368}}.
 \end{dmath}
After the preparations to establish the connection between the graph geometries and ordinary
hyperbolic geometry, we will proceed to perform the integrals $\text{d}a_2\text{d}s_2\text{d}a_1\text{d}s_1$ in equation \eqref{prt}. According to equations \eqref{airy} and \eqref{ribairy}, the $y_i$'s can be represented as functions of $(a_1,s_1,a_2,s_2,b_1,\tau_1,b_2,\tau_2,b)$. 
As shown in figure \ref{gin}, there are four distinct closed geodesics, where none of them intersect with $\ell$. Figure \ref{aa} shows geodesics $a_1$ and $a_2$, which  do not intersect each other, nor do they intersect the $\ell$ geodesic depicted in figure \ref{doubler}.  By cutting the ribbon graph along these curves, one can obtain figure \ref{ribon5}.
It is evident that the $a_1$ geodesic goes once
 through  the $y_1, y_2, y_4$, and $y_6$ edges, while it passes through the $y_3$ edge twice, leading to:
\begin{equation}\label{.2}
	a_1 = y_1 + y_2 + 2y_3 + y_4 + y_5 + y_6.
\end{equation}
Similarly, for $a_2$:
\begin{equation}\label{.4}
	a_2 = y_4 + y_5 + y_6 + 2y_7 + y_8 + y_9.
\end{equation}
The identification of the $a_1$ geodesics and $a_2$ geodesics in a five-holed sphere  can be done up to twists $s_1$ and $s_2$, respectively (see figure \ref{ribont}). One can specify twist $s_1$ ($s_2$) by stating that the $A_1$ ($A_2$) point is identified with the $B_1$ ($B_2$) point, where $s_1$ ($s_2$) represents the distance between the $B_1$ ($B_2$) point and the ``mirror image'' of the $A_1$ ($A_2$) point, denoted as $A_{1}^{\prime}$ ($A_{2}^{\prime}$).
As shown in figure \ref{aa}, the $a_1$ ($a_2$) geodesic  passes through the very thin $y_3$ ($y_7$) edge twice. In this edge, the $a_1$ ($a_2$) geodesic does not intersect itself but closely follows a nearly identical path. By examining figure \ref{ribont}, it becomes evident that within this retraced section of the $a_1$ ($a_2$) geodesic, a specific point, $A_{1}^{\prime}$ ($A_{2}^{\prime}$), is becoming very close to another point, $B_1$ ($B_2$), located at a distance of $s_1$ ($s_2$) along the $a_1$ ($a_2$) geodesic.
So, to determine $s_1$ ($s_2$) in terms of the $y_i$'s, the distance along the $a_1$ ($a_2$) geodesic at which pairs of points are brought close must be found. The two indicated points on the $a_1$ ($a_2$) geodesic are close in the
bulk, but to travel from one to the other along the $a_1$ ($a_2$) geodesic, the indicated path must be followed, which has a length of $y_1 + y_3 + y_5$ $(y_4 + y_5 + y_7 + y_9)$. So we have:
\begin{equation}\label{.5}
 s_1 = y_1 + y_3 + y_5.
\end{equation}
and
\begin{equation}\label{.6}
	s_2 = y_4 + y_5 + y_7 + y_9.
\end{equation}
In addition to  $a_1$ and $a_2$ geodesics, there are two more simple closed curves, $b_1$ and $b_2$. These curves do not intersect with each other as well as $\ell$, $a_1$ and $a_2$ geodesics. However, certain segments of these curves overlap with segments of the $a_1$ and $a_2$ geodesics. According to figures \ref{j} and \ref{jj}, the lengths of these curves are:
\begin{equation}\label{.7}
	b_1 = y_1 + y_2 + 2y_3 + 2y_4 + 2y_5 + 2y_7 + y_8 + y_9,
\end{equation}
and
\begin{equation}\label{.8}
	b_2 = y_1 + y_2 + 2y_3 + 2y_4 + 2y_6 + 2y_7 + y_8 + y_9.
\end{equation}
 By applying the same procedure used to determine the twists of  $a_1$ and $a_2$ geodesics, we can determine the twists of these curves as follows:
\begin{equation}\label{.9}
 \tau_1=y_1+y_3+y_4+y_5+y_7,
\end{equation}
and
\begin{equation}\label{.10}
  \tau_2=y_3+y_4+y_6+y_7+y_8.
\end{equation}

\begin{figure}[h]
	\centering
	\begin{overpic}
		[width=.7\textwidth,tics=6]{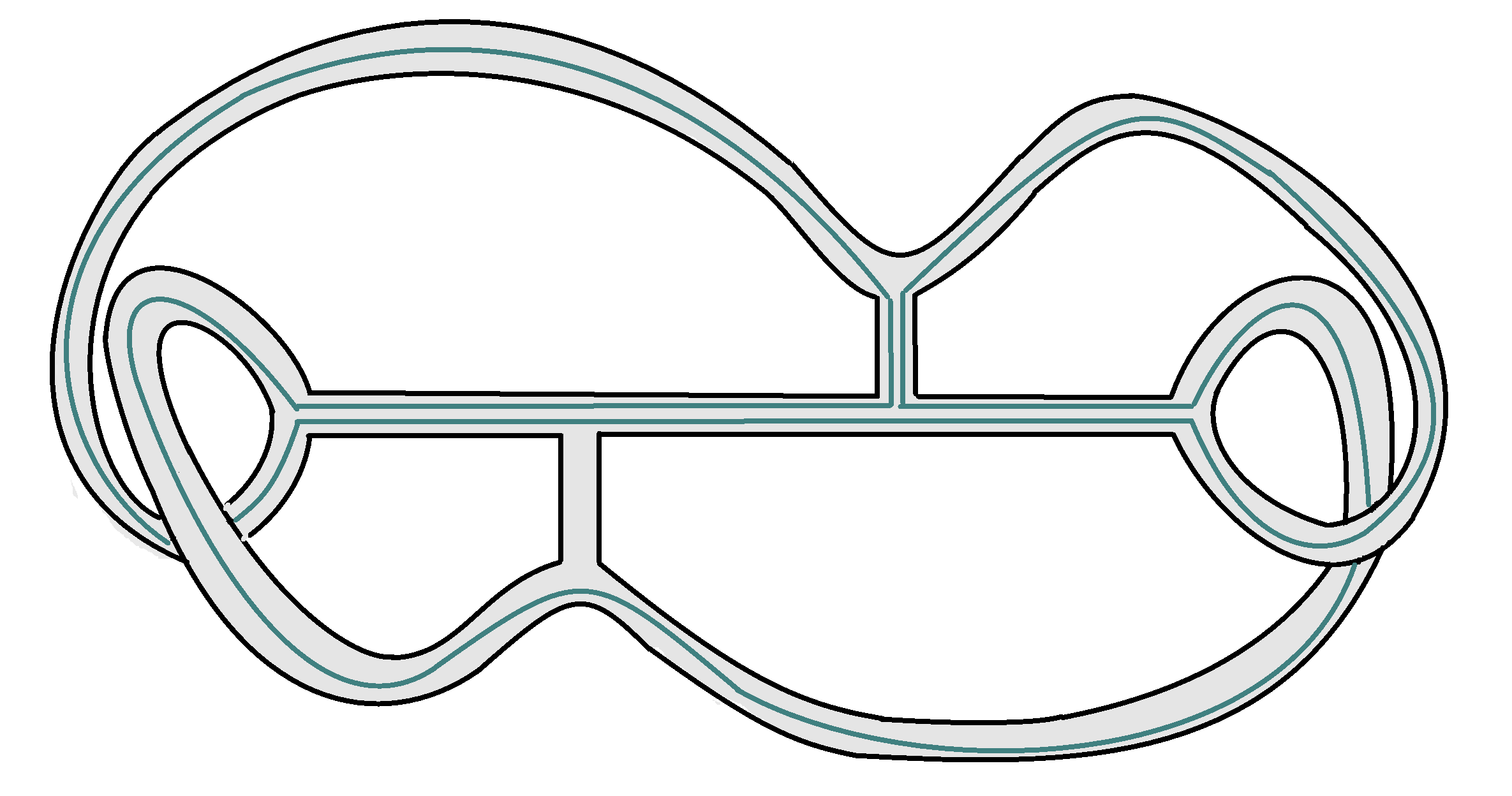}
		\put (80,47) {$\displaystyle y_{1}$}	
		\put (33,19) {$\displaystyle y_6$}
		\put (26,21) {$\displaystyle y_7$}
		\put (54,30) {$\displaystyle y_5$}
		\put (46,21) {$\displaystyle y_4$}
		\put (10,49) {$\displaystyle y_9 $}
		\put (72.6,0) {$\displaystyle y_2 $}
		\put (27,4) {$\displaystyle y_8 $}
		\put (67.3,21) {$\displaystyle y_3$}		
		
	\end{overpic}
	\caption{ The geodesic $b_1$ is depicted by the dark green curve.}
	\label{j}
\end{figure}
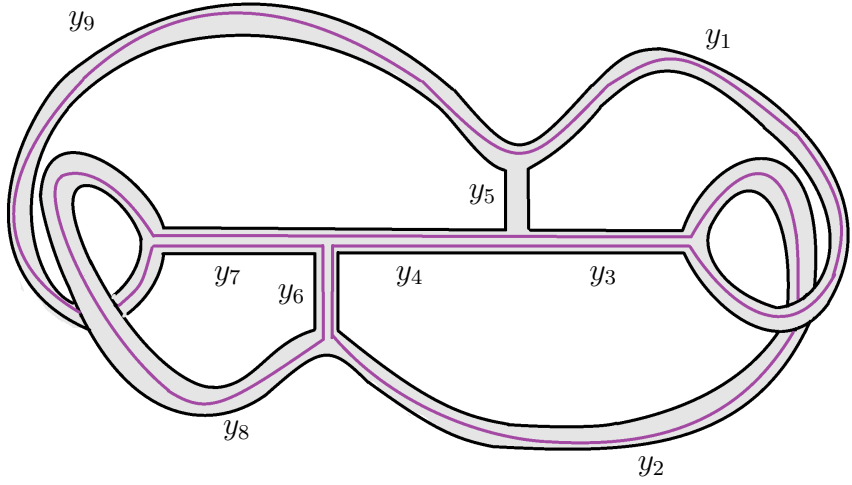
\begin{figure}[h]
	\centering
	\begin{overpic}
		[width=.7\textwidth,tics=6]{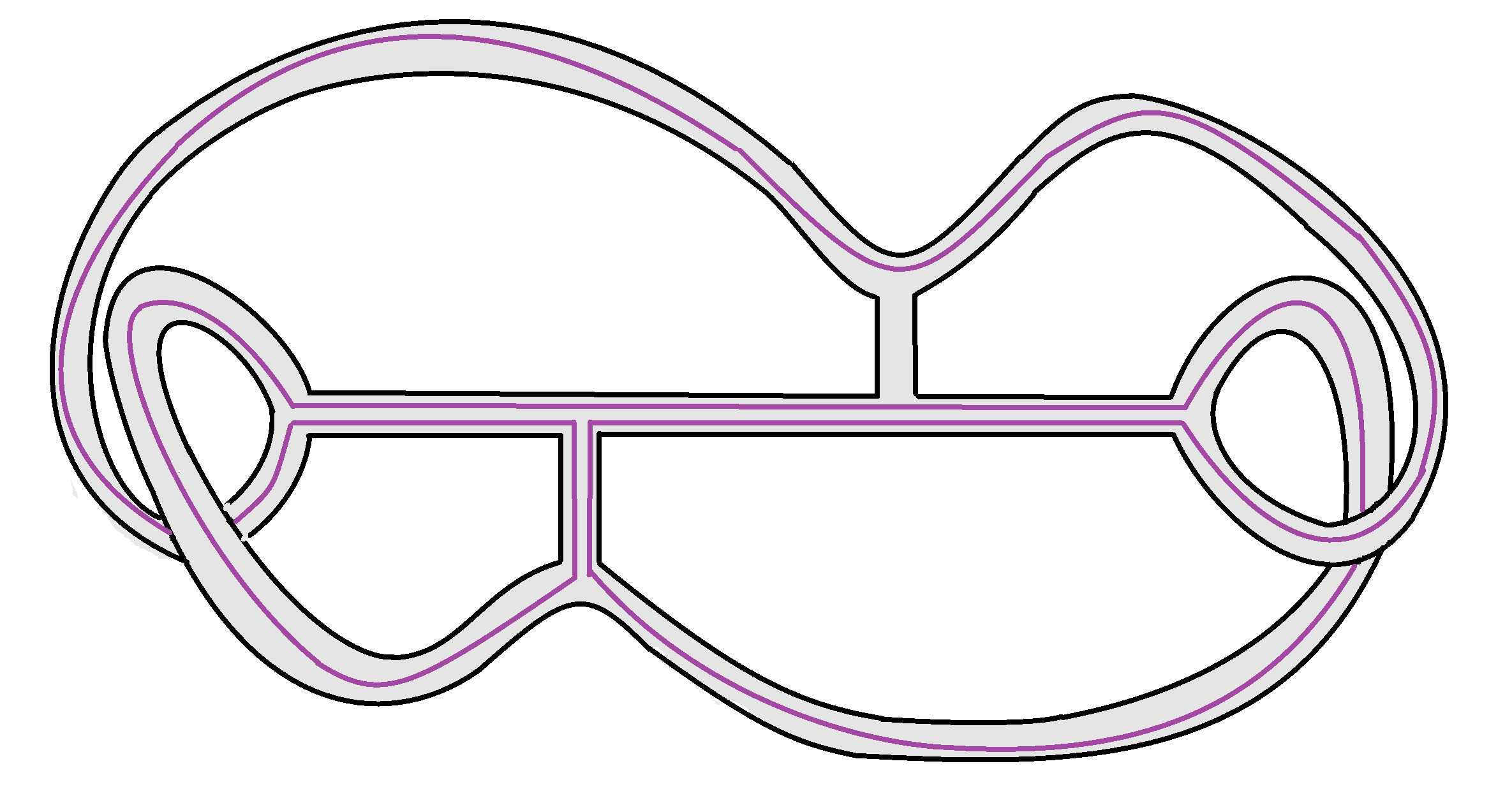}
		\put (80,47) {$\displaystyle y_{1}$}	
		\put (33,19) {$\displaystyle y_6$}
		\put (26,21) {$\displaystyle y_7$}
		\put (54,30) {$\displaystyle y_5$}
		\put (46,21) {$\displaystyle y_4$}
		\put (10,49) {$\displaystyle y_9 $}
		\put (72.6,0) {$\displaystyle y_2 $}
		\put (27,4) {$\displaystyle y_8 $}
		\put (67.3,21) {$\displaystyle y_3$}		
		
	\end{overpic}
	\caption{The geodesic $b_2$ is depicted by the purple curve.}
	\label{jj}
\end{figure}
 Equation \eqref{.1} along with eight equations from \eqref{.2} to \eqref{.10} can be combined to obtain:
\begin{align}
&	y_1= \frac{1}{2} \left(-a_1-a_2+b-b_1\right)+\tau _1,\quad\qquad\quad {y_2= \frac{1}{2} \left(a_1+a_2+b_1-2 \left(s_1+s_2+\tau
		_2\right)\right)},
\nonumber\\
&	y_3= \frac{1}{2} \left(-b+b_1+b_2+2 s_1-2 \tau _1\right),\quad\quad \quad y_4=
	-a_2+\frac{b}{2}-\frac{b_1}{2}-\frac{b_2}{2}-s_1+s_2+\tau _1+\tau _2,
\nonumber\\
&	y_5= \frac{1}{2} \left(a_1+a_2-b_2\right),\quad\quad\qquad\qquad\qquad \hspace{2mm}y_6=
	\frac{1}{2} \left(a_1+a_2-b_1\right),
\\
&	y_7= \frac{1}{2} \left(2 a_2-b+b_1+b_2-2 s_2-2 \tau _2\right),\quad\hspace{1mm} y_8=\frac{1}{2}
	\left(-a_1-a_2+b-b_2\right)+\tau _2,
\nonumber\\
&	y_9= \frac{1}{2} \left(-a_1-a_2+b_2+2 \left(s_1+s_2-\tau
	_1\right)\right).\nonumber
\end{align}
Now, we  would like to enforce the condition that $\ell$ geodesic should be free of shortcuts. This implies that on the trivalent ribbon geometry, segments of the $\ell$ geodesic should not overlap.
So, according to figure \ref{doubler} we require:
\begin{equation}\label{x1}
	|x_1|<y_2+y_6=a_1+a_2-s_1-s_2-\tau _2,
\end{equation}
and
\begin{equation}\label{x2}	
	 |x_2|<y_8+y_6=\frac{1}{2} \left(b-b_1-b_2+2 \tau _2\right).
\end{equation}
From relations \eqref{x1} and \eqref{x2} one can write:
\begin{equation}
	|x_1|+|x_2|<\frac{1}{2} \left(2 a_1+2 a_2+b-b_1-b_2-2 s_1-2 s_2\right).
\end{equation}
Assuming the replacement of the $|x|$ functions in the strip approximation with their corresponding $2 \log \cosh(x/2)$ functions, and subsequently employing \eqref{mwod}, along with the delta functions in \eqref{beta} and \eqref{delt}, we have:
\begin{equation}\label{upb}
	s_1+ s_2	<  a_1+ a_2-\ell.
\end{equation}
Given that all $y_i$'s are greater than 0, we can write the following inequality:
\begin{align}
	0	<  y_1+ y_9=\frac{1}{2} \left(-2 a_1-2 a_2+b-b_1+b_2+2 s_1+2 s_2\right),
\end{align}
which leads to:
\begin{equation}\label{up}
	a_1+a_2-\ell_t	< s_1+ s_2.
\end{equation}
Furthermore, it can be observed that:
\begin{equation}\label{upd}
0<y_1+y_2+y_5+y_6+y_8+y_9\approx a_1+a_2-2\lambda-2\ell,
\end{equation}
and using the delta function, setting $\ell = a_1 + a_2 - 2\lambda - \ell_t$, we have: 
\begin{align}\label{gir}	
	a_1+a_2<2\ell_t+2\lambda.
\end{align}
By integrating the five-holed sphere answer \eqref{f5}, over the moduli space restricted by relations \eqref{upb}, \eqref{up} and \eqref{gir}  as follows:	
	\begin{align}
	\hat{P}(\ell, \lambda) \approx& \int_{0}^{\infty} \text{d}a_1\text{d}a_2 ~\hat{p}(\ell, \textbf{a})\theta\left(a_1+a_2-2\lambda-\ell_t \right)\theta\left(2\ell_t+2\lambda-a_1-a_2 \right)\nonumber\\ &\int_{0}^{a_2} \int_{0}^{a_1} \text{d}s_1\text{d}s_2 ~\theta\left(a_1+a_2-\ell-s_1-s_2 \right) \theta\left(s_1+s_2+\ell_t-a_1-a_2 \right),
	\end{align}
expression \eqref{pro2} is derived.	
\section{Probability distributions in higher genus}\label{6}
In this section, the previous calculation will be expanded to include the emission of $n$ baby universes. The volume $V_{g, n+1}(b,\textbf{a})$ in \eqref{amp} is a symmetric polynomial
function in $ b^2, a_1^2, a_2^2,\ldots, a_{n}^2$ of degree $3g - 3 + n+1$, and can  be written
as  \cite{mir,mir1}:
\begin{align}\label{mir}
	V_{g,n+1}(b,\textbf{a})=\sum_{|\alpha|\leq 3g-3+n+1}c_{g,n+1}\left( \alpha\right)
	\frac{b^{2\alpha_{n+1}}}{2^{2\alpha_{n+1 }}(2\alpha_{n+1}+1)!} 
	\prod_{j=1}^{n}\frac{a_{j}^{2\alpha_{j}}}{2^{2\alpha_{j}}(2\alpha_{j}+1)!}.
\end{align}
Simplifying by setting $g=0$, we approximate relation \eqref{mir} for $b\gg1$ as:
\begin{align}\label{ff}
	{	V_{0,n+1}(b,\textbf{a})\approx 	b^{2(n-2)}\mathcal{F}\left(\textbf{a}\right) },
\end{align}
where
\begin{align}
\mathcal{F}\left(\textbf{a}\right)=\sum_{|\alpha|\leq n-2} \frac{c_{0,n+1}\left( \alpha\right)}{2^{2(n-2)}(2(n-2)+1)!}
\prod_{j=1}^{n}\frac{a_{j}^{2\alpha_{j}}}{2^{2\alpha_{j}}(2\alpha_{j}+1)!}.
\end{align}
The equation:
\begin{equation}\label{pro}
	\hat{P}_{n}\left( \ell\right)\approx e^{S_{0}\chi} \int \text{d}\textbf{a}\text{d}\textbf{s}\int \text{d}\omega\left(2\sqrt{E}\right) \frac{b_1^{2n-3}\sin \left( b_{1}\sqrt{E_{1}}\right) }{2\pi E_{1}}\frac{b_2^{2n-3}\sin \left( b_{2}\sqrt{E_{2}}\right) }{2\pi E_{2}}\mathcal{F}^{2}\left(\textbf{a}\right)e^{-i2t\sqrt{E}\omega }\langle E_{2}|\ell\rangle\langle\ell|E_1\rangle,
\end{equation}
represents the probability of emitting $n$ baby universes $(n\geq2)$ and obtaining a wormhole with length $\ell$. This expression can be derived using the same steps that were taken to obtain equation \eqref{step}. In this case, equation \eqref{amp2} needs to be modified by substituting \eqref{ff}, and in equation \eqref{int}, $k$ should be replaced by ``$2n-3$''.
It is worth mentioning that,
$d\textbf{a}d\textbf{s}=\prod_{j=1}^{n}da_{j}\prod_{j=1}^{n}ds_{j}$.
The integration over $\omega$ in equation \eqref{pro} yields the following result:
\begin{align}\label{nh}
\hat{P}_{n}\left( \ell\right)\approx e^{\chi S_{0}-2 \pi  \sqrt{E}}\frac{1}{2 E^2} \int& \text{d}\textbf{a}\text{d}\textbf{s}~\mathcal{F}^{2}\left(\textbf{a}\right)b_1^{2n-3}b_2^{2n-3}\bigg\lbrace \delta\left( \frac{1}{2}   (b_1+b_2-2( \ell- \ell_t))\right) \nonumber\\&+ \delta\left( \frac{1}{2}(b_1+b_2-2 (\ell+\ell_t))\right) +\delta\left( \frac{1}{2}(b_1+b_2+2 (\ell-\ell_t))\right) 
\bigg\rbrace,
\end{align}
which is similar to equation \eqref{proo}.

\subsection{Firewall-free geometries}
The probability of finding a wormhole with a length of $\ell$ in a firewall-free geometry, after the emission of more than two baby universes is:
\begin{align}\label{ex}	
	\hat{P}_{n,\text{smooth}}\left( \ell\right)&\approx e^{\chi S_{0}-2 \pi  \sqrt{E}}\frac{1}{2  E^2} \int \text{d}\textbf{a}\text{d}\textbf{s}\mathcal{F}^{2}\left(\textbf{a}\right)b_1^{2n-3}b_2^{2n-3}
	\delta\left(\frac{1}{2}(b_1+b_2+2 (\ell-\ell_t))\right)\nonumber\\&\approx e^{-\chi S_{0}-2 \pi  \sqrt{E}}\frac{\left(\ell_t-\ell\right) ^{4n-6}}{2  E^2} \int \text{d}\textbf{a}\text{d}\textbf{s}~\mathcal{F}^{2}\left(\textbf{a}\right)   \delta\left( \sum_{j=1}^{n}a_j-2\lambda-(\ell_t-\ell)\right). 
\end{align}
Using the thin strip approximation $(b_1+b_2)/2$ has been replaced with $\sum_{j=1}^{n}a_j-2\lambda$, as illustrated in figure \ref{f}.
The
$\lambda$ corresponds to the region between all baby universes. 
In this
 geometry, similar to the geometry described in subsection \eqref{i}, the no shortcut criterion does not impose any restrictions on the moduli space. Therefore, the domain of each twist is $0<s_j<a_j$. Due to the presence of the delta function, we can assume that the domain of each baby universe is within the range of $0 < a_j < \infty$.
By expressing $\mathcal{F}^{2}\left(\textbf{a}\right)$ in the following form:
\begin{align}
	\mathcal{F}^{2}\left(\textbf{a}\right) =\sum_{|\alpha,\beta|\leq n-2} 
	\prod_{j=1}^{n}	\mathcal{A}^{(n)}_{\alpha_{j},\beta_{j}}\left(\alpha,\beta\right)a_{j}^{2(\alpha_{j}+\beta_{j})},
\end{align}
where:
\begin{align}
	\mathcal{A}^{(n)}_{\alpha_{j},\beta_{j}}\left(\alpha,\beta\right) = \frac{1}{2^{4(n-2)}\left( (2n-3 )!\right)^{2} }
	\frac{c_{0,n+1}\left( \alpha\right)c_{0,n+1}\left( \beta\right)}{2^{2(\alpha_{j}+\beta_{j})}(2\alpha_{j}+1)!(2\beta_{j}+1)!},
\end{align}
the expression \eqref{ex} after integrating over all twists becomes:
\begin{align}\label{hi}	
	\hat{P}_{n,\text{smooth}}\left( \ell\right)\approx e^{ -S_{0}(2n-1)-2 \pi  \sqrt{E}}\frac{\left(\ell_t-\ell\right) ^{4n-6}}{2  E^2}\sum_{|\alpha,\beta|\leq n-2} 
P^n_{(\alpha,\beta)}\left( \ell\right),
\end{align}
where $P^n_{(\alpha,\beta)}\left( \ell\right)$ is defined as:
\begin{align}\label{z}
P^n_{(\alpha,\beta)}\left( \ell\right)= \int_{0}^{\infty} 		\prod_{j=1}^{n}\mathcal{A}^{(n)}_{\alpha_{j},\beta_{j}}\left(\alpha,\beta\right)\text{d}a_ja_{j}^{2(\alpha_{j}+\beta_{j})+1}    \delta\left( \sum_{k=1}^{n}a_k-2\lambda-(\ell_t-\ell)\right). 
\end{align}
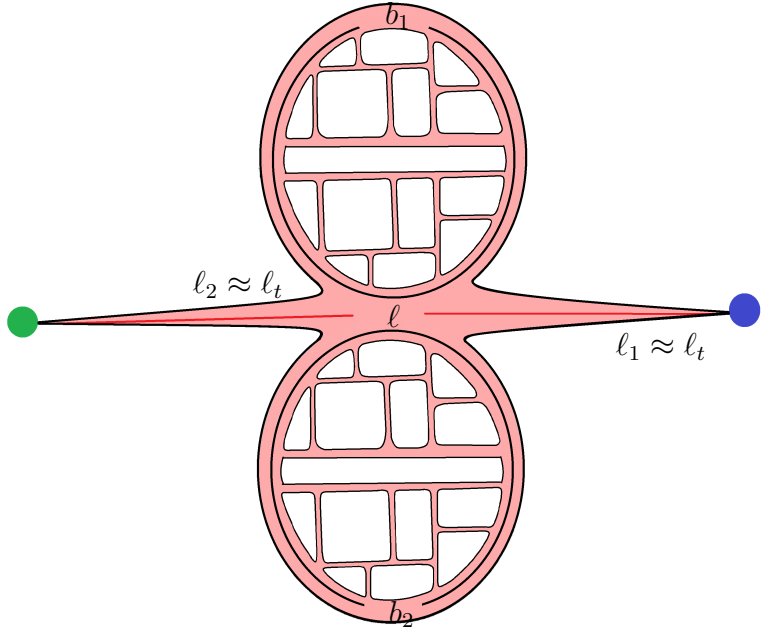
\begin{figure}[h]
	\centering
	\begin{overpic}
		[width=0.74\textwidth]{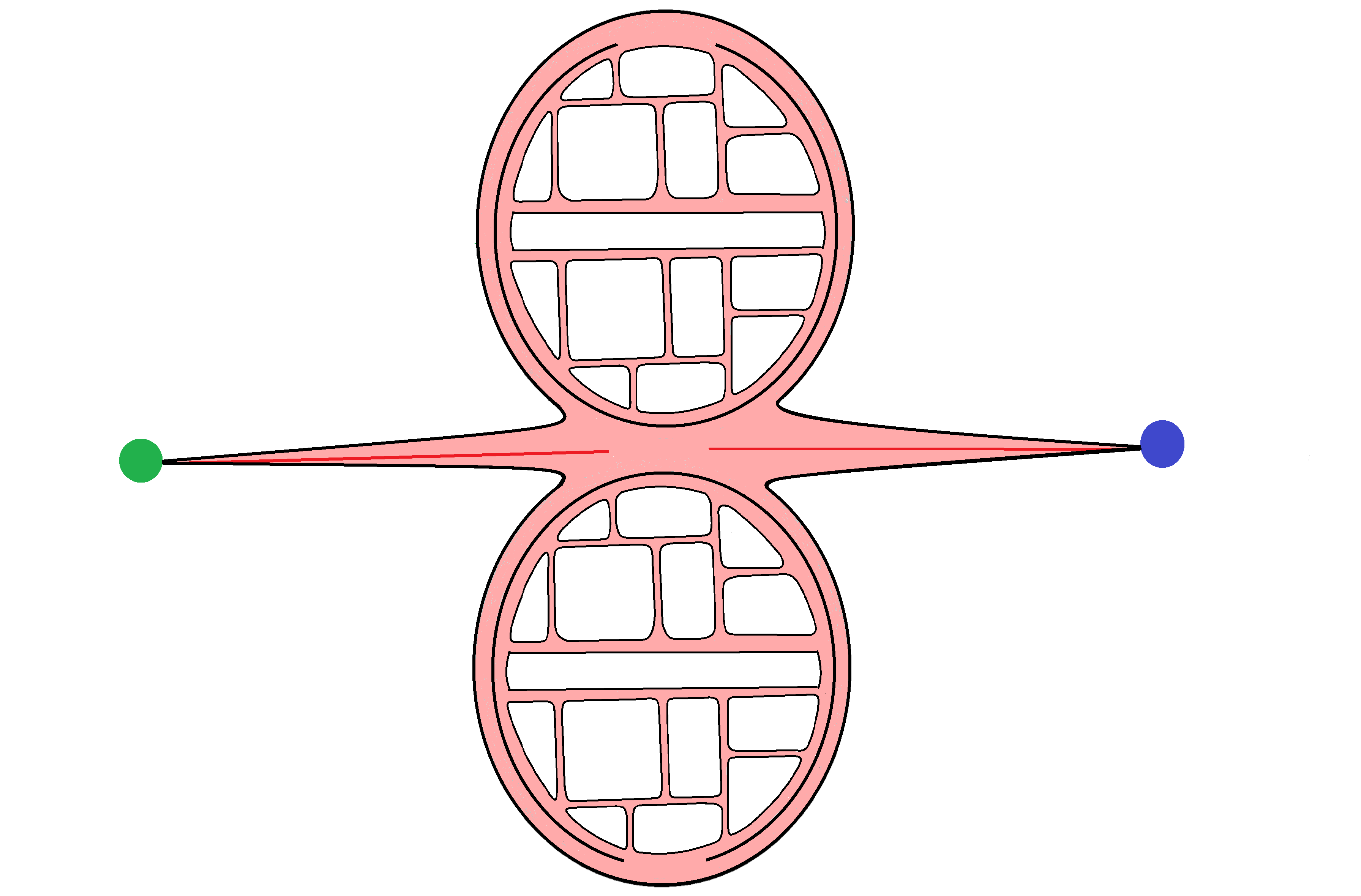}
		\put (48,63) {$\displaystyle b_{1}$}	
		\put (48.22,31.5) {$\displaystyle\ell$}	
		\put (72,28.5) {$\displaystyle\ell_{1}\approx\ell_{t}$}	
		\put (28,35) {$\displaystyle\ell_{2}\approx\ell_{t}$}	
		\put (48.4,0.7) {$\displaystyle b_{2}$}
	\end{overpic}
	\caption{
		The geometry corresponding to $\ell=\ell_{t}-\tilde{b}$, approximated with a ribbon graph, features a growing wormhole that emits $n$ baby universes within the black hole regions.}	
	\label{f}
\end{figure}
It is convenient to multiply \eqref{z} by a factor $``1"$ written using the definition of gamma function,
\begin{equation}
	\frac{1}{\Gamma\left( 2(n+ \sum_{j=1}^{n}\alpha_j+\beta_j)+1\right)  }\int_{0}^{\infty}\zeta^{2(n+ \sum_{j=1}^{n}\alpha_j+\beta_j)}e^{-\zeta}\text{d}\zeta=1,
\end{equation}
and perform a change of variables to $q_k=\zeta a_k$:
\begin{align}\label{l}
\nonumber	P^n_{(\alpha,\beta)}\left( \ell\right)=	\frac{1}{\Gamma\left( 2(n+ \sum_{j=1}^{n}\alpha_j+\beta_j)+1\right)  }&\int_{0}^{\infty}	\prod_{j=1}^{n}\mathcal{A}^{(n)}_{\alpha_{j},\beta_{j}}\left(\alpha,\beta\right)\text{d}q_jq_{j}^{2(\alpha_{j}+\beta_{j})+1}  \\& \int_{0}^{\infty} 	\text{d}\zeta e^{-\zeta} \delta\left( \sum_{k=1}^{n}\frac{ q_k}{\zeta}-2\lambda-(\ell_t-\ell)\right). 
\end{align}
By integrating over $\zeta$, the delta function is eliminated, resulting in the following expression:
\begin{align}\label{zs}
	P^n_{(\alpha,\beta)}\left( \ell\right)= 	\frac{(\ell_t-\ell+2 \lambda )^{-2}}{\Gamma\left( 2(n+ \sum_{j=1}^{n}\alpha_j+\beta_j)+1\right)  }\sum_{k=1}^{n}\int_{0}^{\infty}	\prod_{j=1}^{n}\mathcal{A}^{(n)}_{\alpha_{j},\beta_{j}}\left(\alpha,\beta\right)\text{d}q_jq_{j}^{2(\alpha_{j}+\beta_{j})+1}\exp\left( \frac{-q_j}{\ell_t-\ell+2 \lambda }\right) q_k.
\end{align}
Now, the definition of the gamma function can be used to obtain the following expression: 
\begin{align}\label{zqa}
	P^n_{(\alpha,\beta)}\left( \ell\right)= \ell_t ^{2\sum_{j=1}^{n}(\alpha_j+\beta_j+1)-1}\mathcal{K}_{(\alpha,\beta)}\left(\lambda/\ell_t,\ell/\ell_t\right),
\end{align}
where:
\begin{align}\label{zsrs}
\mathcal{K}_{(\alpha,\beta)}\left(\lambda/\ell_t,\ell/\ell_t\right):=\frac{\left( 1-\frac{\ell}{\ell_t}+\frac{2\lambda}{\ell_t}\right) ^{\sum_{j=1}2(\alpha_j+\beta_j+1)-1}}{\Gamma\left( 2(n+ \sum_{j=1}^{n}\alpha_j+\beta_j)+1\right)  }	\sum_{k=1}^{n} 2(\alpha_k+\beta_k+1) \prod_{j=1}^{n}\mathcal{A}^{(n)}_{\alpha_{j},\beta_{j}}\left(\alpha,\beta\right)\Gamma\left( 2(\alpha_j+\beta_j+1)\right).
\end{align}
  If we choose the highest power of $\ell_t$ in \eqref{hi}, i.e. $\alpha, \beta=n-2 $, after integrating over $\lambda$, the probability of having a smooth geometry with a wormhole length of $\ell$ is given by:
\begin{align}	
	\hat{P}_{n,\text{smooth}}\left( \ell\right)\approx  e^{ -S_{0}(2n-1)-2 \pi  \sqrt{E}}\frac{\left(\ell_t-\ell\right) ^{4n-6}}{2  E^2} 
	\ell_t ^{4n^2-6n}\mathcal{K}_{n}\left(\ell/\ell_t\right),
\end{align}
where
\begin{align}	
\ell_t\mathcal{K}_{n}\left(\ell/\ell_t\right)=\int_{0} ^{\ell_t}\text{d}\lambda \mathcal{K}_{(n-2,n-2)}\left(\lambda/\ell_t,\ell/\ell_t\right).
\end{align}
In the above relation, we assumed that $\lambda$ varies in the range $0 < \lambda < \ell_t$.
Integrating over $\ell$ gives the following normalized probability distribution:
\begin{align}	
P_{n,\text{smooth}}=\mathcal{J}_{n} e^{-2nS(E)}e^{4 \pi  \sqrt{E}(n-1)}E^{2n^2-n-9/2} t ^{4n^2-2n-5},
\end{align}
and the numerical coefficient  $\mathcal{J}_{n}$ is:
\begin{align}\label{srs}
\mathcal{J}_{n}=\frac{2^{4n^2-8n-7}\left( 3^{4 n^2-6 n+1}-2^{4 n^2-6 n+1}-1\right)(2n-3) }{\pi^{4n}(8n^3-24n^2+20n-3)  \Gamma\left(4n^2-6n+1\right)  }  \left( \mathcal{A}^{(n)}_{n-2,n-2}\left(n-2,n-2\right)\Gamma\left( 4n-6\right)\right)^{n} .
\end{align}

\subsection{Firewall geometries}
The probability of encountering a wormhole with a length of $\ell$ in firewall geometry, following the emission of more than two baby universes, can be expressed as:
\begin{align}\label{hhi}	
	\hat{P}_{n,\text{firewall}}\left( \ell\right)\approx e^{ -S_{0}(2n-1)-2 \pi  \sqrt{E}}\frac{\left(\ell_t+\ell\right) ^{4n-6}}{2  E^2}\sum_{|\alpha,\beta|\leq n-2} 
	Q^n_{(\alpha,\beta)}\left( \ell\right),
\end{align}
where $Q^n_{(\alpha,\beta)}\left( \ell\right)$ is defined as:
\begin{align}\label{saq}
	Q^n_{(\alpha,\beta)}\left( \ell\right)= \int	\prod_{j=1}^{n}\mathcal{A}^{(n)}_{\alpha_{j},\beta_{j}}\left(\alpha,\beta\right)\text{d}a_j\text{d}s_ja_{j}^{2(\alpha_{j}+\beta_{j})}   \delta\left( \sum_{k=1}^{n}a_k-2\lambda-(\ell_t+\ell)\right), 
\end{align}
 \begin{figure}[h]
	\centering
	\begin{overpic}
		[width=0.37\textwidth]{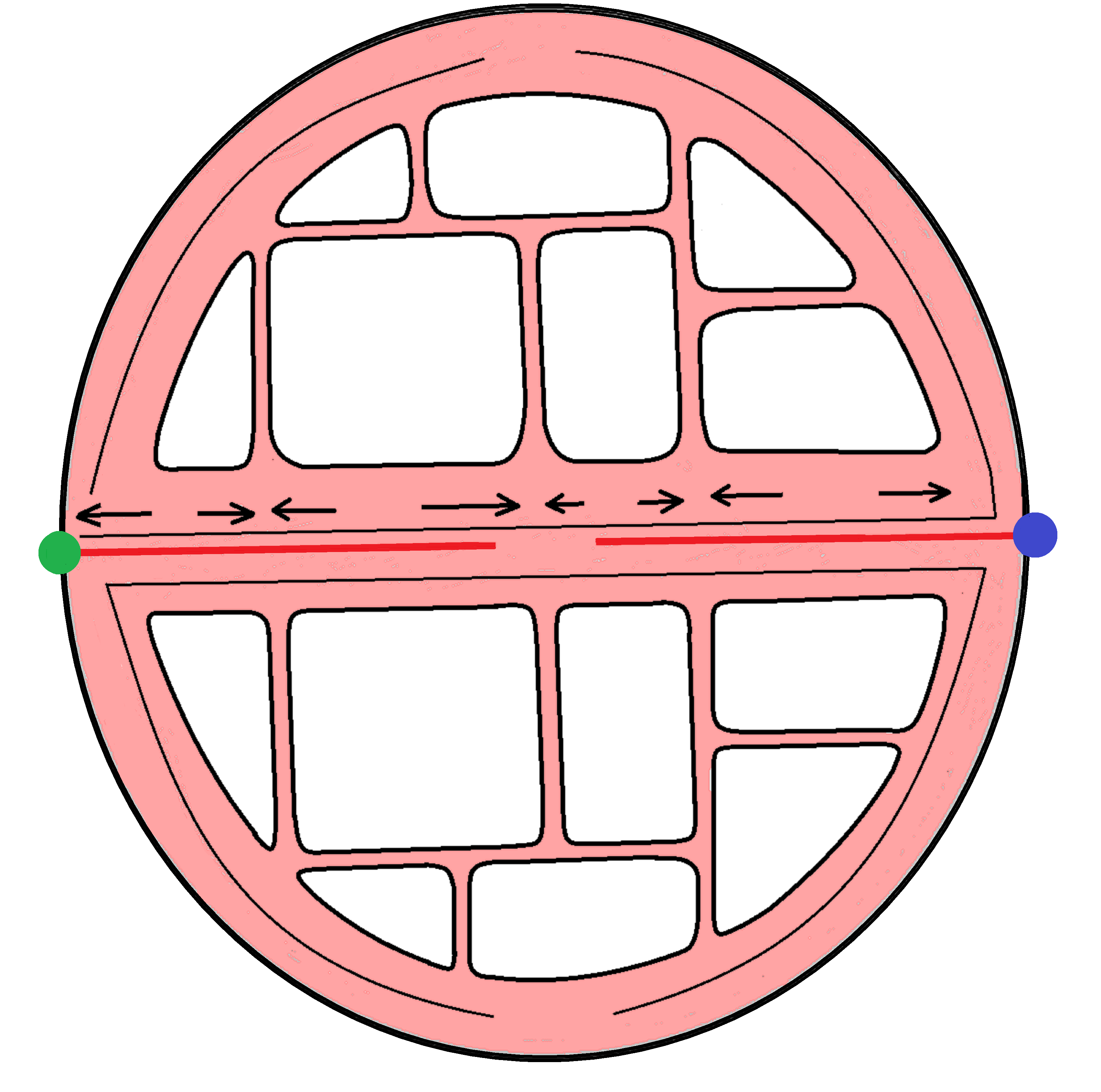}
		\put (47,92) {$\displaystyle b_{1}$}
		\put (52.5,53) {$\displaystyle\gamma_3$}
		\put (73,53.5) {$\displaystyle\gamma_4$}
		\put (31,52.5) {$\displaystyle\gamma_2$}	
		\put (13,52.5) {$\displaystyle\gamma_1$}		
		\put (47.22,46.8) {$\displaystyle\ell$}	
		\put (47,-5) {$\displaystyle\ell_1\approx\ell_{t}$}	
		\put (47,102) {$\displaystyle\ell_2\approx\ell_{t}$}	
		\put (49,5) {$\displaystyle b_{2}$}
	\end{overpic}
	\caption{The geometry, represented by $\ell=\tilde{b}-\ell_{t}$ and approximated with a ribbon graph, features a diminishing wormhole within the white hole regions, leading to the emergence of a firewall. This geometric configuration includes the emission of $n$ baby universes, with some of them sharing a common section with the wormhole.}	
	\label{p}
\end{figure}
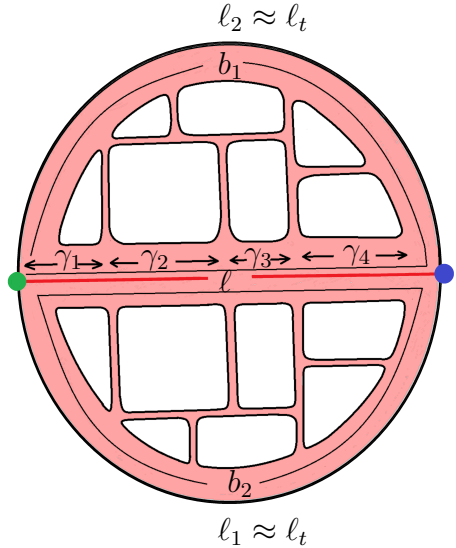
The intermediate steps in deriving the above relations are omitted as they follow a similar procedure to the derivation of relations \eqref{hi} and \eqref{z}.
In this geometry, the no shortcut criterion imposes  restrictions on the moduli space. As discussed in subsection \ref{three}, the existence of overlapping portions between the baby universes and wormholes is crucial for the application of the no shortcut criterion. Let us assume that each baby universe has a parameter $\gamma_i$ representing its common share with the wormhole, as indicated in figure \ref{p}. The domain of $\gamma_i$ is $(0, \ell)$ with the condition:
$	\sum_{j=1}^{n}\gamma_j=\ell.$
After imposing the no shortcut criterion, the twist $s_i$ of the $i$-th baby universe is constrained within the range $\gamma_i + \lambda_i + \lambda_{i+1} < s_i < a_i - \gamma_i$, where $\lambda_i$ is a common region  between baby universe $a_i$ and $a_{i+1}$. 
So, one can expect that:
\begin{equation}
\ell+2\lambda=\sum_{i=1}^{n}a_i-\ell_t<\sum_{i=1}^{n}s_i<\sum_{i=1}^{n}a_i-\ell=\ell_t+2\lambda.
\end{equation}
  Therefore, the relation \eqref{saq} after 
 imposing constraints \eqref{p} takes the following form:

\begin{align}\label{sq}
	Q^n_{(\alpha,\beta)}\left( \ell,\lambda\right)=& \int_{0}^{\infty}\prod_{j=1}^{n}\mathcal{A}^{(n)}_{\alpha_{j},\beta_{j}}\left(\alpha,\beta\right)	\text{d}a_ja_{j}^{2(\alpha_{j}+\beta_{j})}     \theta\left( \sum_{i=1}^{n}a_i-2\lambda-\ell_t\right)\theta\left(2\ell_t+2\lambda-\sum_{i=1}^{n}a_i \right)\nonumber\\&\delta\left(  \sum_{i=1}^{n}a_i-2\lambda-(\ell_t+\ell)\right)\int_{0}^{a_j}\text{d}s_j~ \theta\left(\ell_t+2\lambda -\sum_{i=1}^{n}s_i\right) \theta\left( \sum_{i=1}^{n}s_i-\ell-2\lambda\right). 
\end{align}
 After the change of variables $a_k \rightarrow \ell_tq_k$ and $ s_k \rightarrow \ell_t b_k$ one can express the relation \eqref{sq} as following:
\begin{align}\label{zsx}
	Q^n_{(\alpha,\beta)}\left( \ell,\lambda\right)= \ell_t ^{2\sum_{j=1}^{n}(\alpha_j+\beta_j+1)-1}
	\tilde{\mathcal{K}}_{(\alpha,\beta)}\left(\lambda/\ell_t,\ell/\ell_t\right),
\end{align}
where $\tilde{\mathcal{K}}_{(\alpha,\beta)}\left(\lambda/\ell_t,\ell/\ell_t\right)$ is a polynomial of $\lambda/\ell_t$ and $\ell/\ell_t$.\footnote{The method of calculating \eqref{sq} is provided in Appendix \ref{bqw}.} If the highest power of $\ell_t$ is chosen in equation \eqref{hhi}, i.e. $\alpha, \beta=n-2 $, the probability of encountering a firewall with a wormhole of length $\ell$  is given by:
\begin{align}	
	\hat{P}_{n,\text{firewall}}\left( \ell\right)\approx  e^{ -S_{0}(2n-1)-2 \pi  \sqrt{E}}\frac{\left(\ell_t+\ell\right) ^{4n-6}}{2  E^2} 
	\ell_t ^{4n^2-6n}\tilde{\mathcal{K}}_{(n-2,n-2)}\left(\ell/\ell_t\right).
\end{align}
Here $\tilde{\mathcal{K}}_{n}\left(\ell/\ell_t\right)$ is:
\begin{align}
	\ell_t\tilde{\mathcal{K}}\left(n-2,n-2,\ell/\ell_t\right)&=\int_{0}^{\ell_t}	\tilde{\mathcal{K}}_{(n-2,n-2)} \left(\lambda/\ell_t,\ell/\ell_t\right) \text{d}\lambda. 
\end{align}
Notice the no shortcut condition imposes no restrictions on $\lambda$ and in the above relation, we assumed that $\lambda$ varies in the region $0 < \lambda < \ell_t$.
After integrating over $\ell$, the normalized probability of encountering a firewall after emitting $n$ baby universes is:
\begin{align}\label{genera}	
	P_{n,\text{firewall}}= \tilde{\mathcal{J}}_{n}e^{-2nS(E)}e^{4 \pi  \sqrt{E}(n-1)}E^{2n^2-n-9/2} t ^{4n^2-2n-5},
\end{align}
and the $\tilde{\mathcal{J}}_{n}$ is a numerical coefficient.


\subsection{$``2n+1"$-holed sphere}
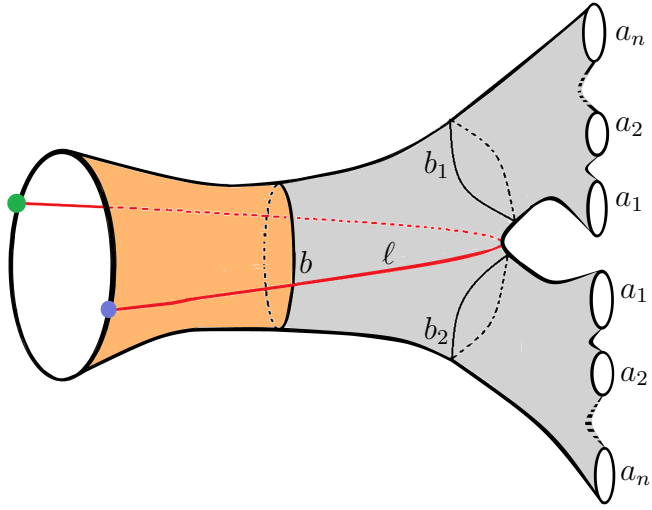
\begin{figure}[h]
	\centering
	\begin{overpic}
		[width=.64\textwidth,tics=6]{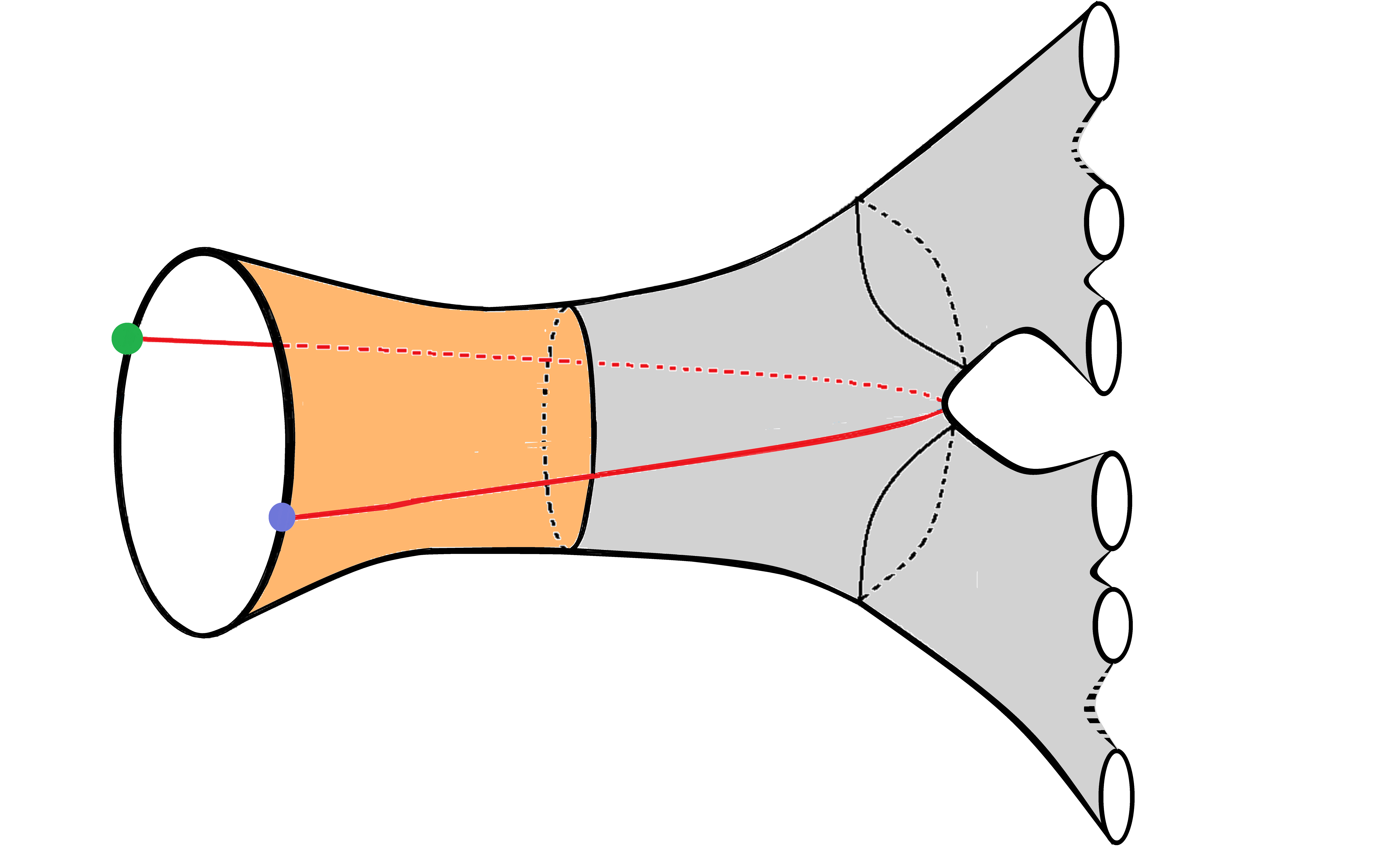}	
		\put (81,56) {$\displaystyle a_{n}$}	
		\put (53,29) {$\displaystyle\ell$}
		\put (58,20) {$\displaystyle b_2$}
		\put (58,40) {$\displaystyle b_1$}
		\put (43,28) {$\displaystyle b$}
		\put (81,45) {$\displaystyle a_2$}
		\put (81,36) {$\displaystyle a_1$}
		\put (81.6,25) {$\displaystyle a_1$}
		\put (81.6,15) {$\displaystyle a_2$}
		\put (81.6,3) {$\displaystyle a_n $}
	\end{overpic}
	\caption{A trumpet glued to a $2n+1$-holed sphere. The endpoints of the geodesic $\ell$  are at the boundary of the
		trumpet region.}
	\label{fi}
\end{figure}

Now, we will explore the additional terms that arise from the small $b$ region.
Let us  consider the geometry, which consists of a trumpet with a closed geodesic of length $b$ and $``2n+1"$-holed sphere with $\sum_{i=1}^{n}a_i>\ell_t$, as shown in figure \ref{fi}.
The  length probability distribution  is described
 by the following expression:	
\begin{align}\label{exp1}
	\hat{p}(\ell,\textbf{a})=  e^{ -S_{0}(2n-1)}\int& \frac{\text{d}\beta}{2\pi i E}\text{d}\ell_1\text{d}\ell_2\text{d}b \text{d}\tau  \text{d}b_2\text{d}\tau_2\text{d}b_1\text{d}\tau_1 e^{\beta E}\langle\frac{\beta}{2}+it|\ell_{2}\rangle\langle \ell_{2},b|\ell_{1}\rangle\langle \ell_1|\frac{\beta}{2}+it\rangle 	V_{0,n+1}(b_1,\textbf{a})	\nonumber\\&V_{0,n+1}(b_2,\textbf{a})\delta\left(\ell-\ell\left(\text{moduli} \right)  \right) \delta\left(b_1
	+2\lambda-\sum_{i=1}^{n}a_i \right)\delta\left(b_2+2\lambda-\sum_{i=1}^{n}a_i \right).
\end{align}
In the above relation, $V_{0,n+1}(b_1,\textbf{a})$ and $V_{0,n+1}(b_2,\textbf{a})$ denote the volumes of moduli spaces corresponding to $``n+1"$-holed spheres attached to geodesics $b_1$ and $b_2$, respectively. 
In the expression \eqref{exp1}, analogous to the case of the five-holed sphere given by \eqref{exp}, two delta functions constrain the lengths of $b_1$ and $b_2$ in terms of all baby universes and their shared sections. Integrating over $ \ell $ removes the delta function, and one can use the orthogonality of wave functions
 and expression \eqref{ff} to get:
\begin{align}
	\int_{-\infty}^{\infty}	\hat{p}(\ell,\textbf{a})\text{d}\ell	&=  \frac{-e^{ -S_{0}(2n-1)}}{4\pi E^{5/2} }\left(\sum_{i=1}^{n}a_i -2\lambda\right)^{2}V_{0,n+1}^{2}\left( \sum_{i=1}^{n}a_i -2\lambda,\textbf{a}\right)\nonumber\\&\approx\frac{-e^{ -S_{0}(2n-1)}}{4\pi E^{5/2} }\mathcal{F}^{2}\left(\textbf{a}\right) \left(\sum_{i=1}^{n}a_i -2\lambda\right)^{2(2n-1)},
	\end{align}
 which is also negative.	
By following the same procedure described in section \ref{sec},  the expression \eqref{exp1} leads to:
\begin{align}\label{fy5}
	\hat{p}(\ell,\textbf{a})&\approx
	e^{ -S_{0}(2n-1)} \frac{e^{-2\pi\sqrt{E}}}{ 2E^{2}}\left(\ell+\ell_t\right)^{2} V_{0,n+1}^{2}(\ell+\ell_t,\textbf{a})\delta\left(\ell+\ell_t+2\lambda -\sum_{i=1}^{n}a_i \right)\nonumber\\&\approx
	e^{ -S_{0}(2n-1)} \frac{e^{-2\pi\sqrt{E}}}{ 2E^{2}}\left(\ell+\ell_t\right)^{4n-6} \mathcal{F}^{2}\left(\textbf{a}\right)\delta\left(\ell+\ell_t+2\lambda -\sum_{i=1}^{n}a_i\right),
\end{align}
this corresponds to the delta function in \eqref{nh}, which characterizes the firewall geometry. 
	
\section{Discussion} \label{4}

An old black hole may emit baby universes and undergo a tunneling process, transforming into a white hole with a firewall at its horizon. The probability of this transition, which involves the emission of a single baby universe corresponding to a genus one surface, was computed in \cite{Stanford:2022fdt}. This probability increases with time and reaches order one when the age of the black hole approaches $e^{S(E)}$.
Furthermore, it was  demonstrated that the probability of having a smooth horizon is equal to the probability of encountering a firewall. This observation was not apparent during intermediate stages since the computation of $P_{1,\text{smooth}}$ did not involve any mapping class group issues.
In this note, an attempt was made to extend these calculations to higher genus scenarios. The results show that the probability distributions of finding  firewall  and smooth horizon, after emitting $n$ baby universes, exhibit a similar behavior:
\begin{equation}
	e^{-2nS(E)}e^{4 \pi \sqrt{E}(n-1)}E^{2n^2-n-9/2} t ^{4n^2-2n-5},
\end{equation}
up to a numerical coefficient.
 In our calculations, we employed the thin strip approximation and, using the ``no shortcut'' condition as defined in \cite{Stanford:2022fdt}, we constrained the moduli space of higher genus geometries. However, it appears that this condition alone does not determine all parameters, requiring additional physical constraints.  By precisely defining the domain of the moduli space, particularly by specifying constraints on the region between baby universes denoted as ``$\lambda$'', it may be feasible to determine numerical coefficients more accurately.
For genus two, the requirement that $P_{2,\text{firewall}}(t) = P_{2,\text{smooth}}(t)$ implies that $0 < \lambda < \ell_t$ in smooth geometry and $0 < \lambda < \ell_t/2$ in firewall geometry. Strictly speaking, the probability of having a firewall-free geometry with a wormhole length of $\ell$ is given by:
\begin{align}
	\hat{P}_{2,\text{smooth}}\left( \ell\right) =\int^{\ell_t }_{0}		\hat{P}_{2,\text{smooth}}\left( \ell,\lambda\right)  \text{d}\lambda=e^{-3S_{0}-2 \pi  \sqrt{E}}\frac{(\ell_t-\ell)^2}{12E^2}\ell_t (2 \ell_t-\ell)  \left(\ell^2-4 \ell \ell_t+5 \ell_t^2\right).
\end{align}
The probability of finding a firewall-free geometry is: 
\begin{align}\label{smo}
	\hat{P}_{2,\text{smooth}}=\int_{0}^{\ell_t}\hat{P}_{2,\text{smooth}}\left(\ell\right)\text{d}\ell= e^{-3S_{0}-2 \pi  \sqrt{E}}\frac{73 \ell_t^7}{360E^2},
\end{align}
when multiplied by the normalization factor $e^{-S(E)}$, as defined in equation \eqref{se}, one can obtain:
\begin{align}\label{smowth}
	P_{2,\text{smooth}}= e^{-4S(E)+4\pi\sqrt{E}}\frac{146 E^{3/2}}{45(2\pi)^{6}}t^7.
\end{align}
The probability of encountering a firewall with a wormhole of length $\ell$ is:
\begin{align}\label{firo}		
	\hat{P}_{2,\text{firewall}}(\ell)=\int_{0}^{\ell_t/2}	\hat{P}_{2,\text{firewall}}(\ell,\lambda)\text{d}\lambda= 
	e^{-3S_{0}-2 \pi  \sqrt{E}}\frac{\left(\ell+\ell_t \right)^{2}}{48 E^2} \ell_t(5\ell_t^3-2\ell^3-9\ell_t\ell^2+6\ell\ell_t^2),	
\end{align} 
and the probability of finding a firewall geometry is: 
\begin{align}
	{	\hat{P}_{2,\text{firewall}} =\int^{\ell_t}_{0 }		\hat{P}_{2,\text{firewall}}\left( \ell\right)  \text{d}\ell= e^{-3S_{0}-2 \pi  \sqrt{E}}\frac{277\ell_t^{7}}{ 1440E^2}}.
\end{align}
After normalization we obtain:
\begin{align}\label{fmire}
	{	P_{2,\text{firewall}}=e^{-4S(E)+4\pi\sqrt{E}}\frac{277 E^{3/2}}{90(2\pi)^{6}}t^7}.
\end{align}
By comparing equation \eqref{fmire} with \eqref{smowth}, it can be seen that  $P_{2,\text{firewall}}(t)=  0.95P_{2,\text{smooth}}(t)$.
If we set the upper bound of $\lambda$ to $\ell_t$ for both geometries, we find that $P_{2,\text{firewall}}(t)\approx3P_{2,\text{smooth}}(t)$.

The calculation of the probability of tunneling is reminiscent of the spectral form factor, $|\langle \beta+i t^{\prime}|\beta+it\rangle|^{2}$, which is an important tool for the geometrical interpretation of the discrete energy spectrum of black holes.   After some initial non-universal decay, the spectral form factor has a linearly growing ramp for  a long amount of time $2\pi e^{S(E)}$ and then it flattens out into a plateau. This late time behavior of the spectral form factor is universal and depends on density of states, $\rho(E)\approx  e^{S(E)}$, and symmetry classification.
Cylindrical topology gives the linear ramp and there is no perturbative correction to it and quadratic answer for the probability of finding  firewall geometry for genus one  arises from its integration. Plateau is non-perturbation in genus expansion, although  there have been some attempts to find a perturbative approach to the late-time plateau  \cite{Blommaert:2022lbh,Weber:2022sov,Saad:2022kfe}, see also \cite{Yan:2022nod}. So, an intriguing question would be to find a convergent sum over the genera of firewall geometries.

From the second method of computing the probability of encountering a firewall geometry for genus two, similar to the case with firewall geometry of genus one, an additional negative term at small $b$ was observed. These terms are expected to contribute something proportional to $\delta\left(\ell-\ell_t \right)$, with the coefficient being determined by the requirement that the total probability equals one ($\int_{0}^{\ell_t} P(\ell)d\ell=1$). A detailed analysis of the negative terms was not carried out due to the belief that these terms effectively subtract a probability mass equivalent to that of the positive terms from the disk answer.
To provide a specific formula, after combining the disk answer ($P_{\text{disk}}(\ell)=\delta (\ell-\ell_t)$) with the contributions from the firewall and smooth geometries, the probability distribution for the physical length should take the following form:
\begin{align}\label{pe}
P\left(\ell \right) =&\left( 1-\frac{e^{-2S\left( E\right) }}{(2\pi)^2}t^2-e^{-4S(E)+4\pi\sqrt{E}}\frac{569 E^{3/2}}{90(2\pi)^{6}}t^7\right) \delta\left(\ell-\ell_t \right)+\bigg( \frac{e^{-2S\left( E\right) }}{2E(2\pi)^2}+\frac{e^{-4S(E)+4\pi\sqrt{E}}}{6(4\pi)^{6}E^2}\ell_t \nonumber\\ &\hspace{1cm}\bigg\lbrace 6\ell^4-13 \ell^3 \ell_t+105 \ell^2 \ell_t^2-71 \ell \ell_t^3+45 \ell_t^4\bigg\rbrace \bigg) \left( \ell_t-\ell\right)\theta\left( \ell_t-\ell\right) +O\left(e^{-6S\left( E\right) } \right).
\end{align}
The first term signifies the contribution of the disk, including  possible negative contributions from the handle disk and double handle disk. The second term includes $P_{1,\text{firewall}}$ and an identical term for $P_{1,\text{smooth}}$, along with $P_{2,\text{firewall}}$ and $P_{2,\text{smooth}}$, i.e. \eqref{smo} and \eqref{firo}. 
Using the relation \eqref{pe}, the complexity of the dual state can be computed as the expectation value of the wormhole length. In \cite{Iliesiu:2021ari, Alishahiha:2022exn}, complexity was calculated by determining the expectation value of all possible geodesics without applying the no shortcut criterion. It was observed that, following a period of linear growth at early times, the complexity saturates at late times. If linear growth is considered an essential feature of complexity, it suggests that the no shortcut criterion may be disregarded  for computing complexity \cite{Belin:2021bga}.
It is worth noting that the volume of wormholes  in all classical geometries does not exhibit linear growth forever. For instance, in multi-black hole geometries \cite{Skenderis:2009ju}, it does not follow a linear growth pattern and also saturates at late times \cite{Zolfi:2023bdp}. It would be interesting to examine these geometries and study the tunneling process.

\section*{Acknowledgments}
Special thanks to Ali Naseh for the discussions and encouragement. The author would like to
extend his appreciation to,  Mohsen Alishahiha, Alessandro Giacchetto, and Phil Saad for constructive discussions on related subjects.
\appendix
\section{The approximation  length
	of $ \ell\left(\text{moduli} \right)  $}\label{apen}

\begin{figure}[h]
	\centering
	\begin{overpic}
		[width=.53\textwidth,tics=0]{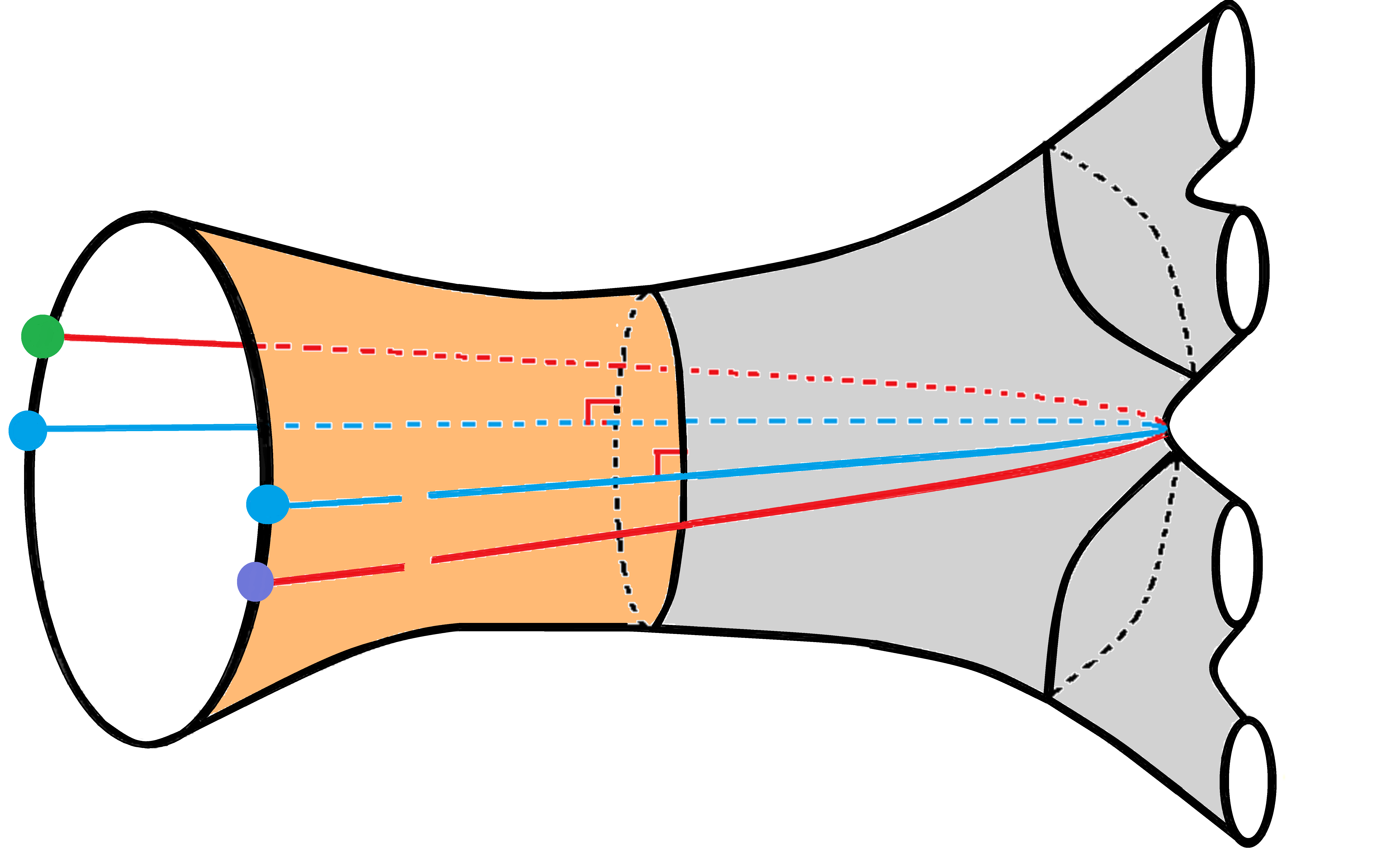}	
		\put (28.9,19) {$\displaystyle\ell$}
		\put (27.99,25) {$\displaystyle\gamma$}
		\put (51,27) {$\displaystyle b,\tau$}
		\put (65,40) {$\displaystyle b_1,\tau_1$}
		\put (65,17) {$\displaystyle b_2,\tau_2$}
		\put (-22,37) {$\displaystyle \color{red}(\sigma_{2},\frac{b}{2}+x_2)$}
		\put (-13,27) {$\displaystyle \color{blue}(\sigma_{2},\frac{b}{2})$}
		\put (5.8,24) {$\displaystyle \color{blue}(\sigma_{1},0)$}
		\put (3.6,15) {$\displaystyle\color{red} (\sigma_{1},x_{1})$}
		\put (92,21) {$\displaystyle a_2$}
		\put (92,41) {$\displaystyle a_2$}
		\put (92,55) {$\displaystyle a_1$}	
		\put (93,5) {$\displaystyle a_1$}						
	\end{overpic}
	\caption{A five holed sphere glued to a trumpet. The blue geodesic $\gamma$ is perpendicular to geodesic (loop) $b$, with its endpoints positioned at the boundary of the trumpet region.}
	\label{gamma}
\end{figure}

\begin{figure}[h]
	\centering
	\begin{overpic}
		[width=.3\textwidth,tics=6]{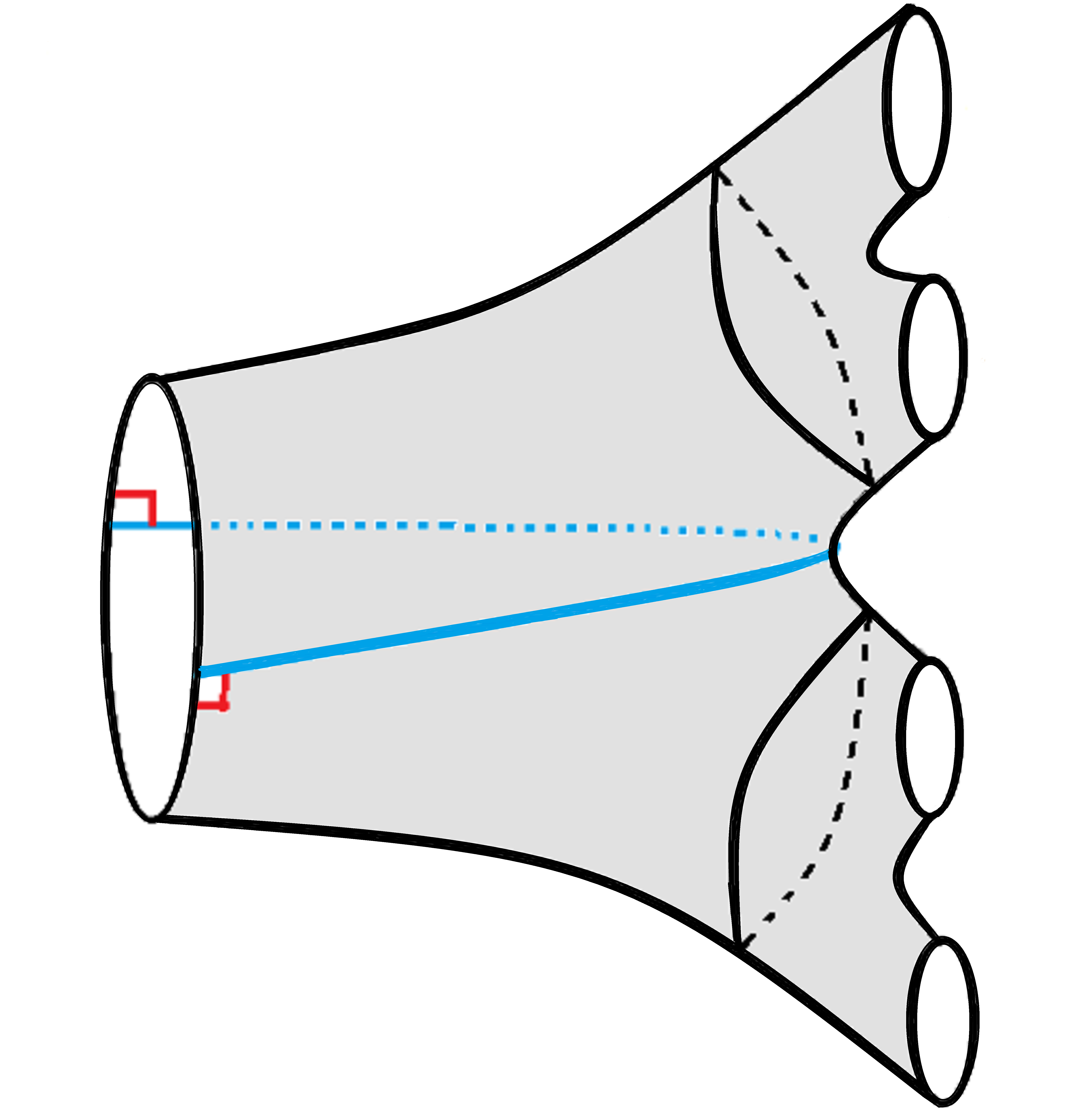}	
		\put (91,8) {$\displaystyle a_{1}$}	
		\put (43,38) {$\displaystyle\gamma_2$}
		\put (90,87) {$\displaystyle a_1$}
		\put (90,28) {$\displaystyle a_2$}
		\put (92,64) {$\displaystyle a_2 $}
		\put (3,40) {$\displaystyle b $}
		\put (57,25) {$\displaystyle b_2 $}
		\put (57.5,65) {$\displaystyle b_1 $}
	\end{overpic}
	\caption{In five holed sphere, the $\gamma_{2}$ is a portion of geodesic $\gamma$ which is perpendicular to $ b $.}
	\label{gamma2}
\end{figure}

	The $  \ell\left(\text{moduli} \right)$ can be derived by representing the geometry in figure~\ref{five} as a quotient of the full hyperbolic space. But it is quite complicated, so one can approximate $  \ell\left(\text{moduli} \right)$ by another geodesic which is perpendicular to $ b $ and  its endpoints situated on the boundary of the trumpet region, as shown in figure \ref{gamma}.
	The length of the portion
	of $ \gamma $  that lies within the five-holed sphere  is denoted by $`` \gamma_2 "$ and it is perpendicular to the geodesic  $ b $ as shown in figure \ref{gamma2}.
	The $\gamma_2$ as a function of $b$, $b_1$ and $b_2$ is:
	\begin{align}\label{gd}
		\gamma_2 = 2\sinh ^{-1}\left(\text{csch}\frac{b}{2}\sqrt{\cosh ^2\left(\frac{b_1}{2}\right)+\cosh
			^2\left(\frac{b_2}{2}\right)+2 \cosh \frac{b}{2}
			\cosh \frac{b_2}{2} \cosh
			\frac{b_1}{2}}
		\right).
	\end{align}		
	The remaining part of $ \gamma $ which is located in the trumpet region  is denoted by $ 	\gamma_1 $ and using  \eqref{metric}  we have:
	\begin{equation}\label{gam1}
		\gamma_1=\sigma_1+\sigma_2=\sigma_{+}.
	\end{equation}	
The portion of $\ell$ located between geodesic $b$ and the boundary is approximately $\sigma_1+\sigma_2+2\log\cosh\frac{x_1}{2}+2\log\cosh\frac{x_2}{2}$. If we approximate the portion of $\ell$ within the five-holed sphere with $\gamma_2$, the $\ell$ geodesics after holographic regularization becomes:
\begin{align}\label{mwod}
	\ell\left(\text{moduli} \right)\approx \hat{\sigma}_{+}+2\log\cosh\frac{x_1}{2}+2\log\cosh\frac{x_2}{2}+ \gamma_2.
\end{align}
\begin{figure}[H]
	\centering
	\begin{overpic}
		[width=.32\textwidth,tics=9]{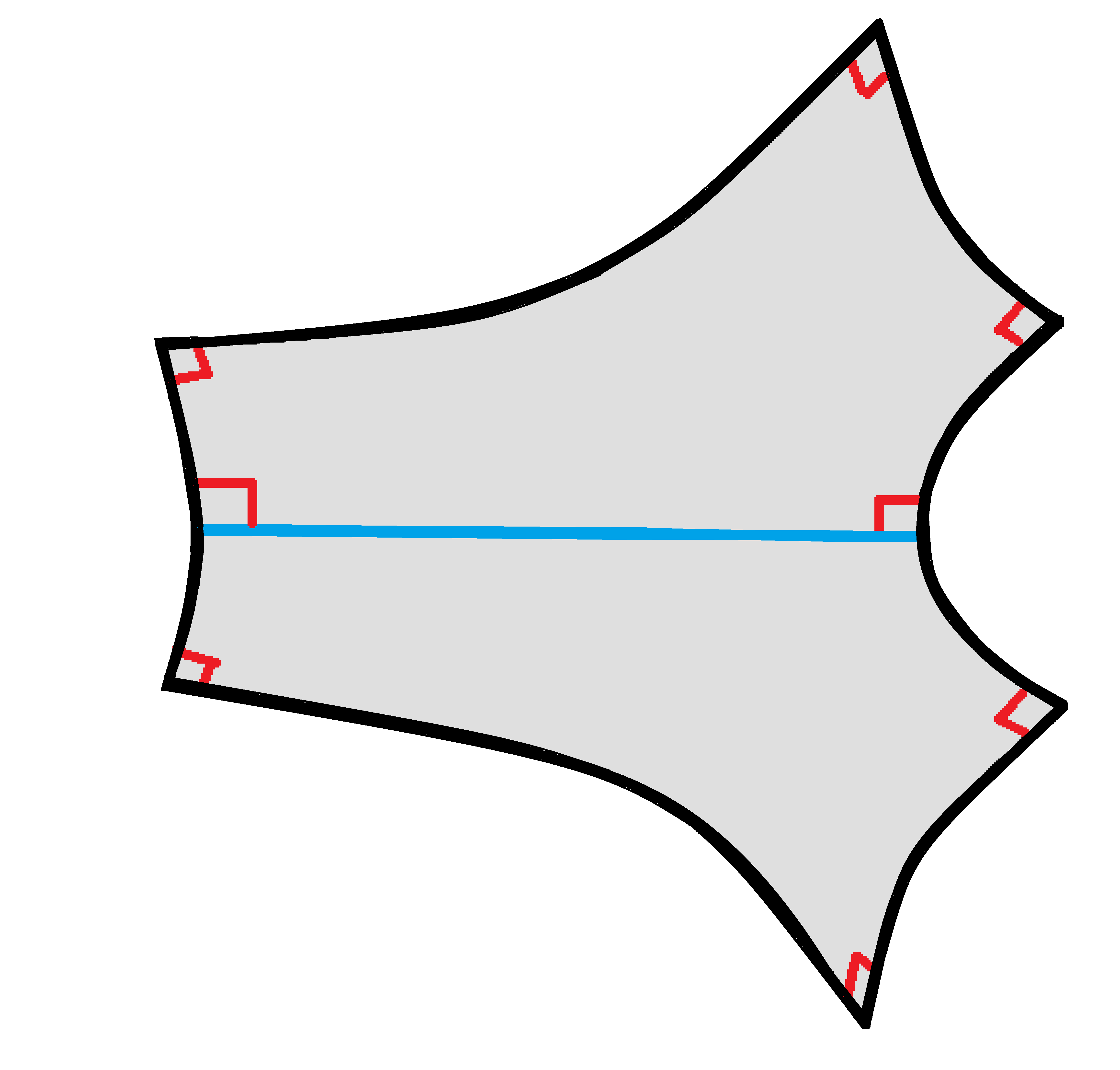}	
		\put (43,37) {$\displaystyle\frac{\gamma_{2}}{2}$}
		\put (19,40) {$\displaystyle \beta $}
		\put (19,55) {$\displaystyle \alpha $}
		\put (7,47) {$\displaystyle \frac{b}{2} $}
		\put (88,14) {$\displaystyle \frac{b_2}{2} $}
		\put (88.5,76) {$\displaystyle \frac{b_1 }{2}$}
	\end{overpic}
	\caption{Two hexagons can be obtained by symmetrically cutting a pair of pants with geodesic boundaries $(b, b_1, b_2)$.}
	\label{gamq2}
\end{figure}
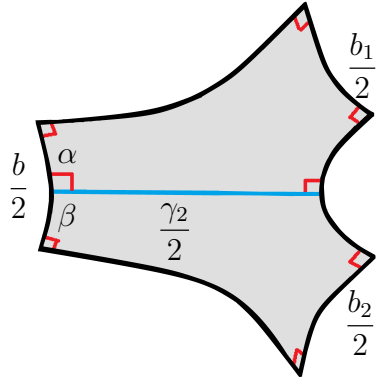
Now we calculate the length of $\gamma_2$.  Referring to figure \ref{gamma2}, we choose the pair of pants on the five-holed sphere with geodesic boundaries $(b, b_1, b_2)$ that contains the geodesic $\gamma_2$. After cutting this pair of pants so that each geodesic boundary is divided in half, two hexagons are obtained, with one of them represented in figure \ref{gamq2}.
The geodesic with a length of $\gamma_2/2$ bisects this hexagon, creating two pentagons. In the upper and lower pentagons, the lengths of a side are denoted by $\alpha$ and $\beta$ respectively, and their sum equals $b/2$.
By a standard hyperbolic trigonometric formula for pentagons (for example,
consider the references \cite{th,bu}) we have  the following relations:
\begin{equation*}
	\alpha+\beta=\frac{b}{2}
\end{equation*}
\begin{equation}
	\sinh\alpha\sinh\frac{\gamma_{2}}{2}=\cosh \frac{b_1}{2}
\end{equation}
\begin{equation*}
	\sinh\beta\sinh\frac{\gamma_{2}}{2}=\cosh \frac{b_2}{2},
\end{equation*}	 
and by combining these equations, we arrive at equation \eqref{gd}.
\section{Integrals with  step functions }\label{bqw} 
To perform integrals in \eqref{sq}, let us consider the following integral:
\begin{align}\label{sqy}
& \int_{0}^{a_j}\text{d}s_j~ \theta\left(\ell_t+2\lambda -\sum_{i=1}^{n}s_i\right) \theta\left( \sum_{i=1}^{n}s_i-\ell-2\lambda\right). 
\end{align}
Using the integral representation of the  step function:
\begin{align}
	\theta(x)=\lim_{\epsilon\rightarrow0^{+}}\frac{1}{2\pi i}\int_{-\infty}^{\infty}\text{d}\tau\frac{e^{ix\tau}}{\tau-i\epsilon},
\end{align}
one can rewrite \eqref{sqy} as:
\begin{align}
&	\lim_{\epsilon\rightarrow0^{+}}\frac{1}{\left( 2\pi i\right)^{2} }\int_{0}^{a_j}\text{d}s_j\int_{-\infty}^{\infty}\text{d}\tau\text{d}\tau^{\prime}\frac{e^{i\tau(\ell_t+2\lambda-s_j)}}{\tau-i\epsilon}\frac{e^{i\tau^{\prime}(s_j-\ell-2\lambda)}}{\tau^{\prime}-i\epsilon}\nonumber\\&=	\lim_{\epsilon\rightarrow0^{+}}\frac{-i}{\left( 2\pi i\right)^{2} }\int_{-\infty}^{\infty}\text{d}\tau\text{d}\tau^{\prime}\frac{e^{i\tau(\ell_t+2\lambda)}}{\tau-i\epsilon}\frac{e^{i\tau^{\prime}(-\ell-2\lambda)}}{\tau^{\prime}-i\epsilon}\frac{e^{ia_j(\tau^{\prime}-\tau)}-1}{\tau^{\prime}-\tau}.
\end{align}
So, for $\alpha$ and $\beta$ both equal to $n-2$, \eqref{sq} becomes:
\begin{align}\label{wsqy}
		&	\lim_{\epsilon\rightarrow0^{+}}\frac{(-i)^{n}}{\left( 2\pi i\right)^{2} }\int_{-\infty}^{\infty}\text{d}\tau\text{d}\tau^{\prime}\frac{e^{i\tau(\ell_t+2\lambda)}}{\tau-i\epsilon}\frac{e^{i\tau^{\prime}(-\ell-2\lambda)}}{\tau^{\prime}-i\epsilon} \frac{1}{(\tau^{\prime}-\tau)^{n}}\nonumber\\
		&\hspace*{2.5cm}\int_{0}^{\infty}\prod_{j=1}^{n}	\text{d}a_ja_{j}^{4(n-2)}\left(e^{ia_j(\tau^{\prime}-\tau)}-1 \right)     \delta\left(  \sum_{i=1}^{n}a_i-2\lambda-(\ell_t+\ell)\right).
\end{align}
Using the integral representation of the  delta function:
\begin{align}
	\delta(x)=\frac{1}{2\pi }\int_{-\infty}^{\infty} e^{ix\tau^{\prime\prime}}\text{d}\tau^{\prime\prime},
\end{align}
relation \eqref{wsqy} becomes:
\begin{align}\label{wsqyu}	\lim_{\epsilon\rightarrow0^{+}}\frac{-(-i)^{n}}{\left( 2\pi \right)^{3} }\int_{-\infty}^{\infty}\text{d}\tau\text{d}\tau^{\prime}\text{d}\tau^{\prime\prime}\frac{e^{i\tau(\ell_t+2\lambda)}}{\tau-i\epsilon}\frac{e^{i\tau^{\prime}(-\ell-2\lambda)}}{\tau^{\prime}-i\epsilon} \frac{e ^{-i\tau^{\prime\prime}\left(  2\lambda+\ell_t+\ell\right)}}{(\tau^{\prime}-\tau)^{n}}\int_{0}^{\infty}\prod_{j=1}^{n}	\text{d}a_ja_{j}^{4(n-2)}\left(e^{ia_j(\tau^{\prime}+\tau^{\prime\prime}-\tau)}-e^{i\tau^{\prime\prime}a_j} \right),	  
\end{align}
and integrating over $a_j$'s can be performed as following:
\begin{align}
\int_{0}^{\infty}\prod_{j=1}^{n}	\text{d}a_ja_{j}^{4(n-2)}\left(e^{ia_j(\tau^{\prime}+\tau^{\prime\prime}-\tau)}-e^{i\tau^{\prime\prime}a_j} \right)&=\left\lbrace\int_{0}^{\infty}\text{d}a_j\frac{\partial^{4(n-2)}}{\partial\tau^{\prime\prime{4(n-2)}}}	\left(e^{ia_j(\tau^{\prime}+\tau^{\prime\prime}-\tau)}-e^{i\tau^{\prime\prime}a_j} \right) \right\rbrace ^{n}
\nonumber\\&=\left\lbrace\frac{\partial^{4(n-2)}}{\partial\tau^{\prime\prime{4(n-2)}}}\left(\frac{i}{\tau^{\prime}+\tau^{\prime\prime}-\tau}-\frac{i}{\tau^{\prime\prime}} \right) \right\rbrace^{n} 
\nonumber\\&=\left(\frac{i}{(\tau^{\prime}+\tau^{\prime\prime}-\tau)^{4(n-2)+1}}-\frac{i}{\tau^{\prime\prime4(n-2)+1}} \right)^{n}. 
\end{align}
Therefore \eqref{wsqy} simplifies to:
\begin{align}\label{wsyu}	&\lim_{\epsilon\rightarrow0^{+}}\frac{-1}{\left( 2\pi \right)^{3} }\int_{-\infty}^{\infty}\text{d}\tau\text{d}\tau^{\prime}\text{d}\tau^{\prime\prime}\frac{e^{i\tau(\ell_t+2\lambda)}}{\tau-i\epsilon}\frac{e^{i\tau^{\prime}(-\ell-2\lambda)}}{\tau^{\prime}-i\epsilon} \frac{e ^{-i\tau^{\prime\prime}\left(  2\lambda+\ell_t+\ell\right)}}{(\tau^{\prime}-\tau)^{n}}\left(\frac{\Gamma(4(n-2)+1)}{(\tau^{\prime}+\tau^{\prime\prime}-\tau)^{4(n-2)+1}}-\frac{\Gamma(4(n-2)+1)}{\tau^{\prime\prime4(n-2)+1}} \right)^{n}.
\end{align}

\bibliographystyle{C:/Users/Hamed/Desktop/draft/utphys.bst}
\bibliography{C:/Users/Hamed/Desktop/draft/ref.bib}

\end{document}